\documentclass{aa}  

\usepackage{tablefootnote}
\usepackage{graphicx, color}
\usepackage[space]{grffile}
\usepackage{latexsym}
\usepackage{float}
\usepackage{booktabs}
\usepackage{amsfonts,amsmath,amssymb}
\usepackage{url}
\usepackage{fancyref}
\usepackage{multirow}
\usepackage{xcolor}
\usepackage{txfonts}
\begin{document} 

\title{One- and Two-point Source Statistics from the LOFAR Two-metre Sky Survey First Data Release}
   \titlerunning{LoTSS -- Statistics}

\author{T.~M.~Siewert\inst{1}\thanks{E-mail: t.siewert@physik.uni-bielefeld.de} \and C.~Hale\inst{2,3} \and N.~Bhardwaj\inst{1} \and M.~Biermann\inst{1} \and
D.~J.~Bacon\inst{4} \and M.~Jarvis\inst{2,5} \and H.~J.A.~R\"ottgering\inst{6} \and D.~J.~Schwarz\inst{1} \and
T.~Shimwell\inst{7,6} \and P.~N.~Best\inst{8} \and K.~J.~Duncan\inst{6} \and M.~J.~Hardcastle\inst{9} \and J.~Sabater\inst{8} \and C.~Tasse\inst{10,11} \and G.~J.~White\inst{12,13} \and W.~L.~Williams\inst{6} }
	\authorrunning{Siewert et al.}

\institute{$^1$Fakult\"at f\"ur Physik, Universit\"at Bielefeld, Postfach 100131, 33501 Bielefeld, Germany \\
         $^2$Astrophysics, University of Oxford, Denys Wilkinson Building, Keble Road, Oxford, OX1 3RH, UK \\
         $^3$CSIRO Astronomy and Space Science, PO Box 1130, Bentley, WA 6102, Australia\\
         $^4$Institute of Cosmology\! \&\! Gravitation, University of Portsmouth, Dennis Sciama Building, Burnaby Road, Portsmouth PO1 3FX, UK \\
         $^5$Department of Physics \& Astronomy, University of the Western Cape, Private Bag X17, Bellville, Cape Town, 7535, South Africa\\
         $^6$Leiden Observatory, Leiden University, PO Box 9513, NL-2300 RA Leiden, The Netherlands \\
         $^7$ASTRON, the Netherlands Institute for Radio Astronomy, Postbus 2, 7990 AA, Dwingeloo, The Netherlands \\
         $^8$SUPA, Institute for Astronomy, Royal Observatory, Blackford Hill, Edinburgh, EH9 3HJ, UK\\
         $^9$Centre for Astrophysics Research, School of Physics, Astronomy and Mathematics, University of Hertfordshire, College Lane, Hatfield AL10 9AB, UK\\
         $^{10}$GEPI \& USN, Observatoire de Paris, Université PSL, CNRS, 5 Place Jules Janssen, 92190 Meudon, France\\
         $^{11}$Department of Physics \& Electronics, Rhodes University, PO Box 94, Grahamstown, 6140, South Africa\\
         $^{12}$RAL Space, The Rutherford Appleton Laboratory, Chilton, Didcot OX11 0NL, UK\\
         $^{13}$Department of Physical Sciences, The Open University, Walton Hall, Milton Keynes MK7 6AA, UK\\
        }

\date{\today}

\abstract{
The LOFAR Two-metre Sky Survey (LoTSS) will eventually map the complete Northern sky and 
provide an excellent opportunity to study the distribution and evolution of the large-scale structure of 
the Universe.}
{We test the quality of LoTSS observations through statistical comparison of the LoTSS first data release (DR1) catalogues to 
expectations from the established cosmological model of a statistically isotropic and homogeneous Universe.
}
{We study the point-source completeness and define several quality cuts, in order to determine the count-in-cell 
statistics and differential source counts statistic and measure the angular two-point correlation function.
We use the photometric redshift estimates which are available for about half of the LoTSS-DR1 radio sources, to compare the 
clustering throughout the history of the Universe.}
{For the masked LoTSS-DR1 value-added source catalogue we find 
point-source completeness of 99\% above flux densities of $0.8$ mJy.  
The counts-in-cell statistic reveals that the 
distribution of radio sources cannot be described by a spatial Poisson process. Instead, a good fit is provided by a compound 
Poisson distribution. The differential source counts are in good agreement with previous 
findings in deep fields at low 
radio frequencies and with simulated catalogues from the SKA design study sky and the Tiered Radio Extragalactic Continuum Simulation.
Restricting the value added source catalogue to low-noise regions and applying a flux density threshold of 2 mJy provides our most reliable estimate 
of the angular two-point correlation. 
Based on the distribution of photometric redshifts and 
the Planck 2018 best-fit cosmological model, the theoretically predicted angular two-point correlation between 0.1 deg and 6 deg agrees 
reasonably well with the measured clustering for the subsample of radio sources with redshift information.}
{The deviation from a nmhg distribution 
might be a consequence of the multi-component nature of a 
large number of resolved radio sources and/or of uncertainties on the flux density calibration. 
The angular two-point correlation function is $<10^{-2}$ at angular scales 
$> 1$ deg and up to the largest scales probed. At 2~mJy flux density threshold and at an pivot angle of 1 deg 
we find a clustering amplitude of $A=(5.1\pm 0.6)\times10^{-3}$ with a slope parameter of $\gamma=0.74\pm0.16$. For smaller flux density 
thresholds 
systematic issues are identified, most likely related to the flux density calibration of the individual pointings.
We conclude that we find agreement with the expectation of large-scale statistical isotropy of the radio sky at the per cent level. 
The angular two-point correlation agrees well with the expectation of the cosmological standard model.}


\maketitle

\section{Introduction}

The LOFAR Two-metre Sky Survey (LoTSS)\footnote{\url{www.lofar-surveys.org}} will provide the deepest and 
best resolved inventory of the radio sky at low frequencies over the coming decades \citep{LoTSS2017}. 
Having already produced high fidelity images and catalogues over $424$ square degrees at a central 
frequency of $144$~MHz  \citep{LoTSS2019A}, LoTSS will continue to produce a catalogue that is estimated 
to contain about 15 million radio sources over all of the Northern hemisphere. 
A large fraction of those sources will come with optical identifications \citep{LoTSS2019B} and photometric 
redshifts \citep{LoTSS2019C}. Already, for the first data release, about half of the radio sources have measured
photometric redshifts.
In addition to this, the WEAVE-LOFAR survey \citep{LOFARWEAVE2016} will 
measure spectroscopic redshifts for about a million sources from the LoTSS catalogue.  
The survey is therefore expected to provide a rich resource not only 
for astrophysics, but also for cosmology, see
e.g.~\citet{Raccanelli2012}, \cite{Camera2012}, \cite{Jarvis2015} and \cite{AASKA2015}. 
Together with photometric redshifts and, at a later stage, spectroscopic 
redshifts, we will be able to measure the luminosity and number density 
evolution directly, and through a clustering analysis will also be able to measure 
the relative bias between the different radio source populations.

Extragalactic radio sources are tracers of the large scale structure of 
the Universe. The evolution of the large scale structure 
in turn depends on many fundamental parameters; for example it depends on the 
model of gravity, the proportion of visible and dark matter 
as well as dark energy, and the primordial curvature fluctuations. Unfortunately, 
these dependencies are blended with unknowns from astrophysics such as the bias factors for 
active galactic nuclei (AGN) and starforming galaxies (SFG), their number density 
and luminosity evolutions. The purpose of this work is to make a first step towards the 
cosmological analysis of LoTSS.

For cosmological studies, surveys must cover a sizeable fraction of the sky
and sample the sky fairly homogeneously, down to some minimal flux density.
Currently available radio surveys in the LoTSS frequency range are the
TIFR GMRT Sky Survey (TGSS-ADR1; \citealt{TGSS2017}) and GaLactic and
Extragalactic All-sky MWA survey (GLEAM; \citealt{GLEAM2017}).
The first alternative data release of the TGSS covers 36\,900 square degrees
of the sky at a central frequency of $147.5$~MHz and at an 
angular resolution of $25\arcsec$. A $7$-sigma detection limit with a 
median rms noise of $3.5$~mJy/beam results in 623\,604 sources.
Comparing the measured TGSS source counts to SKADS 
(SKA Design Study, \citealt{SKADS2008}) sky simulations 
shows good agreement for flux density thresholds above $100$~mJy.
The GLEAM catalogue covers $24\,831$ square degres and contains $307\,455$ 
sources with 20 separate flux density measurements between $72$~MHz and 
$231$~MHz, centred at $200$~MHz at an angular resolution of $2\arcmin$.
The catalogue is estimated to be $90\%$ complete at a flux density threshold 
of $170$~mJy in the entire survey area for a $5$-sigma detection limit.
The rms noise varies between $10$~mJy/beam and 
$23$~mJy/beam along four declination ranges, 
which complicates the measurements of cosmic structures on large angular scales.   

As LoTSS will eventually cover all of the Northern sky and detect about 15 
million radio sources, it will allow us to overcome statistical
limitations due to shot noise and substantially reduce cosmic variance 
in cosmological analysis, two issues from which contemporary wide area radio continuum 
catalogues suffer.

In this work we study the one- and two-point statistics for the sources in the LoTSS data 
release 1 (DR1).
Covering an area of $424$~square degress over the HETDEX spring field, DR1 contains $325\,694$ radio sources, detected 
by means of {\sc PyBDSF} (Python Blob Detector and Source 
Finder\footnote{\url{http://www.astron.nl/citt/pybdsf/}}, \citealt{PYBDSF2015}) 
with a peak flux density of at least five times the local 
rms noise. The median rms noise in the observed area is $71~\mu$Jy/beam 
at an angular resolution of $6\arcsec$. The LoTSS-DR1 value-added 
catalogue, as described by \citet{LoTSS2019B} removes artefacts 
and corrects wrong groupings of Gaussian components. 
It contains $318\,520$ sources of which $231\,716$ have 
optical/near-IR identifications in Pan-STARRS/WISE.

Before the LoTSS catalogues can be used for cosmological analyses, 
the consistency of the flux density and the 
completeness and reliability of the detected sources must be 
carefully examined. For  
cosmological analysis we are interested in the large scale features on 
the sky, and large scale instrumental or calibration effects must 
be identified and accounted for, before we can draw credible cosmological conclusions. 

The goal of this work is therefore to recover and re-establish the well known 
and tested properties of large-scale structure in the radio sky. The study 
of the one- and two-point number count statistics of 
the LoTSS-DR1 value-added catalogue offers an excellent opportunity to do 
so, and the cleaning and quality control methods presented in this work 
will provide a good basis for future cosmological exploitation of LoTSS.

The potential of radio continuum surveys for cosmology has been studied 
in detail in the context of the SKA, see e.g.\ \citet{Jarvis2015, SKARedBook2018}
and its precursors, among them LOFAR \citep{Raccanelli2012}.
Some of the cosmological SKA science cases can already be tackled by 
LoTSS, even well before regular SKA surveys will start. In the pre-SKA era, 
a key topic of investigation will be to improve our understanding of dark energy 
and modified gravity; these can be parametrized so that we can constrain 
e.g. the equation of state of dark energy and its evolution, the deviation of the 
relationship between density and potential from that expected in the Poisson 
equation, and the ratio of the space- and time-parts of the metric. 
These parameters have observable consequences via their effect on the 
expansion history and/or structure growth history of the Universe. This in turn 
affects the predictions for observable cosmological probes including the 
auto-correlation of source counts, the cross-correlation of source counts 
with the CMB (integrated Sachs-Wolfe effect, \citealt{BallardiniMaartens2019}), and the cross-correlation 
of source counts at different redshifts (which is activated by gravitational 
lensing magnification effects). The radio sky also provides an opportunity 
to constrain primordial non-Gaussianity in the distribution of density 
modes in the Universe \citep{Ferramacho2014,Raccanelli2015}; this is observable as an enhanced autocorrelation 
at large angular scales. 
In addition, very wide surveys can probe the 
kinematic and matter radio dipole \citep{Bengaly2019}, which can act as a fundamental test of the 
cosmological principle. Here we focus on the simplest statistical tests, in 
particular the two-point source count statistics.

In Sect. \ref{sec:lss} we summarize the theoretical expectation for 
the one- and two-point number counts. In Sect. \ref{sec:data} we 
describe how we identify the survey regions that are most reliable, estimate the 
completeness of LoTSS-DR1 and describe the masks and flux density cuts that we apply to the 
data. In order to compare expectation and data 
we generate mock catalogues, which are described in 
Sect.~\ref{sec:mocks}.
The properties of the one-point statistics are discussed in 
Sect.~\ref{sec:onepoint}. For this, we ask if the radio sources in a pixel 
on the sky are drawn from a Poisson process and we investigate the 
differential number counts and then compare them to other surveys and 
to simulations.
In Sect.~\ref{sec:twopoint}, we estimate the two-point 
statistics, 
the angular correlation function, which we fit to a phenomenological model and compare them to findings from previous surveys,
 as well as to the theoretically expected 
angular two-point correlation function based on the Planck 2018 best-fit cosmological model, 
the photometric redshift distribution found for LoTSS-DR1 radio sources and a bias function 
from the literature. We present our conclusions in Sect. \ref{sec:conclusions}.

This work is complemented by four Appendices. In App.~\ref{app:A} we describe a 
masking procedure for the TGSS-ADR1 catalogue that is used for comparison and estimate 
the corresponding angular two-point correlation function.
Five common estimators for the angular two-point correlation function are described 
and compared in the context of LoTSS-DR1 in App.~\ref{app:B}. 
We also test the accuracy of the software package 
{\sc TreeCorr} \citep{TreeCorr2004} that we use 
for the computation of the angular two-point correlation function by means of an independent,  
computationally slow but presumably exact brute force algorithm (App. \ref{app:C}). 
In App. \ref{app:dipole} we show that the contribution of the kinematic radio dipole to the angular 
two-point correlation function is negligible for the angular scales probed in this work.
 
\section{Large scale structure in radio continuum surveys \label{sec:lss}}

Before we investigate the data, we first discuss what the standard model of cosmology 
predicts for the statistical tests that we will consider throughout this work.

\subsection{Source counts in cells}

The cosmological principle is fundamental to modern cosmology, 
stating that on large enough scales the distribution of matter and light is isotropic and 
homogeneous on spatial sections of space-time. Isotropy on 
large scales is observed at a wide range of frequencies, from the 
distribution of radio sources, to the distribution of 
gamma-ray bursts, and is most precisely tested by means of the 
cosmic microwave sky (see e.g.~\citealt{Peebles1993, PlanckIsotropy2015,PlanckIsotropy2018}). 
Therefore we also expect to find an statistically isotropic distribution of 
extragalactic radio sources for LoTSS, i.e.\ the expectation value of 
the number of radio sources per unit solid angle, or surface density 
$\sigma$, with flux density above a certain threshold $S_\mathrm{min}$, is 
independent of the position on the sky $\mathbf{e}$. The number counts 
in a pixel (or cell) of solid angle $\Omega_\mathrm{pix}$ centred at $\mathbf {e}$ 
are
\begin{equation}
  N(\mathbf{e},S_\mathrm{min}) = \int_{\Omega_{\mathrm{pix}}} \!\!
      \sigma(\mathbf{e},S_\mathrm{min}) \mathrm{d}\Omega, 
\end{equation}
with (ensemble) expectation value
\begin{equation}
  \langle N(\mathbf{e},S_\mathrm{min}) \rangle = \bar{N}(S_\mathrm{min}) 
      = \bar{\sigma}(S_\mathrm{min})\Omega_{\mathrm{pix}}.
\end{equation}

The simplest model for the distribution of radio sources assumes that they 
are (i) identically and (ii) independently distributed, and (iii) pointlike (i.e. it is possible to  
reduce the pixel size until each pixel would contain at most one fully contained source).
These assumptions define what is called a homogenous Poisson process 
(see e.g.~\citealt{Peebles1980}). Thus the naive expectation is that the probability 
of finding $k$ sources above a flux density threshold $S_\mathrm{min}$ in any cell 
of fixed size is given by a Poisson distribution with intensity parameter $\lambda$, i.e., 
\begin{equation}
  p_k^{\mathrm{P}} = \frac{\lambda^k}{k!}e^{-\lambda},\label{eq:poisson}
\end{equation}
with expectation $\bar{N} \equiv \mathrm{E}[k] = \lambda$ and variance $\mathrm{Var}[k] = \lambda = \bar{N}$.

Deviations from a Poisson distribution are expected due to 
effects from gravitational clustering of large-scale structure 
[a violation of condition (ii)], resolved sources [a violation of 
condition (iii)], and multi-component sources, such as
FRII radio galaxies in which the radio lobes are not 
statistically independent from each other [violation of condition (ii)]. 
Different types of radio sources could follow different statistical distributions, 
which would then violate condition (i). 
These effects and additional observational systematics are expected in radio continuum surveys, 
and thus we must expect that radio sources should not be perfectly Poisson distributed. 

Let us consider the expected modifications due to multiple radio components and show that 
this effect can be modelled by means of a compound Poisson distribution \citep{James2006}, 
i.e.~the distribution that follows from adding up $n$ identically distributed and mutually 
independent random counts $n_i$, with $i= 1$ to $n$, and $n$ itself follows a Poisson 
distribution with mean $\beta$. Let us first assume that the number of radio components 
is also Poisson distributed. Then the probability $p$  to find $k$ sources in a cell follows 
from $p(k) = \sum_{n=0}^\infty p(k | n) p(n)$, where the first factor is the conditional 
probability to find $k$ radio components, like distinct hot spots and the core, associated with $n$ galaxies and the second 
factor is the probability to have $n$ galaxies. We further assume 
$\gamma$ is the mean number of components per galaxy and thus the mean of the conditional 
probability is $n\gamma$. This results in
\begin{equation}
	p_k ^{\mathrm{CP}}= \sum_{n=0}^{ \infty}\left[\frac{(n\gamma)^k e^{-n\gamma}}{k!}
	\frac{\beta^n e^{-\beta}}{n!}\right],\label{eq:compoundpoisson}
\end{equation}
with expectation and variance now given by
\begin{equation}
\bar{N} \equiv \mathrm{E}[k]=\beta \gamma, 
\qquad      
\mathrm{Var}[k]=\beta \gamma(1+\gamma) = \bar{N} (1+\gamma). \label{eq:CPmean}        
\end{equation}
Thus we see that unidentified multiple radio components can increase the variance of the source 
counts, e.g. for a textbook FRII with a detected core we would see three components which would immediately lead to 
an increase of the variance. 
This statement is independent of the size of the cell, 
but how many radio components can be identified does depend on the angular resolution and 
completeness of the radio continuum survey. 

It is useful to define the clustering parameter \citep{Peebles1980}
\begin{equation}
n_c \equiv \frac{\mathrm{Var}[k]}{\mathrm{E}[k]},
\label{eq:nc}
\end{equation} 
which is a proxy for the number of sources per `cluster'.
For the Poisson distribution $n_c = 1$, while 
$n_c = 1 + \gamma$ for a compound Poisson distribution.   
Groups of radio sources, like a group of SFGs, also contribute to $n_c$, and thus $n_c$ is also a tracer of clustering at small angular scales. 
The measurement of $n_c$ alone can not distinguish between galaxy groups, multi-component sources and imaging artefacts.

Whilst we believe assuming a Poisson distribution for the number of radio components per physical source will be appropriate for this work, 
we can chose another distribution, which will result in another compound distribution. To give a second example, 
assuming a logarithmic distribution results in a negative binomial distribution \citep{James2006}, which 
interestingly provides the best fit to three dimensional counts-in-cell in the 
Sloan digital sky survey \citep{Hurtado-Gil2017}.

\subsection{Differential source counts}

While counts in cells provides information on the spatial distribution of radio sources, it is also interesting to study 
their distribution in flux density. 
The number of sources per solid angle and per flux density observed at radio frequency $\nu$, or 
the so-called differential source count is given by
\begin{eqnarray}
\lefteqn{\frac{\mathrm{d}N}{\mathrm{d}\Omega \mathrm{d}S} (S|\nu)
= \frac{\mathrm{d}\sigma}{\mathrm{d}S} (S|\nu)} \\
&=&  \int_0^\infty \! \mathrm{d}z\, \left(\frac{\mathrm{d}L}{\mathrm{d}S} 
\frac{\mathrm{d}\sigma}{\mathrm{d}L \mathrm{d}z}\right)(S,z|\nu)  \label{eq:numbercounts1}\\
&=& 4 \pi c\! \int_0^\infty\! \mathrm{d}z \frac{d_m^4(z)}{H(z)} 
(1+z)^{1+\alpha}
\phi(L_\nu(S,\alpha,z),\alpha;z) , \label{eq:numbercounts2}
\end{eqnarray}
where $\sigma$ is the source density and we assume that the specific luminosity can be written as a power-law, 
$L_\nu \propto \nu^{-\alpha}$, 
with spectral index $\alpha$, and $\phi(L_\nu, \alpha; z)$ is the comoving 
luminosity density of radio sources at redshift $z$. 
In reality radio sources show 
a distribution in $\alpha$, often assumed to be a fixed value $0.7$ to $0.8$.  A LOFAR study 
of radio sources in the Lockman hole compared to NVSS sources measured a median spectral index
$\alpha = 0.78 \pm  0.015$ \citep{Mahony2016}, with errors obtained by bootstrapping. In a study of spectral indices comparing 
NRAO VLA Sky Survey (NVSS, \citealt{NVSS1998}) and TGSS-ADR1 sources an averaged $\bar{\alpha} = 0.7870 \pm 0.0003$ \citep{Gasperin2018}
was found, which is comparable to measurements by \cite{GLEAM2017} with median and semi-inter-quartile-range
$\alpha=0.78\pm 0.20$ for flux densities $S<0.16$~Jy at 200~MHz in the GLEAM survey. This also matches the finding by \cite{Tiwari2016}, who estimated a mean spectral index of $\bar{\alpha}= 0.763\pm 0.211$ for sources with flux densities $S_{\mathrm{TGSS}}\geq 100$~mJy and $S_{\mathrm{NVSS}}\geq 20$~mJy. For the sake of simplicity 
we assume here that all radio sources have the same spectral index. The relationship between spectral luminosity and flux density is given by:
\begin{equation}
L_\nu = 4\pi d_m^2(z) (1+z)^{1+\alpha} S.
\end{equation}
In Eq. (\ref{eq:numbercounts2}) we express the surface density by the luminosity 
density and integrate it over the past light-cone. This introduces 
the dependence on the Hubble rate at particular redshift $H(z)$ and an extra factor 
involving the transverse comoving distance (or proper motion distance)
\begin{equation}
d_m(z) = \frac{c}{H_0} \frac{1}{\sqrt{\Omega_k}} \sinh\left( \sqrt{\Omega_k}\int_0^z \!\! \mathrm{d}z' \frac{H_0}{H(z')}\right),
\end{equation}
where $H_0$ denotes today's Hubble rate and $\Omega_k$ denotes the dimensionless curvature parameter, which is positive, zero, or negative for 
hyperbolic, flat, or spherical space, respectively.
If we were to live in a static Universe with Euclidean geometry, the differential source counts 
would be proportional to $S^{-5/2}$ \citep{Condon1988}. Observations of source counts are typically rescaled by this factor to highlight 
the evolution of the Universe and of radio sources.  

\subsection{Angular Two-point correlation function}\label{sec:tpcf_intro}
\begin{figure*}
\centering
\includegraphics[width=0.8\linewidth]{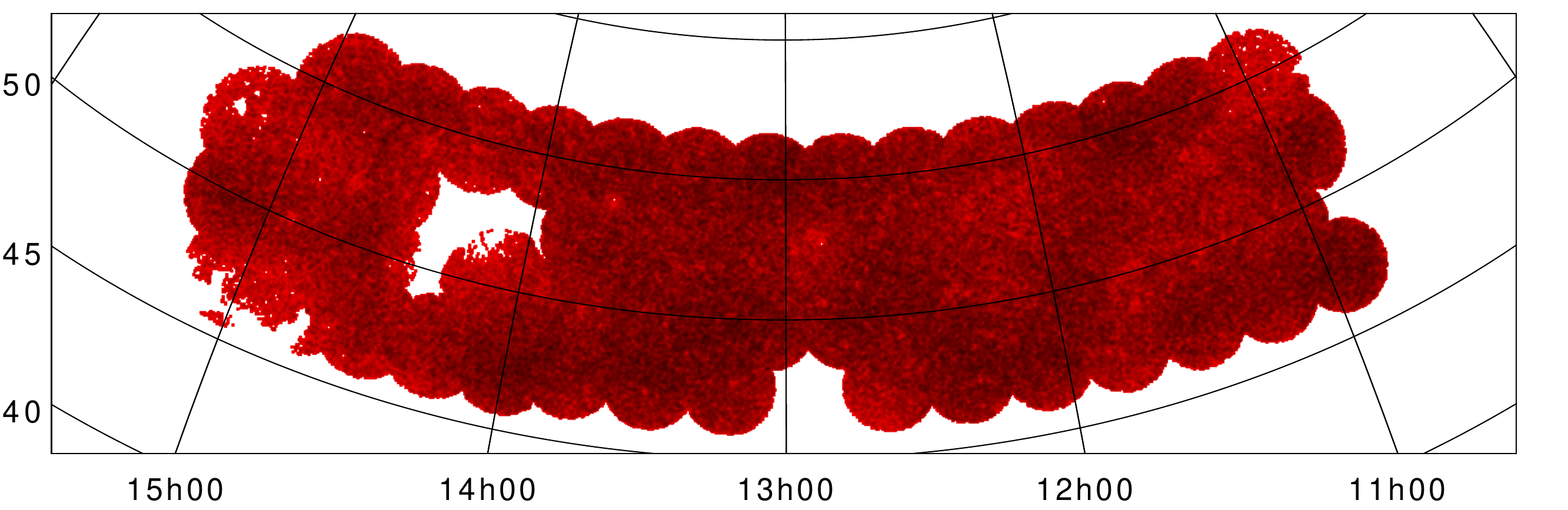}\\[2pt]
\includegraphics[width=0.8\linewidth]{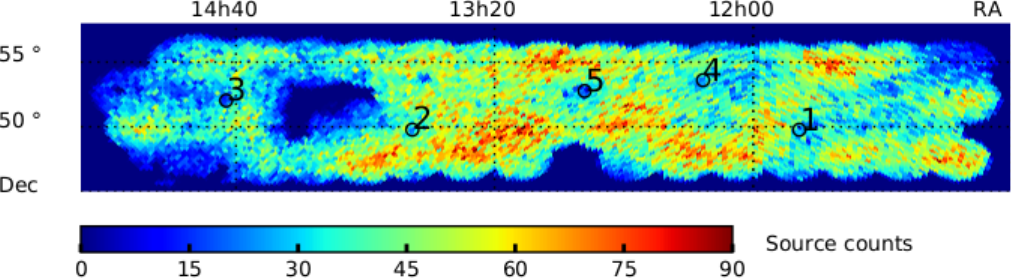}
\caption{The distribution of radio sources observed in the LoTSS-DR1 HETDEX spring field. Plotted are all individual sources (top), as well as the number counts per cell in Cartesian projection at {\sc HEALPix} resolution $N_\mathrm{side}=256$ (bottom). 
Observed are nearly $325\,000$ sources within 58 pointings on the sky covering $424$ square degrees. The positions of the
five brightest radio sources in terms of integrated flux density are indicated in black (see Sect.~\ref{sec:consistencycheck} for details).}
\label{fig:overview}
\end{figure*}

In order to study the clustering of radio sources and to use them as a probe of the 
large-scale structure of the Universe, the third quantity of interest in this work is the 
angular two-point correlation function. 

We denote the angular two-point correlation function of radio sources 
above a given flux density threshold $S = S_\mathrm{min}$ by 
$w(\mathbf{e}_1, \mathbf{e}_2, S)$, which is in principle a 
function of four position angles and the flux density threshold. It measures how likely it is 
to find $k_1$ sources within a solid angle $\Omega$ at position $\mathbf{e}_1$ and at the 
same time find $k_2$ sources around $\mathbf{e}_2$ within $\Omega$ in excess of what 
would be found for a isotropic distribution of sources,  i.e.
\begin{equation}
w(\mathbf{e}_1, \mathbf{e}_2, S_\mathrm{min}) \equiv 
\frac{\langle k_1, k_2 \rangle}{\langle k_1 \rangle \langle k_2 \rangle} - 1 =  
\frac{\langle \sigma(\mathbf{e}_1,S),  \sigma(\mathbf{e}_2,S)\rangle}{\bar\sigma(S)^2} - 1.
\end{equation}

The cosmological principle tells us that the correlation function should be isotropic, 
i.e.\ invariant under rigid rotations of the sky, and thus should 
only depend on the angle $\theta =\arccos(\mathbf{e}_1 \cdot \mathbf{e}_2)$,  such that:
\begin{equation}
  w(\mathbf{e}_1, \mathbf{e}_2, S)
  = w(\theta, S). 
\end{equation}
As a square integrable function on the interval $\cos \theta \in [-1,1]$
can be expressed as a series of Legendre polynomials $P_\ell (\cos\theta)$, 
this can allow $w$ to be rewritten as:
\begin{equation}
 w(\theta, S) = \frac{1}{4\pi} 
 \sum_{\ell = 0}^\infty (2\ell +1 ) C_\ell(S) P_\ell(\cos \theta). \label{eq:twopointfunction}
\end{equation}
The coefficients $C_\ell$ are called the angular power spectrum.

In this work we will parametrise the two-point correlation function by a simple power-law: 
\begin{equation}
w(\theta) = A_* \left(\frac{\theta_*}{\theta}\right)^{\gamma}, 
\label{eq:w_power-law}
\end{equation}
which is the result of several approximations \citep{Totsuji1969,Peebles1980}, including Limber's equation \citep{Limber1953} relating the angular correlation function to its spatial counterpart. $A_*$ is the amount of 
correlation at the pivot angular scale $\theta_*$, which we fix at 1 deg. 
We arrive at the form in Eq. (\ref{eq:w_power-law}) based on the following assumptions: the power spectrum of 
matter density fluctuations the $P(k,z)$ is assumed to be scale free; the bias, $b(k,z)$ \citep{MoWhite1996,ShethTormen1999,SKADS2008, Raccanelli2012, TiwariNusser2016}, is assumed 
to preserve the scale-free spectrum; lensing and other relativistic effects are ignored 
and we consider only small  angular separations, i.e. $\theta \ll 1$~rad. 

While we use the power-law parametrisation (\ref{eq:w_power-law}) in order to compare to the two-point correlation function found in 
other studies of radio surveys \citep{Kooiman1995, Rengelink1999, BlakeWall2002, Overzier2003, 
Blake2004, RanaBagla2019, Dolfi2019}, we would like to note that this approximation is not accurate 
enough to enable the extraction of interesting information on cosmological parameters. 
Studies of the NVSS catalogue measured typical values of $A \sim 10^{-3}$ and $\gamma \sim 1$ \citep{BlakeWall2002, Overzier2003, Blake2004}, while 
first studies of TGSS-ADR1 data revealed much larger amplitudes $A \sim 10^{-2}$ and comparable values of $\gamma$ \citep{RanaBagla2019, Dolfi2019}.  

In order to compare the angular two-point correlation function to the prediction from the 
standard model of cosmology and going beyond the approximations that lead to Eq. (\ref{eq:w_power-law}), we use the publicly available software package {\sc CAMB sources}\footnote{\url{http://camb.info/sources/}} \citep{CAMBsources2011}; more details 
are provided in Sect.~\ref{sec:twopoint}.

The two-point correlation function and angular power spectrum for source counts is of great value in informing us about cosmology. We can fit parametrised theoretical models 
to the data, hence finding the range of acceptable parameters. 
One cannot constrain cosmological parameters individually, but rather a combination of parameters which all affect the observable and include:

(i) bias parameters \citep{MoWhite1996,ShethTormen1999,TiwariNusser2016,Hale2018}, revealing the relationship between source count fluctuations and underlying total density fluctuations, as a function of scale and time.
These can give insight into the astrophysics-cosmology interface, informing us about the range of halo masses that radio sources inhabit.
 Further to this, with Halo Occupation Distribution Modelling \citep[HOD; see descriptions and uses in e.g.][]{Berlind2002,Zheng2005,Hatfield2016}, the properties of how galaxies occupy dark matter haloes can be determined. 
 This will be especially important with deep radio observations, such as the LOFAR deeper tier surveys \citep{Rottgering2010,LOFAR2013}, where it may be possible to observe the `1-halo' clustering \citep[see e.g.][]{Yang2003,Zehavi2004}, which describes the clustering between radio sources in the same parent dark matter halo. 
 By observing both the `2-halo' and `1-halo' term and modelling the observed clustering within a HOD framework, it is possible to determine quantities which describe the distribution of central and satellite galaxies for different radio source populations. 
 Finally, if the cross correlation function is instead investigated, the clustering observed may also be important in investigating how different radio sources within single dark matter haloes may be affected by other galaxies within the same halo \citep[see e.g.][]{Hatfield2017}.

(ii) Parameters describing the total density of matter, $\Omega_m$, and the rms amplitude of 
fluctuations in the matter density in a sphere of $8$~$h^{-1}$~Mpc, $\sigma_8$, which affect $P(k,z)$. 
$\Omega_m$ tells us about the degree to which dark 
matter dominates the matter budget in the Universe, whilst $\sigma_8$ relates to the 
degree to which structures have grown by the present day.

(iii) Dark energy parameters: the equation of state of dark energy at scale factor $a$ is given by $w=w_0+(1-a)w_a$ \citep{ChevallierPolarski2001,Linder2003}, where the present day equation of state is $w_0$, and its time evolution is parameterised by $w_a$. These parameters affect the growth of structure and hence enter into $P(k,z)$. 

(iv) Parameters describing modifications to gravity \citep{Amendola2008,Zhao2010}: we can assess the slip parameter $\eta$, which is the ratio of the space- and time- perturbations in the metric. In addition we can examine the Poisson equation $\nabla^2 \Phi =4\pi G a^2 \mu \rho \delta$, where $\mu$ parametrises deviations from the GR expectation $\mu=1$. These parameters again enter into $P(k,z)$ as they affect the growth of structures.

(v) Finally, primordial non-Gaussianity of density modes affects the measured two-point statistics \citep{Dalal2008,Matarrese2008,Ferramacho2014,Raccanelli2015}. On large scales, the effective bias is greatly increased, leading to a substantial increase in amplitude of the auto-correlation function or power spectrum. Constraints on the non-Gaussianity parameter $f_{NL}$ are expected to improve on constraints by Planck. 

\begin{figure*}
\centering
\includegraphics[width=0.5\linewidth]{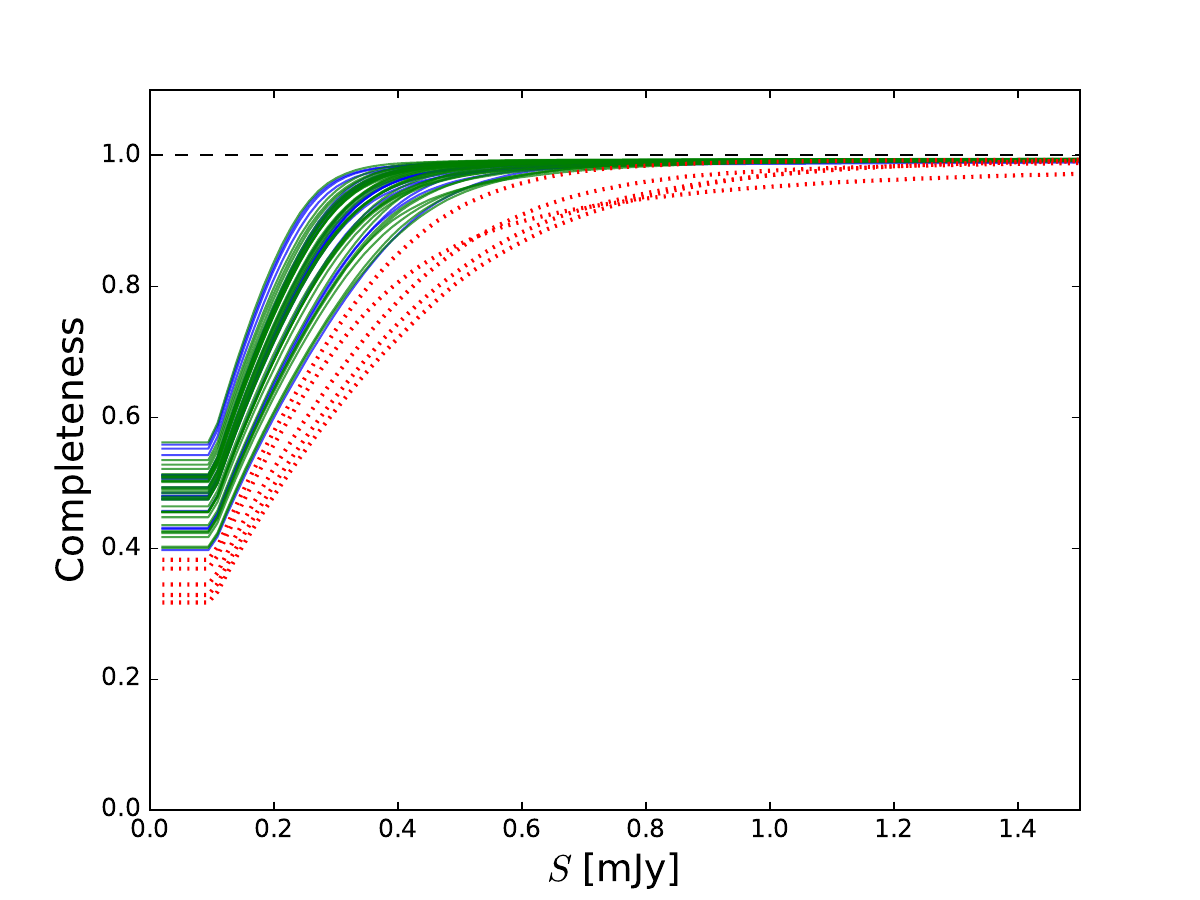}\includegraphics[width=0.5\linewidth]{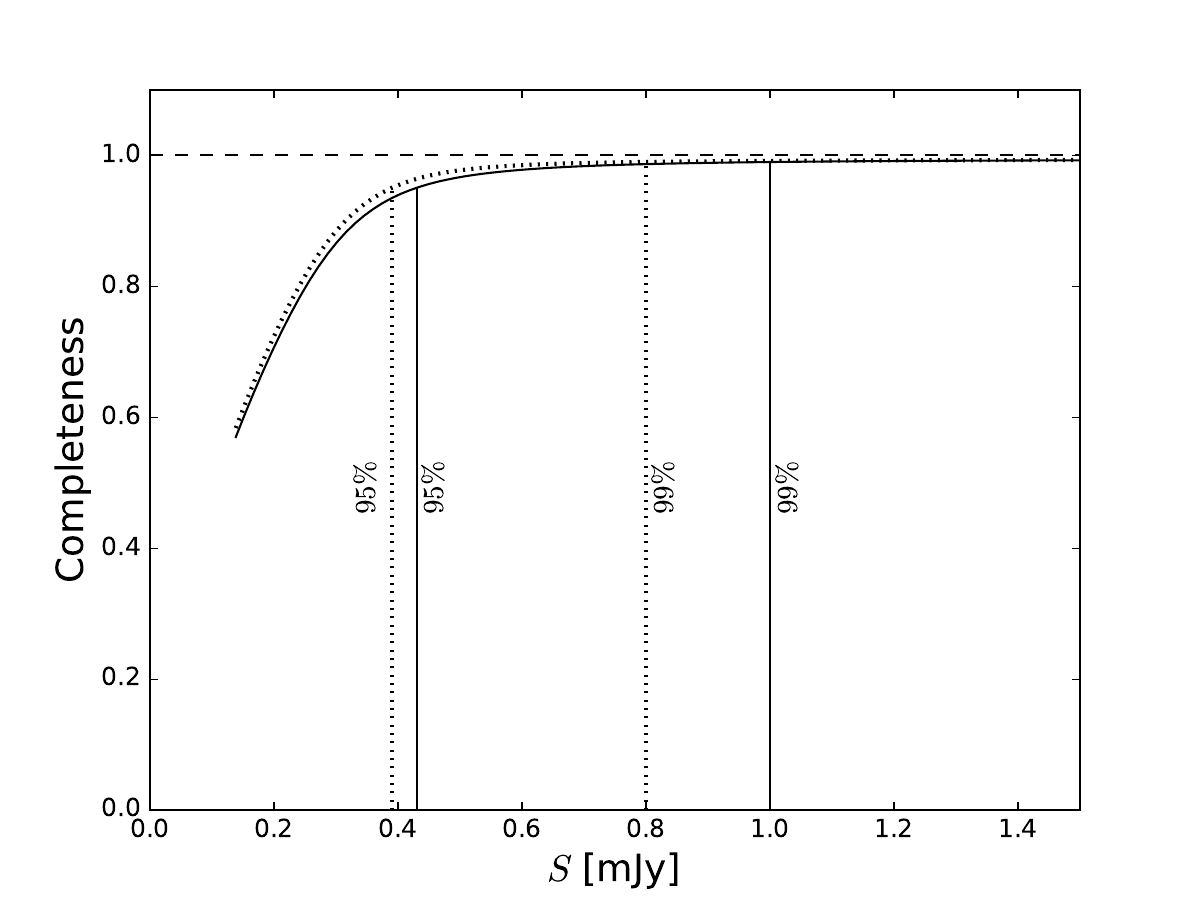}
\caption{Left: Estimated point-source completeness for each of the 58 pointings in the HETDEX field 
as a function of flux density. Blue, green and red (dotted) lines indicate inner, outer and the five most incomplete pointings, respectively. 
Right: Mean point source completeness of all pointings (solid line) and after rejection of the 
five most incomplete pointings (dotted line).}
\label{fig:completeness}
\end{figure*} 

\section{LoTSS-DR1: data quality \label{sec:data}}
  
\subsection{Requirements and cell size}

To study the cosmic large scale structure, we require three 
essential properties of a radio survey. First of all, the survey must 
cover a sizeable fraction of the sky in order to measure 
properties on large angular scales and to ensure that the effects 
of interest are not dominated by cosmic variance.
Secondly, the survey must sample the 
sky fairly homogeneously to some minimal flux density, which then allows for 
reliable and complete source counts. Thirdly, in order to identify 
foreground effects and to classify radio sources, 
identification with an optical or infra-red counterpart and associated photometric or spectroscopic redshift, is essential.

In order to connect number counts with theoretical predictions we must estimate $\sigma(S,\mathbf{e}$) by counting radio sources in 
cells of equal and non-overlapping areas,
a necessary (but not sufficient) condition for the statistical independence of the counts.
Finally, these cells should cover the sky completely.
Thus we need to select a scheme to pixelize the sky and for this pixelisation we need to decide how large 
those cells should be.  
The pixel sizes of the LoTSS imaging pipeline and used by the source finder {\sc PyBDSF} 
are too small to be efficient for cosmological 
tests (most of them contain only noise) and 
it would be computationally expensive to correlate all pixel pairs. 
On the other hand the individual LoTSS 
pointings are too large to define cell sizes that 
are useful for cosmological analysis, as there are about 6000 sources per pointing.

The scheme in {\sc HEALPix}\footnote{\url{http://healpix.sourceforge.net}}\citep{Healpix2005} is one such method that satisfies the above requirements (equal area, no overlap, complete sky coverage) and has been developed for the purpose of the analysis 
of the cosmic microwave background.
We use it in the so-called ring scheme, which numbers the cells in rings of decreasing declination.
In order to avoid confusion with imaging pixels, we will denote {\sc HEALPix} pixels as cells in 
the following. The cell size is specified by means of the parameter $N_\mathrm{side}$, which 
can take values of $2^m$, where $m$ is an integer. The total number of 
cells on the sky is given by $12N_\mathrm{side}^2$.
     
For each cell we count the number of radio sources, either in the 
catalogue originally produced by {\sc PyBDSF} (LoTSS-DR1 radio source catalogue) or in 
the final LoTSS-DR1 value-added source catalogue, where radio components of a single source have been grouped and artefacts removed.
The position of each source was taken as either the output position from {\sc PyBDSF} or the RA and Dec value that was assigned in the value-added catalogue (see \citealt{LoTSS2019B} for a description of how these were generated).

The mean number of sources per cell is 
\begin{equation}
N = \sigma \Omega_\mathrm{cell} 
	= \frac{N_\mathrm{survey}}{\Omega_\mathrm{survey}}
    \frac{4\pi}{12 N_\mathrm{side}^2},
\end{equation}
where $N_\mathrm{survey}$ and $\Omega_{\rm survey}$ denote the total number of sources and the total solid angle covered by the survey. 
We want to find a value of $N_\mathrm{side}$, that guarantees that all cells contain at least 
one source, if 
the cell was properly sampled, i.e.~each cell area should be completely within 
the survey area and we would like to disregard regions with very low completeness. 
We assume that the source counts are Poisson distributed and
estimate the probability that a cell does not contain a source as
\begin{equation}
p_0 =e^{-N} .
\end{equation}

\begin{table*}
\caption{Number of included cells ($N_{\mathrm{cell}}$) and sky coverage ($\Omega$) for different masks and flux density thresholds ($S_{\mathrm{min}}$). 
Unless explicitly stated otherwise, we use the default `mask d' throughout this work. Thus 
we highlight the respective entry in bold font. 
The retained number of sources for each mask are shown for the LoTSS-DR1 radio source ($N_{\text{rs}}$) and 
value-added source ($N_{\text{vas}}$) catalogues. For detailed explanation see Sect. \ref{sec:area}, \ref{sec:noise} and \ref{sec:tpcfredshift}}.
\label{tab:masks}
\centering
\begin{tabular}{ccccccc}
\hline\hline
mask & $N_{\mathrm{cell}}$ & $\Omega$& $S_{\mathrm{min}}$ & $N_{\text{rs}}$ & $N_{\text{vas}}$& Description\\
& & [sr] & [mJy] && &\\ \hline
none     & 8422 & 0.13458 & 0.00 & 325\,694 & 318\,520 & all sources\\ 
p & 7182 & 0.11476 & 0.00 & 306\,684 & 300\,601& sources within union of 53 discs ($\theta=1.7$~deg)\\\hline
\multirow{4}{*}{\textbf{d} } & \multirow{4}{*}{\textbf{ 7176 }} & \multirow{4}{*}{\textbf{0.11467 }} &\textbf{ 0.00 } & 306\,670 &\textbf{ 300\,588 }& \multirow{4}{*}{\& exclude cells with less than five value added sources}\\
& & & 1.00 & 108\,539 & 102\,940 & \\
& & & 2.00 & 55\,459 & 51\,288& \\
& & & 4.00 & 33\,040 & 30\,556 &   \\\hline
\multirow{2}{*}{3} & \multirow{2}{*}{7104} & \multirow{2}{*}{0.11352} 
			& 0.00 & 305\,186 &299\,311& \multirow{2}{*}{\& exclude cells with $S_\mathrm{rms} > 3 \times \mathrm{median}(S_\mathrm{rms})$ } \\
        & & & 1.05 & 101\,714 & 96\,404  &\\ \hline
  \multirow{3}{*}{2}& \multirow{3}{*}{6954} & \multirow{3}{*}{0.11112} 
            & 0.00 & 301\,527 & 295\,903  &\multirow{3}{*}{\& exclude cells with $S_\mathrm{rms} > 2 \times \mathrm{median}(S_\mathrm{rms})$}\\
        & & & 0.70 & 158\,226 & 152\,662 & \\
        & & & 1.05 & 99\,411  & 94\,326  &\\ \hline
        \multirow{7}{*}{1} & \multirow{7}{*}{2957} & 
        \multirow{7}{*}{0.04725} 
            & 0.00 & 152\,498 & 150\,568 &\multirow{7}{*}{\& exclude cells with $S_\mathrm{rms} >  \mathrm{median}(S_\mathrm{rms})$}\\
        & & & 0.35 & 136\,150 &134\,178& \\
        & & & 0.70 & 66\,027 &64\,118  &\\
        & & & 1.00 & 42\,329 & 40\,599 & \\
        & & & 1.05 & 39\,919 &38\,222 & \\  
        & & & 2.00 & 20\,848 & 19\,719 & \\
        & & & 4.00 & 11\,805 & 11\,269 & \\\hline
        \multirow{3}{*}{z} &\multirow{3}{*}{7139} &\multirow{3}{*}{0.11407} & 0.00 & - & 153\,111 & \multirow{3}{*}{ \shortstack{`mask d' \& missing Pan-STARRS information \\ \& only sources with redshift information }}\\
        & & & 2.00 &- & 24\,420 &\\
        & & & 4.00 &- & 14\,506 & \\\hline
        \multirow{3}{*}{z1} & \multirow{3}{*}{2940} & \multirow{3}{*}{0.04698}& 0.00 & - & 76\,602 & \multirow{3}{*}{ \shortstack{`mask 1' \& missing Pan-STARRS information \\ \& only sources with redshift information }}\\
        & & & 2.00 & - & 9505 &\\
        &  & & 4.00 & - & 5432 & \\\hline        
\end{tabular}
\end{table*}

The probability that all cells contain at least one source is then given by 
$P = (1 - p_0)^{N_\mathrm{cell}}$, with $N_\mathrm{cell} = 12 N_\mathrm{side}^2 
\Omega_\mathrm{survey} / 4\pi$ is the number of cells covering the 
survey area. We wish to keep the probability to find empty cells, 
$P_0(N_\mathrm{side}) = 1-P \approx p_0 N_\mathrm{cell}$ well below one, but at the same 
time would like to allow for the best angular resolution.
With $\Omega_\mathrm{survey} = 424$ square degrees (0.12916 sr) and $N_\mathrm{rs} = 325\,694$ we 
find $P_0(256) = 3 \times 10^{-14}$, while $P_0(512)$ is of order unity.
In a resolution of $N_\mathrm{side} = 256$ the cells have a mean spacing of 
$\bar{\theta}_{i,j}=0.229$~deg and a cell covers $\Omega_{pix} \approx 1.60\times 10^{-5}$ sr.
The set of all non-empty cells defines the effective survey area.
The number of cells within the survey area for the chosen $N_\mathrm{side}$ and after 
masking can be seen in Table \ref{tab:masks}.
Figure \ref{fig:overview} shows the cell counts of the LoTSS-DR1 radio source catalogue at a 
resolution of $N_\mathrm{side} = 256$, which is a good compromise between large enough cell size 
to make sure that the shot noise in each cell is not the dominant feature (i.e. all cells contain at least one source) and to retain as 
much angular resolution as possible. 
One can also see that plotting the number counts per cell has advantages over a map 
that shows each radio source as a dot, as such a map quickly saturates when the surface 
density of objects is high (see Fig. \ref{fig:overview}).

\subsection{Completeness}

The LoTSS-DR1 catalogue was generated by combining $58$ individual LOFAR pointings on the sky. 
The current LOFAR calibration and imaging pipeline used in DR1 produces sub-standard images in a 
few places due to poor ionospheric conditions and/or due to the 
presence of bright sources. Such areas are not 
included. Furthermore, in some regions, where the astrometric position offsets from 
Pan-STARRS is large, the LoTSS maps are blanked. This results in an 
inhomogeneous sampling of the HETDEX spring field
as is apparent from the source density map presented in Fig. \ref{fig:overview}.

\begin{table}
\caption{Undersampled pointings with name and position.}
\label{tab:pointings}
\centering
\begin{tabular}{ccc}
\hline\hline
Name & RA & Dec \\
& [deg] &[deg]\\ \hline
P164+55 &164.633	&54.685\\
P211+50 &211.012	&49.912\\
P221+47 &221.510	&47.461\\
P225+47 &225.340	&47.483\\				
P227+53 &227.685	&52.515\\ \hline
\end{tabular}
\end{table}

\begin{figure}
\centering
\includegraphics[width=\linewidth]{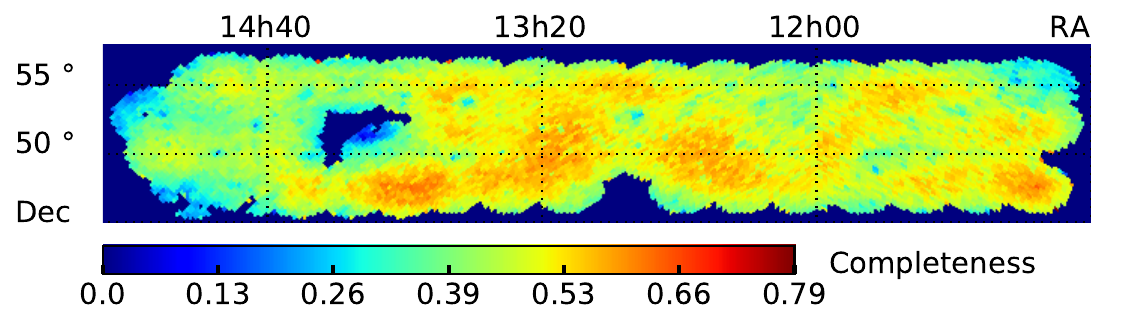} \\
\includegraphics[width=\linewidth]{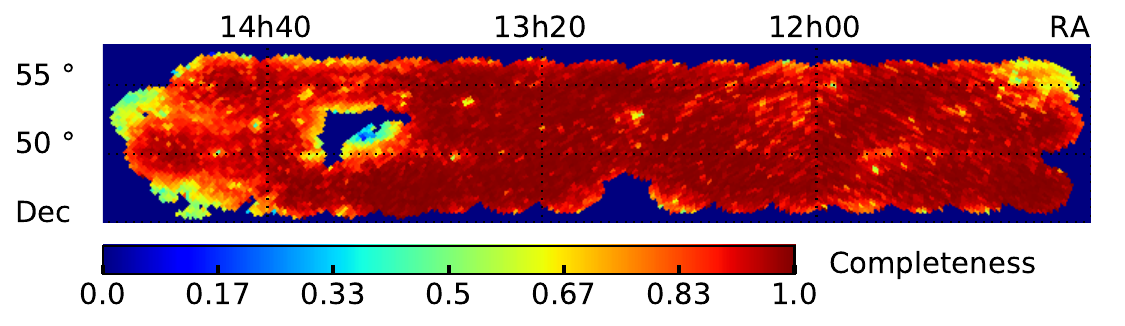}
\caption{Top: Completeness of the LoTSS-DR1 catalogue per {\sc HEALPix} cell. Bottom:
Completeness of cells after applying a flux density threshold of $0.39$~mJy, which corresponds to 
an overall point source completeness of 95\%.}
\label{fig:completenesscell}
\end{figure}

We estimated the point source completeness of all pointings in the HETDEX field by injecting random 
sources in the residual maps and using the same {\sc PyBDSF} set up used for the LoTSS-DR1 radio source catalogue. 
Only sources with flux densities five times greater than the local rms noise are retained. 
The completeness itself is estimated by taking the fraction of recovered sources to the total number 
of injected sources above a certain flux density threshold.
In total we simulated $50$ samples with $6000$ sources each for each of the 58 pointings.
The completeness of each pointing is shown in Fig. \ref{fig:completeness}, where pointings at the 
edge of the survey are marked in green and pointings in the inner field are marked in blue. 
Additionally five pointings are marked in red, which are clearly undersampled, for reference see 
Table \ref{tab:pointings}.
Using all pointings, the survey is $95\% $ point source complete at $0.43$~mJy and reaches $99\%$ 
completeness at $1.0$~mJy. Rejecting the five most incomplete pointings, the $95\% $ level is 
at $0.39$~mJy and the $99\% $ level is reduced to $0.80$~mJy.

As we use HEALPix cells to determine the source count statistics, we estimate the 
completeness for each cell. Without any flux density threshold the completeness per cell is shown in 
Fig. \ref{fig:completenesscell}. The structure of the completeness across the survey 
matches the number density of Fig.~\ref{fig:overview}. Areas with high number densities 
appear to be already more complete without assuming any flux density threshold and underdense regions are comparable to areas with low completeness.
Applying a flux density threshold of $0.39$~mJy, corresponding to a point source completeness of $95\%$
in the region without the five pointings of Table~\ref{tab:pointings}, results in a much improved 
uniformity of the completeness (see also Fig.~\ref{fig:completenesscell}).

\subsection{Consistency of source counts}\label{sec:consistencycheck} 

\begin{figure}
\centering
\includegraphics[width=\linewidth]{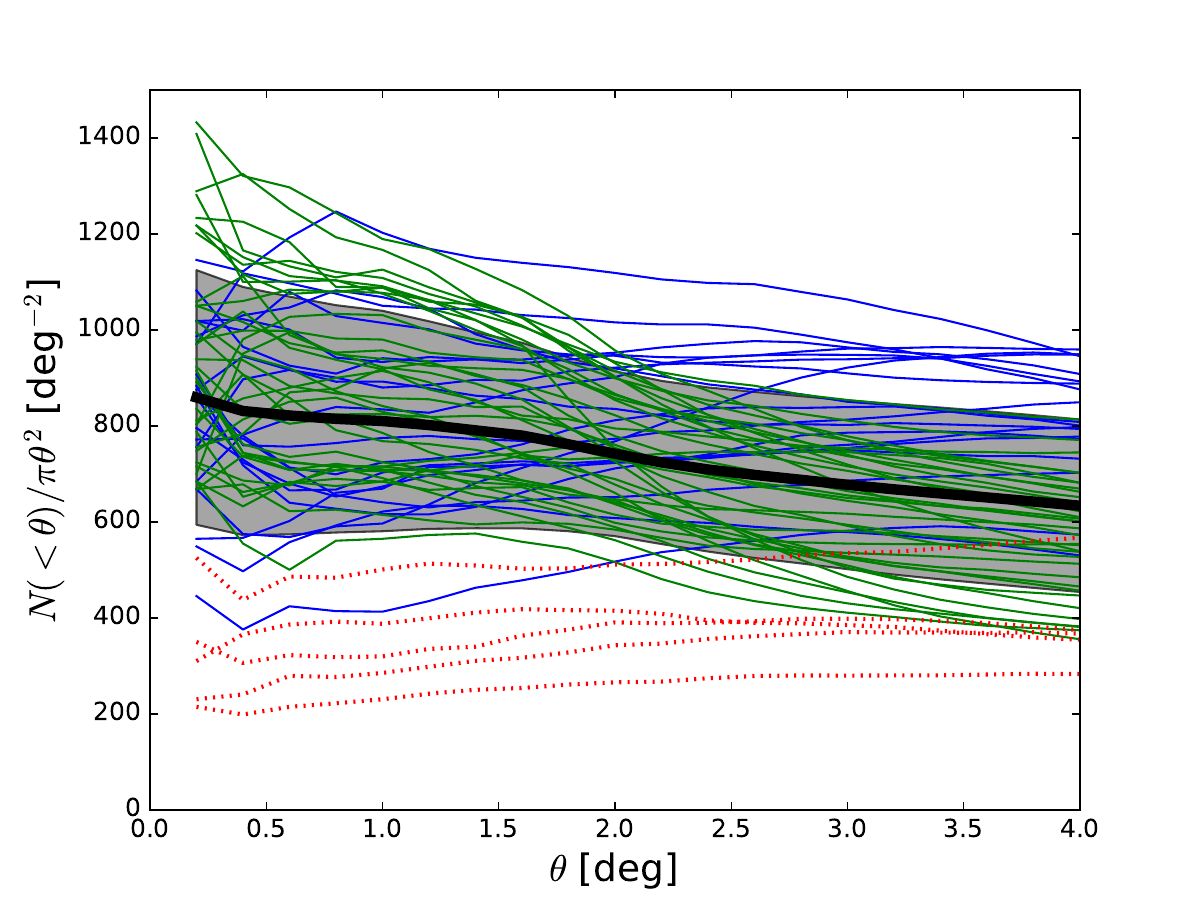}
\includegraphics[width=\linewidth]{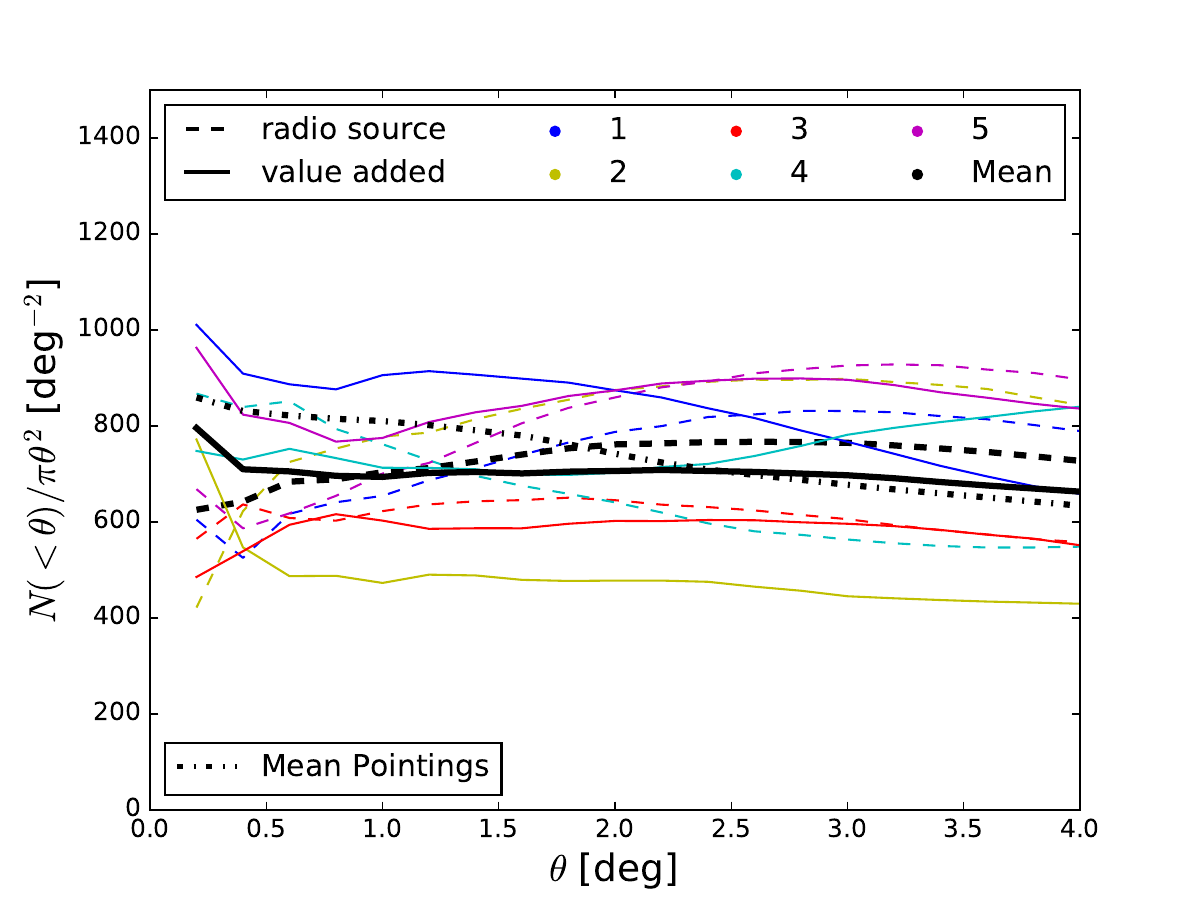}
\caption{Top: Source counts for each pointing within angular distance $\theta$ around the pointing center, normalized by covered area. Pointings are classified by position in the HETDEX field, with pointings on the edge (green), in the inner field (blue) and undersampled ones (red, dotted). The mean is shown in black with standard deviation (grey band) of all pointings. Bottom: Source counts around the five brightest radio sources in terms of integrated flux density from the radio source (dashed lines) and value-added source catalogue (solid lines). The mean number counts around the five brightest sources are shown in black for both catalogues and additionally also the mean over all pointings (dash dotted).}  
\label{fig:countspointing}
\end{figure}

Completeness and total source counts will be a function of the distance from the pointing 
centre, as the sensitivity is not uniform across the primary beam. This is investigated by means 
of radial source counts around the pointing centers. All sources within angular distance, $\theta$,
from the pointing center are counted and the sum is normalized by the solid angle of the 
corresponding disk. We split the pointings into three groups, depending on their position 
and whether they appear undersampled (see Table \ref{tab:pointings}). 
In Fig.~\ref{fig:countspointing} we show source counts for pointings at the edge of the HETDEX 
field (green), inner pointings (blue) and pointings which are excluded from the further analysis (red dotted).
The mean source counts of all pointings is shown in black, with the $1\sigma$ region in grey.
The source counts of green pointings drop after the angular distance reaches regions which are not 
covered by overlapping pointings of the survey any more. Pointings in the inner field have more continuous source counts, as 
they overlap with other pointings. 
The five undersampled pointings from the latter appear in this test also as the undersampled ones.  

Additionally we study the source counts around the five brightest sources.
The five sources are listed in Table \ref{tab:5brightest} and are the same in the LoTSS-DR1 
radio source and value-added catalogues. They are displayed in Fig.~\ref{fig:overview} as 
black circles to show the underlying regions. Comparing both catalogues, the radio source 
catalogue shows a stronger effect on the source counts due to limited dynamic range around 
bright sources. This effect is visible by eye in Fig. \ref{fig:overview} (bottom), where the bright sources are located in underdense regions. 
In contrast, in the value-added catalogue the mean of sources 
becomes flatter, because many sources are matched together. Overall we see a deficit of sources around the five brightest sources compared 
to the overall mean of all pointings, but that deficit is well within the variance of source counts 
and thus we decided to keep regions that include bright sources in our analysis.

\begin{table}
\centering
\caption{The five brightest sources of LoTSS-DR1 in terms of total flux density.}
\begin{tabular}{cccc}
\hline\hline
Name & RA & Dec & $S_\mathrm{int}$\\
& [deg] & [deg] &[Jy]\\\hline
ILTJ114543.39+494608.0 & 176.43 & 49.77& 14.49\\
ILTJ134526.39+494632.4& 206.36& 49.78& 14.13\\
ILTJ144301.53+520138.2&220.76& 52.03& 14.10\\
ILTJ121529.77+533553.6& 183.87& 53.60&11.98\\
ILTJ125208.61+524530.4& 193.04&52.76&8.35\\\hline

\end{tabular}
\label{tab:5brightest}
\end{table}

\begin{figure}
\centering\includegraphics[width=\linewidth]{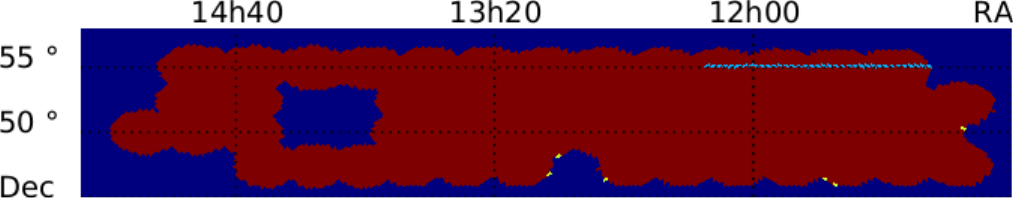}
\caption{LoTSS-DR1 HETDEX spring field masks: `mask p' rejects all cells shown in 
dark blue and includes 53 pointings modelled by disks of radius $1.7$~deg. Our default `mask d' 
additionally rejects cells with less than five sources (yellow cells), see also text in Sec.\ref{sec:area}. 
For analysis that includes redshift information `mask z' additionally rejects a strip shown in light 
blue. For further details, see the text in Sect. \ref{sec:redshift}. }
\label{fig:pointings}
\end{figure}

\subsection{Survey area}\label{sec:area}

A proper definition of the survey area directly affects the one- and two-point statistics, especially 
the mean surface density. 
As we exclude all sources of the five most incomplete pointings (see Table \ref{tab:pointings}), it is therefore important to define the region being investigated throughout this work, excluding these pointings.

To remove the sources of those five pointings and to model the boundaries of the survey
we produce a mask (mask p). We model each pointing as a disc with radius of $1.7$~deg, inferred 
from the (average) radius of pointings in the mosaic and mask all cells 
which are not included in the union of all discs (see Fig. \ref{fig:pointings}). 
We verified that this procedure does not result in a single empty cell,
consistent with the argument that we used to set the value of
$N_\mathrm{side}$. 

We test for the robustness of this method by also masking cells containing fewer than five sources. This results in removing another six cells and 14 sources. 
We adopt this slightly stronger mask (mask d) as the basis of our analysis.
The total number of sources and the effective 
survey area for the various masks and cuts can be found in Table~\ref{tab:masks}. 
Our base mask (mask d) applied to the LoTSS-DR1 catalogue results in a mean 
number of sources per cell of $\bar{n} = 42.0$ and a mean surface density of 
$\bar\sigma = 2.6215 \times 10^6/$~sr $= 798.6/$~deg$^2 = 0.2218/$~arcmin$^2$.

\begin{figure}
\centering
\includegraphics[width=\linewidth]{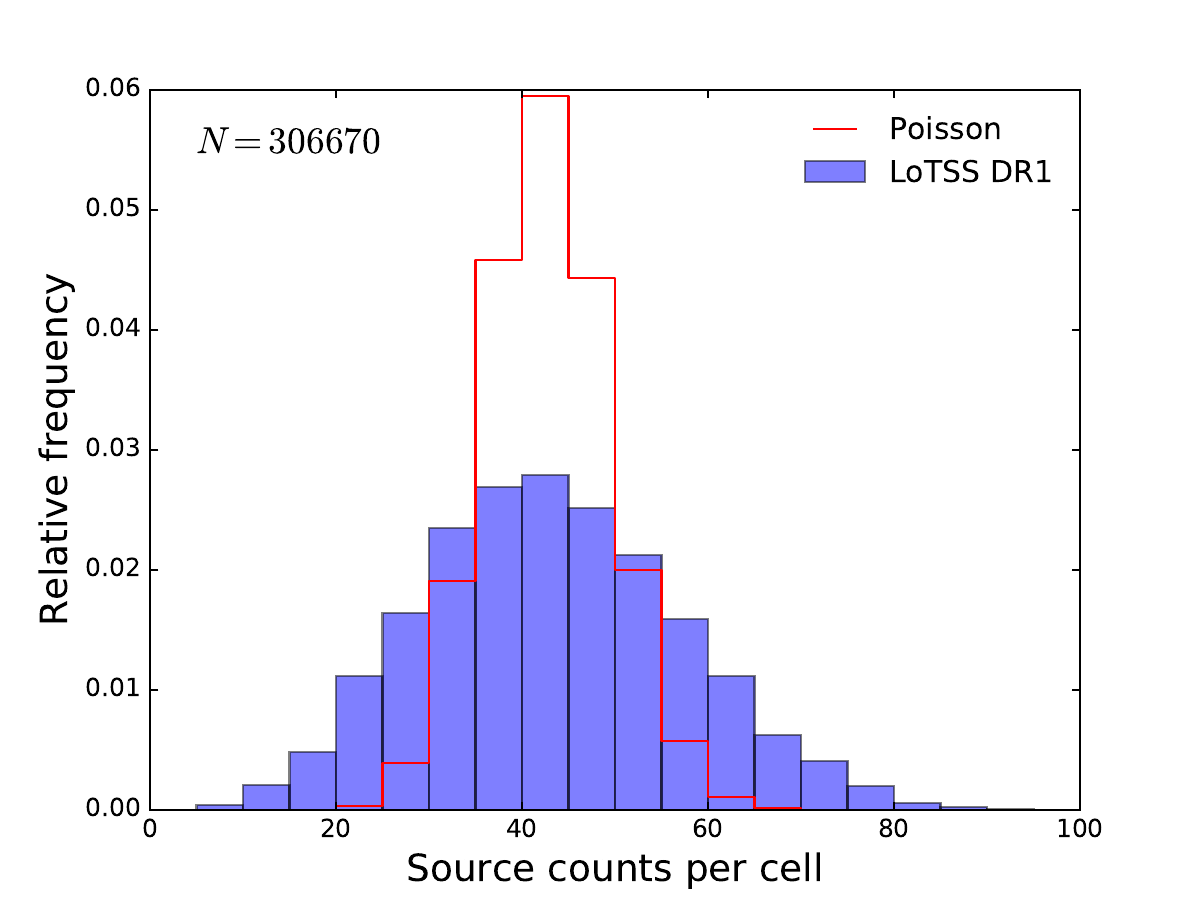}
\caption{Histogram of source counts per cell (blue) and binned Poisson 
distribution with empirical mean (red line) from the LoTSS-DR1 radio source catalogue at $N_\mathrm{side}=256$, masked and including only cells with at least five sources (mask d).}
\label{fig:pointingsge5}
\end{figure}

The histogram for the masking that excludes the 
five bad pointings and all cells with less than five sources is shown 
in Fig.~\ref{fig:pointingsge5}. For comparison we also plot a Poisson 
distribution with identical mean. We observe a broadening of the 
source count distribution when compared to a Poisson distribution, 
which obviously is not a good fit to the data. Thus we see that the naive 
expectation about the number count distribution is not met. 
  
\subsection{Local rms noise}\label{sec:noise}

To further characterize the properties of LoTSS-DR1, we take a closer look 
at the properties of the local rms noise. We define a set of tiered masks to reject cells with 
noise above certain noise thresholds.

\begin{figure}
\includegraphics[width=\linewidth]{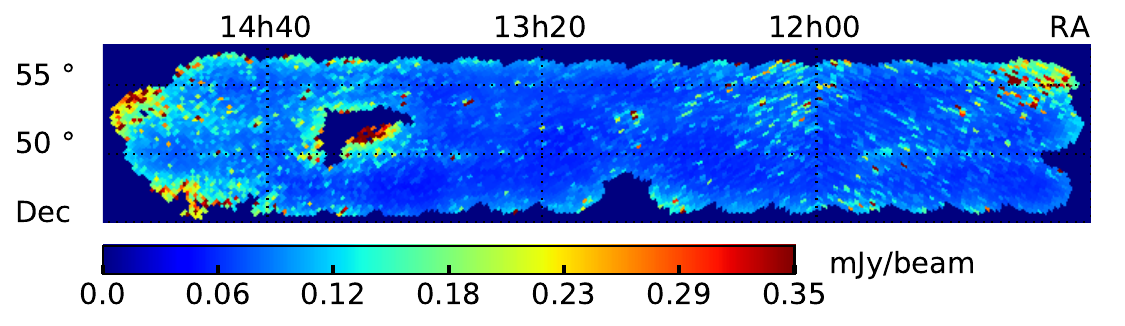}\\
\includegraphics[width=\linewidth]{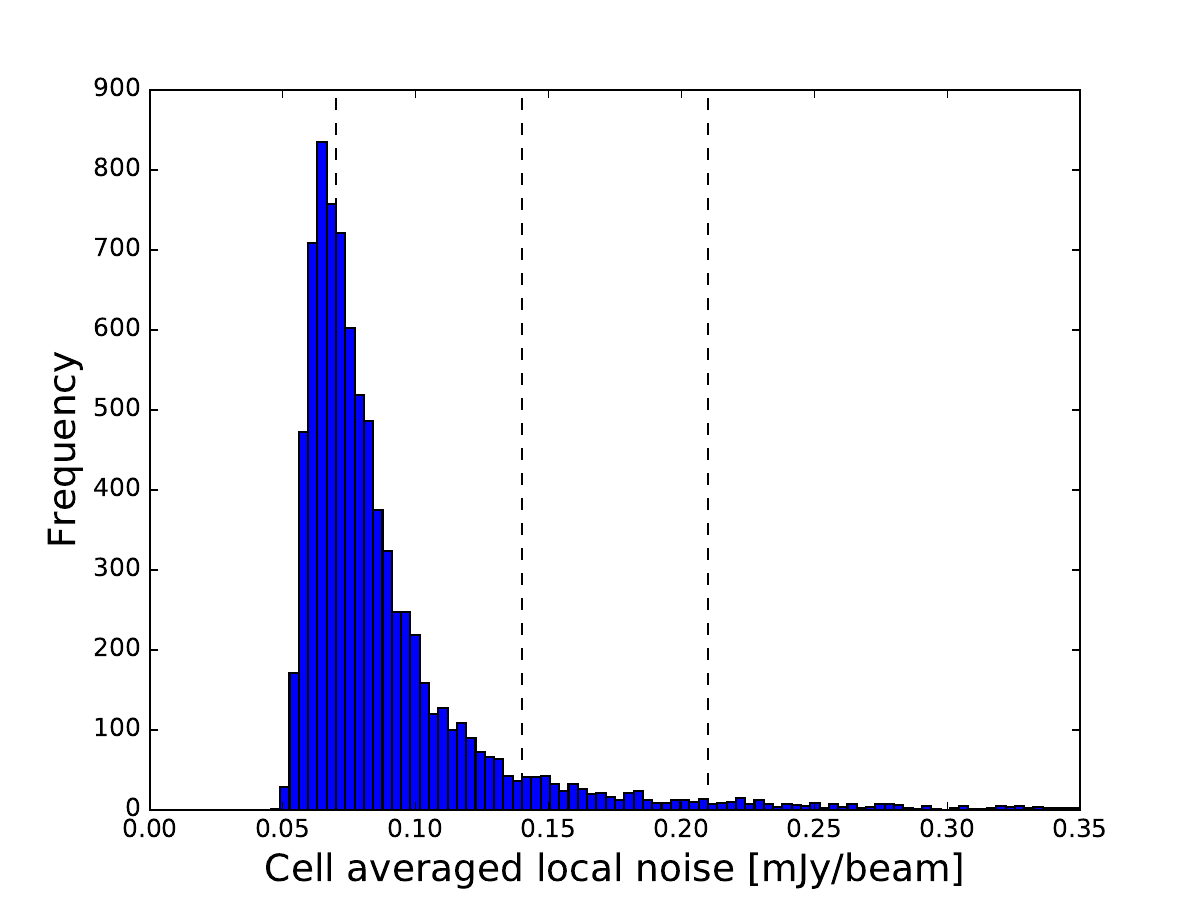}
\caption{Local rms noise per {\sc HEALPix} cell, calculated via the mean 
of the local rms around each LoTSS-DR1 radio source. The heat map (top) and histogram (bottom) 
of the local rms is clipped at an upper limit of five times the median rms noise. 
The median rms noise of $0.07$~mJy/beam, as well as the values of two and 
three times the median rms noise are marked in the histogram with black 
dashed lines.}
\label{fig:localrms}
\end{figure}

Fluctuations in the local rms noise are expected for several reasons. 
In the vicinity of bright sources, limitations of dynamic range give rise 
to an increase of the local rms noise. Directions and epochs with 
unfavorable ionospheric conditions will also result in higher noise 
levels. To find regions of higher noise we therefore produced a 
{\sc HEALPix} map of the local rms per {\sc HEALPix} cell, as well as the corresponding histogram of the 
local rms noise distribution (see Fig. \ref{fig:localrms}). The map is produced by averaging the local rms noise 
associated to each source in the cell, which is defined as the averaged background rms value of the corresponding island, obtained from the 
LoTSS-DR1 catalogue.

Using the local rms noise attached to each source gives rise to a slightly larger cell average,
than doing cell averages on the noise maps themselves. 
This effect is due to bright sources, which increase the noise.
The mean local rms noise of the {\sc HEALPix} cells is $94~\mu$Jy/beam and median local rms noise 
in a cell is $76~\mu$Jy/beam, which is in good agreement with the median rms noise 
$71~\mu$Jy/beam in the total observed area based on the much smaller mosaic 
pixels \citep{LoTSS2019A}.

\begin{figure}
\centering
\includegraphics[width=\linewidth]{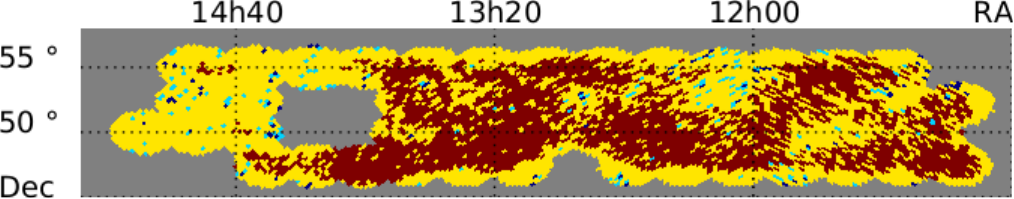}
\caption{The three local rms noise masks. The red cells are included 
for an average noise $<0.07$~mJy/beam in the {\sc HEALPix} cells (`mask 1'), 
red and yellow pixels are included for an average noise of 
$<0.14$~mJy/beam (`mask 2') and red, yellow and light blue cells are 
included for an average noise of $<0.21$~mJy/beam (`mask 3'). Dark blue cells 
are additionally included in `mask d'. Regions in grey are excluded by all masks.
\label{fig:mask123}}
\end{figure}

To produce a tiered set of noise masks we require the local rms noise to be below 
one, two and three times the median rms noise of $0.07$~mJy/beam 
and denote the resulting masks by mask 1, mask 2 and mask 3, respectively. 
Most of the sources are unaffected with the $0.21$~mJy/beam and $0.14$~mJy/beam rms mask, but for the upper limit of $0.07$~mJy/beam rms noise (mask 1), 
we obtained less than 50 percent of the original number of sources 
(see Table \ref{tab:masks}). 
The difference in the masking can also be seen in the remaining number of 
cells $N_\mathrm{cell}$ and sky coverage $\Omega$ (see Table \ref{tab:masks}).
These noise masks are shown in Fig. \ref{fig:mask123}. 

We also checked that the variance of the 
number count distribution becomes smaller with decreasing the upper rms noise limit.
We return to more details of the statistical evaluation in Sect. \ref{sec:onepoint}.

In the analysis below we combine spatial masking with flux density thresholds in order 
to improve the completeness and reliability of the studied sample of radio sources. The faintest, 
at five times signal to local noise, observed radio sources in the LoTSS-DR1 survey have a 
flux density of around $0.1$~mJy, and, as shown above, the survey is certainly not complete 
at such low flux densities. Thus, below we test different flux density thresholds to increase 
the completeness and reliability of the survey. The source counts corresponding to flux 
density thresholds (for unresolved sources) of five, ten and fifteen times the rms noise of the 
masked survey are listed in Table \ref{tab:masks} for both the LoTSS-DR1 radio source and 
the value-added source catalogue. We can easily see that a cosmological data analysis has to find 
a good compromise between high demands on data quality (more aggressive masking and higher flux density thresholds) and the demand for statistics (large number of radio sources).

\section{Mock catalogues \label{sec:mocks}}

\begin{figure}
\centering
\includegraphics[width=\linewidth]{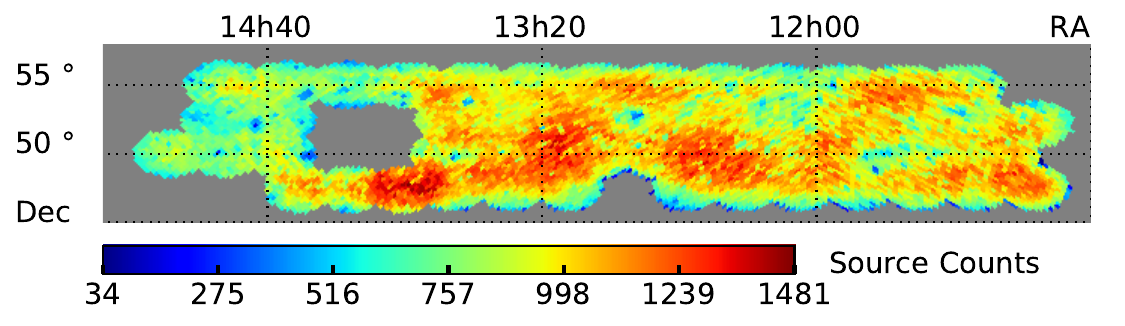}
\caption{Mock catalogue of random sources that are detectable at five times the local rms noise and masked with `mask d'.}
\label{fig:mockcatalogues}
\end{figure}

As discussed in Section \ref{sec:tpcf_intro}, the two-point correlation function quantifies the excess in clustering observed within a galaxy catalogue at different separation scales compared to that of a uniform distribution of galaxies. As such, it is necessary to construct a mock random catalogue which is a realistic distribution of sources that could be observed but has no knowledge of large scale structure. With a uniform noise distribution, this would involve constructing a catalogue where random positions across the observable survey area are selected. However as can be seen in Fig. \ref{fig:localrms}, the noise across the field of view is non-uniform. This will affect how sources of different flux densities can be detected across the field of view. To account for this non-uniform noise, therefore, and its effect on the detection of sources when constructing a random catalogue, we follow the method of \cite{Hale2018}. 

To obtain a mock catalogue that accurately reflects radio sources that could be observed with LOFAR we make use of the SKA Design Study Simulated Skies \citep[SKADS;][]{SKADS2008, SKADS2010}. 
These extragalactic simulated catalogues provide a realistic distribution of sources that could be observed across 100 square degrees, with flux density measurements at five frequencies ranging from 151 MHz to 18 GHz. 
These sources are a mixture of both AGN as well as SFGs and have further information on the type of AGN (\cite{FanaroffRiley1974} Type I/II sources as well as radio quiet quasars) or SFG (i.e. normal star forming galaxy or  starburst). 
As these SKADS catalogues have realistic radio flux density distributions, they are used to construct a mock catalogue by comparing whether the flux density of a randomly generated source from the SKADS catalogue could be observed above the noise within the LoTSS image. 

Therefore, the rms maps from LoTSS were used to determine whether a randomly generated source would be detectable above the noise and could realistically be observed. To generate a mock catalogue, random positions within the observed region were generated and a flux density from the SKADS catalogue were also assigned to the sources. Under the assumption that the source is unresolved, the flux density from SKADS\footnote{Using the 1.4 GHz fluxes scaled to the frequency of LoTSS using $\alpha=0.7$} was combined with a randomly generated flux density to account for the noise at the position \citep[see][]{Hale2018} to form a total “measured" flux density. This noise was selected from a normal distribution centred on zero with a sigma given by the rms at that position. The measured flux density for a source was then compared to the rms noise at the location of the source. A source only remained within the mock catalogue if this measured flux density was at least five times greater than the rms value at its position. We generated enough random positions until we had roughly a total of 20 times the number of detected sources of the LoTSS- DR1 radio source catalogue.  

The distribution of the sources within this mock catalogue (after masking has been applied) can be seen in Fig.~\ref{fig:mockcatalogues}.

\begin{figure*}
\centering
\includegraphics[width=\linewidth]{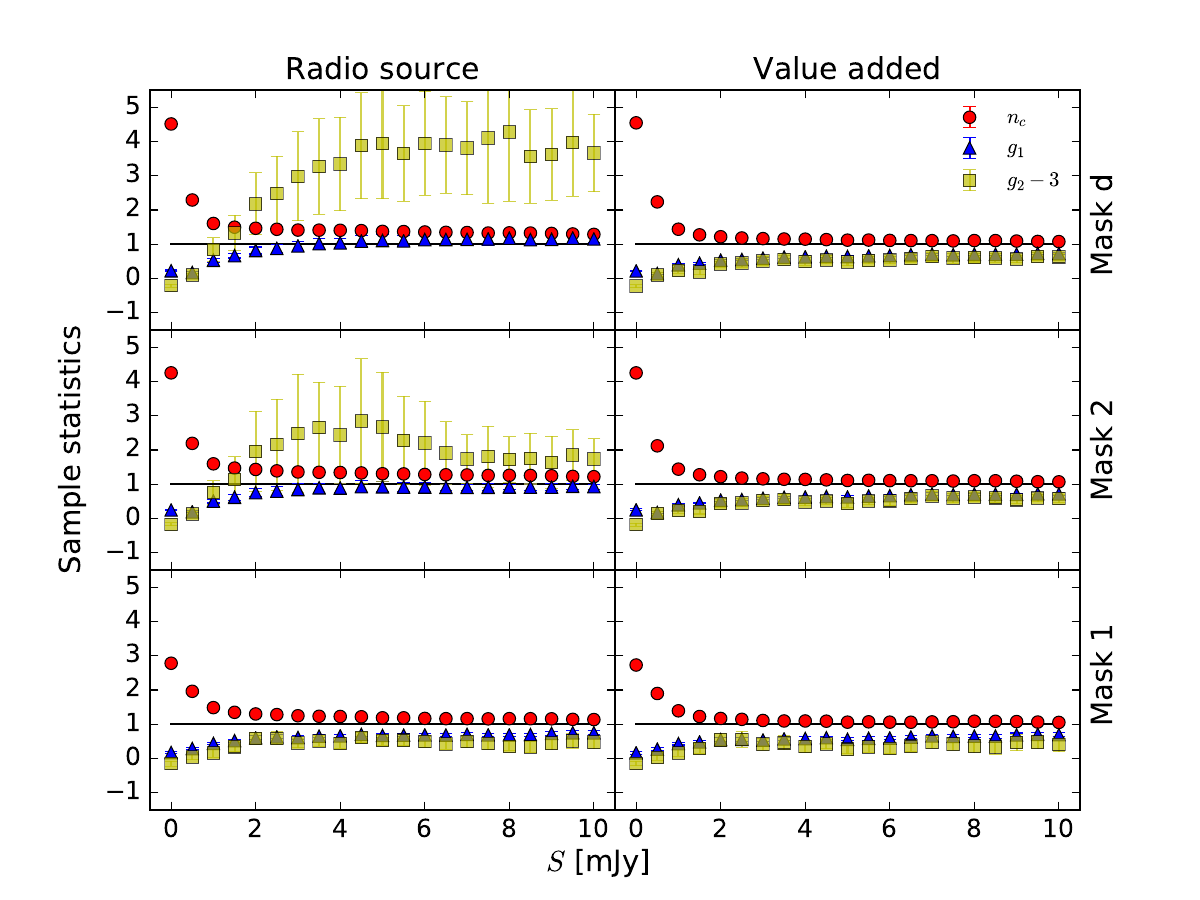}
\caption{Sample statistics of number counts in cells as a function of flux density threshold. Shown are the clustering parameter $n_c$ (variance over mean), which is expected to be one 
for the Poisson distribution, the skewness $g_1$ and excess kurtosis $g_2 - 3$ with error bars calculated from 100 bootstrap samples. On the left hand side for the LoTSS-DR1 radio source catalogue, on the right hand side for the 
LoTSS-DR1 value-added source catalogue. From top to bottom: mask d and masks 2 and 1.}
\label{fig:statistics}
\end{figure*}

\section{One-point statistics \label{sec:onepoint}}
   
\subsection{Distribution of radio source counts}

As shown in Sect. \ref{sec:data}, the distribution of number counts
is broader than expected for a Poisson distribution.
The naive assumption of a Poisson distribution arises from the 
expectation of a homogeneous and isotropic universe and independent, identically distributed  and point-like radio sources. 

There are at least four contributions to a deviation from a homogeneous spatial Poisson process: 
a) multi-component sources \citep{Magliocchetti1998},
b) fluctuations of the calibration,
c) confused sources (several sources are counted as a single source),
d) cosmic structure.
Here we investigate the statistical properties of the counts in cell by 
measuring moments of the empirical counts-in-cell 
distribution and comparing it to theoretical models. 

Let $k_i$ denote the counts in the $i$th cell. Then the 
central moments of a sample map are given by: 
\begin{equation}
m_j =\frac{1}{N_\mathrm{cell}}\sum_{i=1}^{N_\mathrm{cell}}(k_i-\mu)^j, 
\end{equation}
with the sample mean:
\begin{equation}
\mu = \frac{1}{N_\mathrm{cell}} \sum_{i=1}^{N_\mathrm{cell}} k_i.
\end{equation}

To analyse the counts-in-cell statistics, we calculate the clustering parameter 
$n_c$ (see Eq.~\ref{eq:nc}) as a function of the flux density threshold. We also calculate the 
coefficients of skewness ($g_1$) and excess kurtosis ($g_2-3$) \citep{ZwillingerKokoska2000}:
\begin{equation}
g_1 \equiv \frac{m_3}{m_2^{3/2}}, \quad
g_2 - 3 \equiv \frac{m_4}{m_2^2}-3.
\end{equation}
For the Poisson distribution, Eq.~(\ref{eq:poisson}), with $\lambda = \mu$, we find:
\begin{equation}
g_1^{P} = \mu^{-1/2}, \quad
g_2^{P}-3 = \mu^{-1}, 
\end{equation}
and $n_c^P = 1$.

For the compound Poisson distribution (Eq. \ref{eq:compoundpoisson}), 
\begin{equation}
g_1^{CP} = \frac{\gamma^2+3\gamma+1}{(\beta \gamma)^{1/2} (\gamma+1)^{3/2}}, \
g_2^{CP} -3 = \frac{\gamma^3+6\gamma^2+7\gamma+1}{\gamma\beta(\gamma+1)^2},
\end{equation}
and $n_c= 1 + \gamma$. With $\beta\gamma =\mu $ we can rewrite the coefficients as: 
\begin{align}
&g_1^{CP} = \frac{1}{\sqrt[]{\mu}}\left[\frac{n_c^2+n_c-1}{n_c^{3/2}}\right], \\
&g_2^{CP} -3= \frac{1}{\mu}\left[\frac{n_c^3+3n_c^2-2n_c-1}{n_c^{2}}\right].
\end{align}

In Fig.\ \ref{fig:statistics}  we show the clustering parameter $n_c$ (red circles) and the coefficients of 
skewness (blue triangle) and excess kurtosis (yellow squares) for the LoTSS-DR1 radio source and the 
LoTSS-DR1 value-added source catalogues as a function 
of flux density threshold and for three different masks (mask d, mask 2 and mask 1). 
Error bars are computed from 100 bootstrap samples, but are in most cases smaller than the symbol.
It can be seen that for the lowest flux density thresholds $n_c$ is well above unity, 
but at flux density thresholds above $1$~mJy, the clustering parameter 
is almost constant and only slightly above unity. It approaches unity faster for the value 
added catalogue. It is also interesting to observe that the radio source 
catalogue shows a strong evolution of excess kurtosis 
$g_2 - 3$ with increasing flux density threshold, except for noise mask 1, which 
masks all but the cleanest cells. In contrast, the value-added catalogue shows the 
qualitatively expected behaviour for excess kurtosis and skewness for all masks
considered. The value-added catalogue differs from the original radio source catalogue in a 
statistically significant way, especially with respect to higher moments, despite the fact that the number of sources in both catalogues differs by less than 2 per cent.

\begin{table}
	\centering
    \caption{Pearson $\chi^2$-test statistic of counts-in-cell distribution for the masked LoTSS-DR1 value-added source catalogue 
    with `mask d' for four flux density thresholds. We compare a Poisson (P) and a Compound Poisson (CP) distribution to the measured histograms. For each threshold value, we provide 
    the number of sources in the catalogue, the clustering parameter $n_c$, the reduced $\chi^2$-values ($\chi^2$/dof) and the degrees of freedom 
    (dof = number of histogram bins minus number of parameters of distribution) for both statistical 
    models.
     \label{tab:chi2}}
    \begin{tabular}{ccccccc}
    \hline\hline
    $S_\mathrm{min}$ & $N$ & $n_c$&$\frac{\chi^2_\mathrm{P}}{\text{dof}_\mathrm{P}}$ & dof$_\mathrm{P}$ & $\frac{\chi^2_\mathrm{CP}}{\text{dof}_\mathrm{CP}}$& dof$_\mathrm{CP}$ \\
    
       	[mJy] & && & && \\\hline
        1 & 102\,940&1.44&30.67 & 32 & 0.76 & 31 \\
        2 & 51\,288&1.22&11.67 & 20 & 1.12 & 19 \\
        4 & 30\,556& 1.15&7.69 & 14 & 1.38 & 13 \\
        8 &  19\,612&1.11& 3.52 & 11 &  0.46& 10 \\\hline
    \end{tabular}
\end{table}

\begin{figure}
	\centering
    \includegraphics[width=\linewidth]{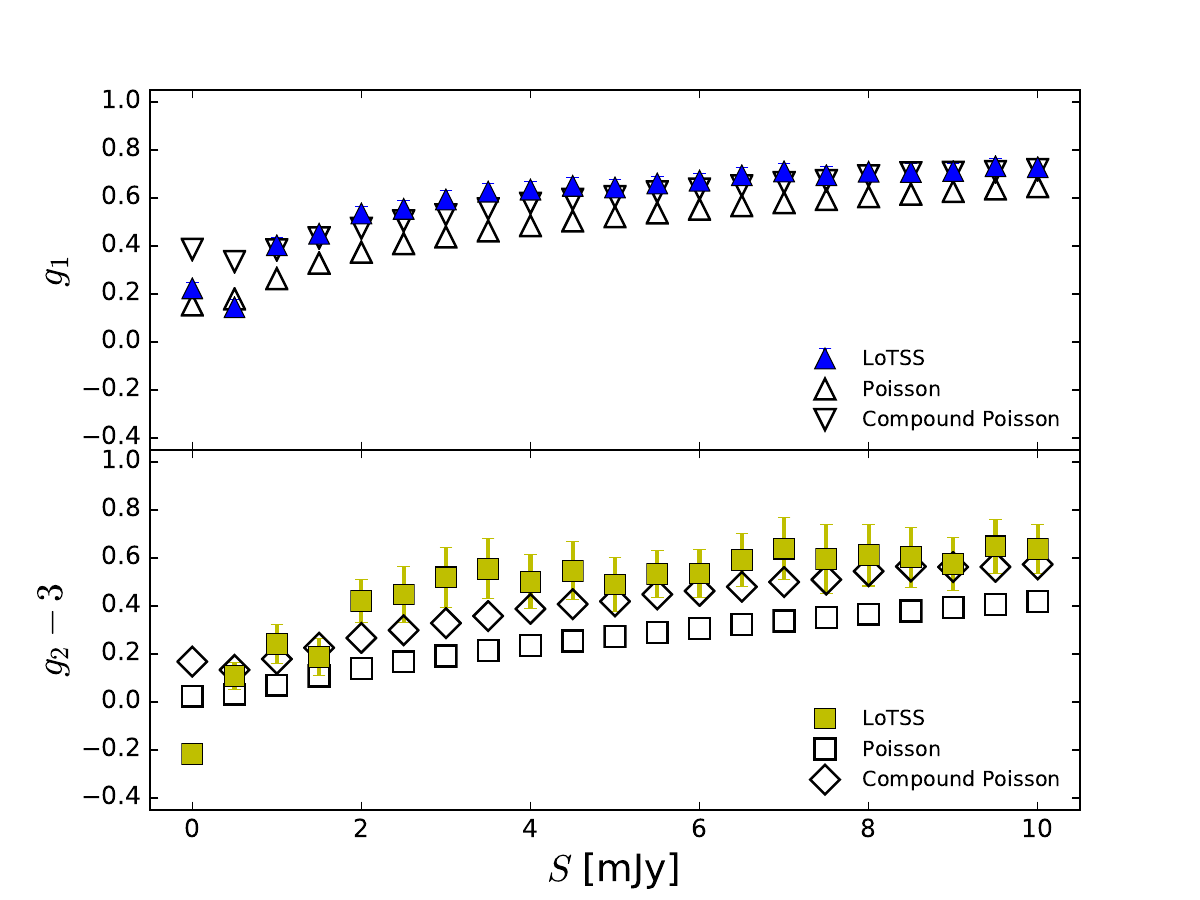}
    \caption{Shown are the skewness ($g_1$) and excess kurtosis ($g_2-3$) of the masked LoTSS DR1 value-added source catalogue (mask d), also plotted are the expected moments of a Poisson and compound Poisson distribution. Errors bars for the data sample are computed from bootstrap sampling. \label{fig:expectedmoments}}
\end{figure}

In Fig.\ \ref{fig:expectedmoments} we compare the observed coefficients of skewness and excess
kurtosis of the LoTSS-DR1 value-added source catalogue with `mask d' to their theoretical expected values for a Poisson and a compound Poisson distribution. 
We observe that the compound Poisson distribution provides a significant improvement over the 
Poisson distribution, which extends to values well into the regime in which we can regard the 
catalogue to be complete.

\begin{figure*}
	\centering
    \includegraphics[width=0.45\linewidth]{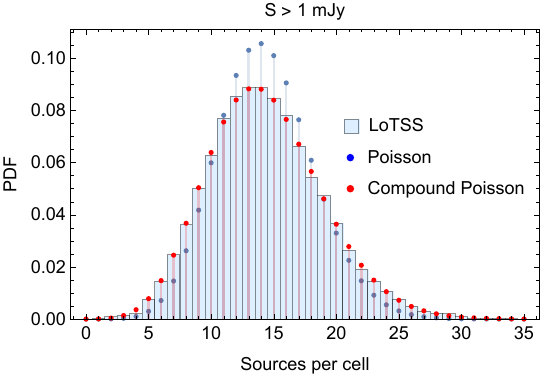} 
     \includegraphics[width=0.45\linewidth]{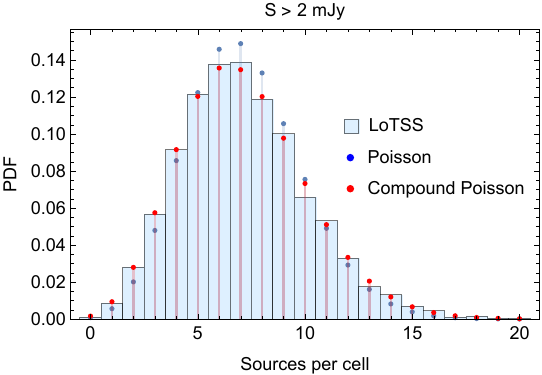} \\[3pt]
     \includegraphics[width=0.45\linewidth]{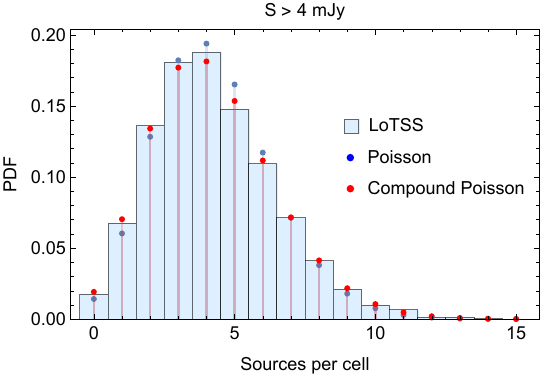}
      \includegraphics[width=0.45\linewidth]{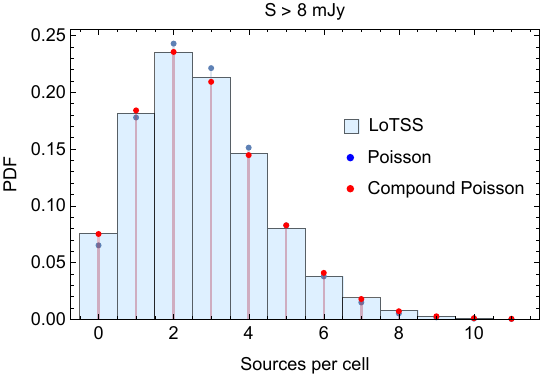}
    \caption{Histograms of LoTSS-DR1 counts-in-cell for the flux density thresholds 1, 2, 4 and 8~mJy.
    Also shown are the best-fit Poisson and compound Poisson distributions.}
    \label{fig:dist_histo}
\end{figure*}

To further quantify the quality of fit, we tested both distributions with a Pearson chi-square test
for four different flux density thresholds applied on the LoTSS-DR1 value-added catalogue with mask d. 
The results of that test are shown in Fig.\ \ref{fig:dist_histo} and Table\ \ref{tab:chi2}.  While the 
coefficient of skewness shows very nice agreement between the compound Poisson distribution and 
the data, the coefficient of excess kurtosis shows better agreement with the compound Poisson 
distribution compared with the Poisson distribution. In terms of the Pearson $\chi^2$-test the 
compound Poisson distribution describes the data significantly better than the Poisson distribution, 
see Table\ \ref{tab:chi2}. Values of $\chi^2/$dof of order unity indicate a good fit. For the 
$1$~mJy sample, this ratio is $30.7$ and $0.76$ for the Poisson and compound 
Poisson distributions, respectively.

We conclude that the counts-in-cell distribution of the LoTSS-DR1 value-added catalogue is 
not Poissonian. The compound Poisson distribution provides an excellent fit to the 
data, but other distributions (not studied in this work) might also provide a good fit to the data.

\begin{figure}
\centering
   \includegraphics[width=\linewidth]{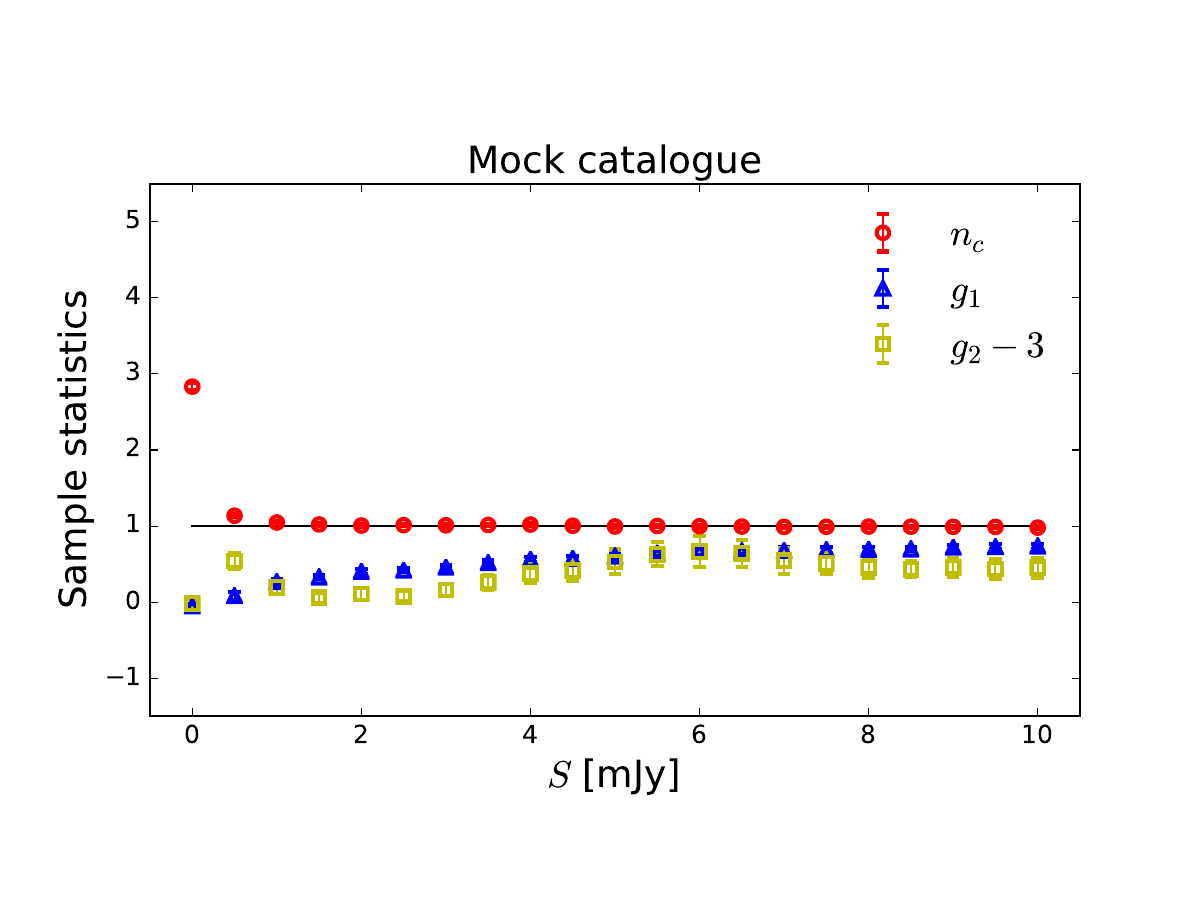}
    \caption{Clustering parameter and coefficients of skewness and kurtosis for a subsample 
    of the mock catalogue, which matches the size of the value-added source catalogue. Error bars are computed from bootstrap sampling.}
    \label{fig:mock_moments}
\end{figure}

We can also test if the mock catalogue shows the same statistical behaviour as the data. 
Their clustering parameter and coefficients of skewness and excess kurtosis are 
shown in Fig.~\ref{fig:mock_moments}. In order to compare the mock catalogue to the 
LoTSS-DR1, we randomly draw subsamples of the mock catalogue that contain the same 
number of data points as the LoTSS-DR1 value-added source catalogue. 
At $S > 1$~mJy, we find that the clustering parameter in the mocks is closer to one and the higher 
statistical moments are closer to a Poisson distribution than the LoTSS-DR1 value-added source 
catalogue. We checked that fitting a compound Poisson distribution to the mocks also improves the 
fits (as there are more free parameters), but not by as much in the case of the LoTSS-DR1 value 
added source catalogue. We thus conclude that there are indeed clustering effects in the LoTSS-DR1 data on top of the effects that are taken care of in the mock catalogue.

\subsection{Differential source counts}

\begin{figure*}
	\centering
	\includegraphics[width=\linewidth]{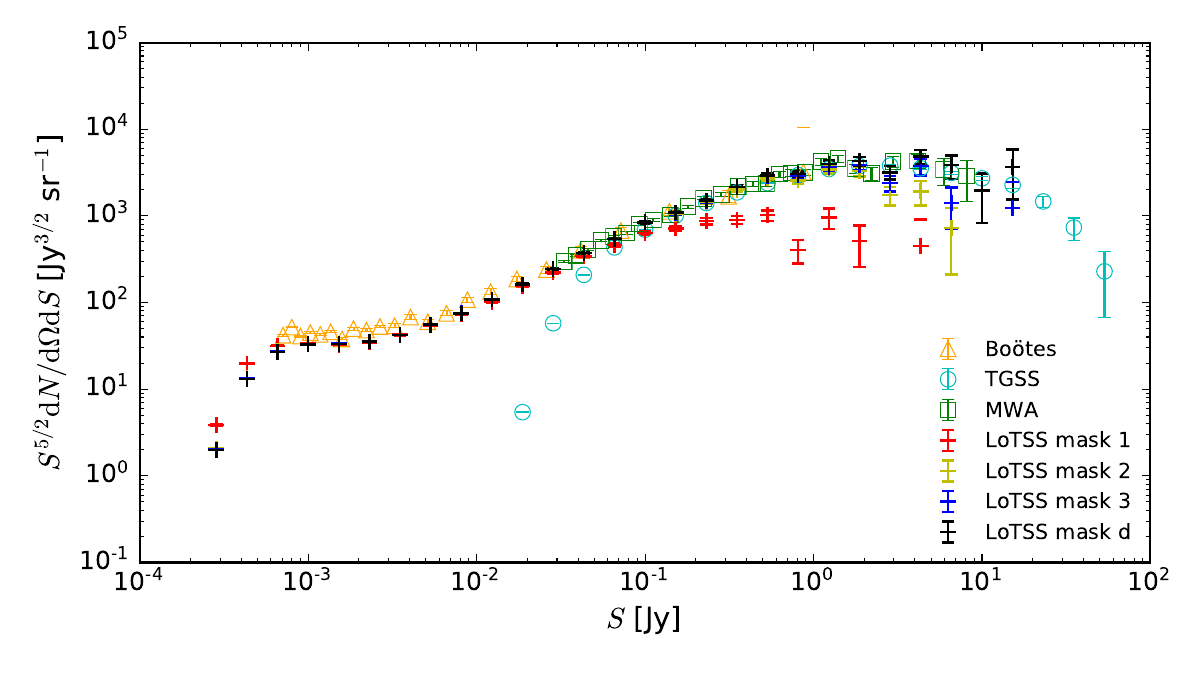}
	\caption{Differential number counts per flux density interval of the masked LoTSS-DR1 value-added source catalogue for four different masks. Additionally the masked 
	TGSS-ADR1 ($147.5$~MHz; this work, blue circle), the LOFAR Boötes field (\citealt{Bootes2016}, orange triangle) 
	and the MWA ($154$~MHz; \citealt{MWA2016}, green box) are shown.
			Error bars for the LoTSS and TGSS counts are due to Poisson noise in each flux density bin, which have equal step width in $\log_{10}(S)$.}
	\label{fig:diffnumbercounts}
\end{figure*}

\begin{figure}
	\centering
	\includegraphics[width=\linewidth]{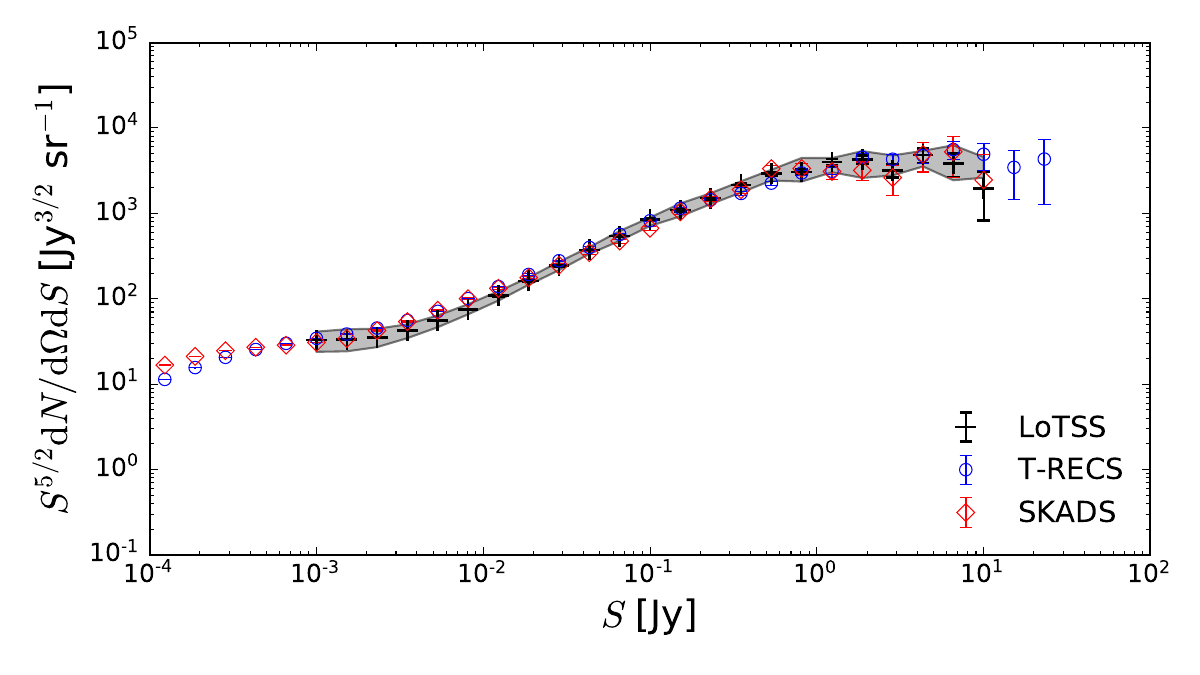}\\
	\includegraphics[width=\linewidth]{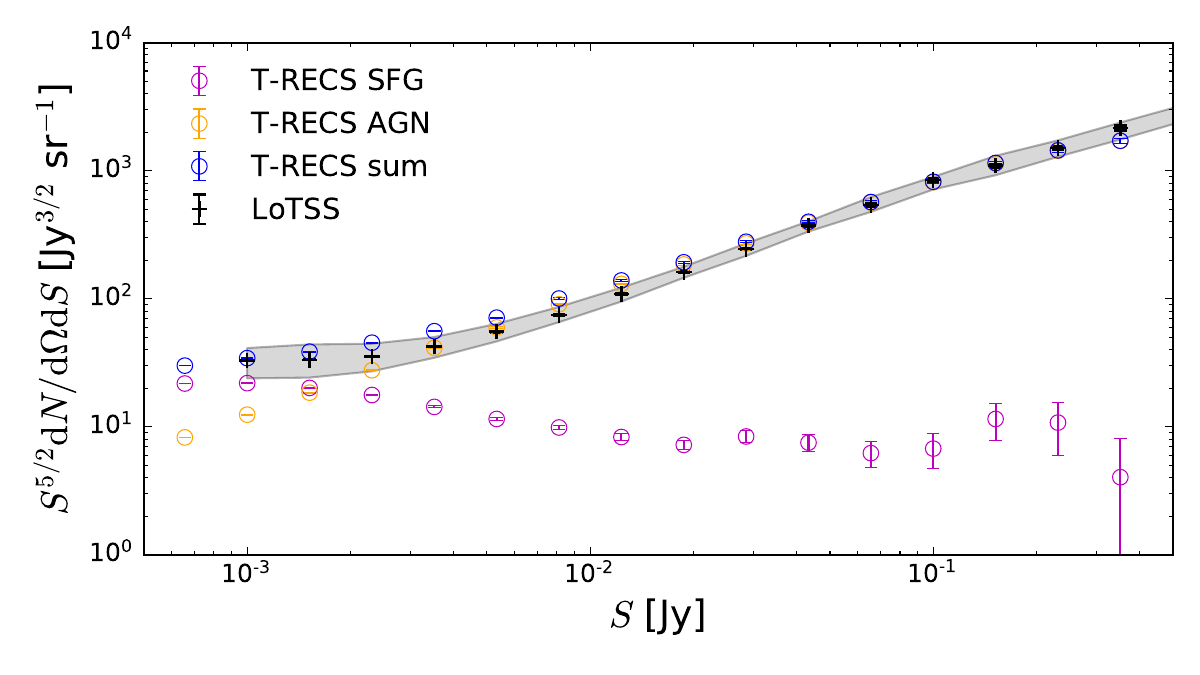}
	\caption{Top: Comparison of LoTSS-DR1 differential source counts using `mask d' and 
	        SKADS $151$~MHz  and T-RECS  `wide' 150 MHz simulations. The grey band 
	        corresponds to a $\pm 20\%$ variation of the LoTSS flux density scale due to uncertainties in 
	        the flux density calibration.
		Bottom: Contributions from AGNs and SFGs in the T-RECS `wide' differential source 
		counts as compared to the total LoTSS-DR1 differential source counts.
		All: Error bars are due to Poisson noise in each flux density bin, 
		which have equal bin width in $\log_{10}(S)$. 
		}
	\label{fig:diffnumbercounts2}
\end{figure}

Let us now turn our attention to the differential source counts as a function of flux density 
(we use the integrated flux density for all sources). 
In Fig.\ \ref{fig:diffnumbercounts} we plot the differential number counts of the LoTSS-DR1 
value-added source catalogue with Euclidean normalisation, i.e.\ in a static, homogeneous 
and spatially flat Universe the normalised counts would be constant as a function of flux 
density. The bins in the differential number counts plot have equal step 
width in $\log_{10}(S)$. We determine the source counts for four masks (masks d, 1, 2, and 3) 
applied.

The errors are assumed to follow Poisson noise in each bin. This assumption seems to be 
in contradiction to our findings from the previous section. Therefore, we alternatively estimated 
the errors by means of 100 bootstrap samples of the masked survey.  Sample mean and 
standard deviation of the 100 bootstrap samples turn out to be in agreement with analysis that 
just assumes Poisson noise for each bin. Surprisingly, the bootstrap sample variance tends to 
be slightly smaller over the complete flux density range. For simplicity and to be on the safe side we thus 
show the Poisson noise only. Stating the fact that the value-added and masked source catalogue 
is $95\%$ point-source complete at $0.39$~mJy (note that this is for the total source counts, the 
differential counts at that flux density are already incomplete), we refrain from applying any 
completeness corrections to the differential number counts, but instead work with flux density 
thresholds.

Figure \ref{fig:diffnumbercounts} shows that noise mask 1, and to lesser extent mask 2, result in a lack of sources at high flux 
densities. This can be easily understood as masking regions with larger rms noise selects regions 
that include the high-flux density sources, since limited dynamic range leads to increased rms 
noise in their neighbourhood. At low flux densities, applying the strongest noise mask (mask 1), 
the differential number counts show increased completeness at low flux densities compared to 
all other masks. This difference shows up below $1$~mJy. This is an independent confirmation that 
the value-added source catalogue has a high degree of completeness at $S > 1$~mJy. This test 
also allows us to argue that it is not only point source complete, but also shows a high degree of 
completeness for extended sources, as this test does not distinguish between point sources 
and resolved sources. Independently from the arguments given in the previous section we arrive 
at the conclusion that we can trust the source counts at $S > 1$~mJy. 

For comparison we also plot the masked source counts for the TGSS-ADR1 radio source catalogue, which agree
very well with the LoTSS-DR1 value-added source counts for flux densities between
$80$~mJy and $20$~Jy. In order to obtain the differential number counts of the 
TGSS-ADR1, we masked the Milky Way with a cut in galactic latitude 
at $|b|\leq 10$~deg, discarded unobserved regions and missing 
pointings with a {\sc HEALPix} mask at $N_\mathrm{side} = 32$. 
On top we applied a noise mask with an upper cut in local rms noise of $5$~mJy/beam (see App. \ref{app:A} and Fig. \ref{fig:TGSS} ).
For the TGSS there are more sources detected at higher flux densities than shown in the differential source counts as we focus on the available flux density range defined by the LoTSS-DR1 sample. The decreasing trend of the source counts at higher flux densities is not physical and can be explained by the masking procedure. Masking with larger cells at the same noise levels will average over larger regions and therefore samples over larger number of sources. Therefore bright and noisy sources will be more often taken into account in the analysis than by masking with higher resolutions. 

Additionally we also plot the differential source counts 
from \cite{MWA2016} obtained with the MWA $154$~MHz survey and 
from \cite{Bootes2016} obtained with LOFAR at $150$~MHz from the Boötes field.
We find that the LoTSS-DR1 value-added source catalogue agrees well with these existing 
studies. Note that no completeness corrections (besides masking) are applied to the LoTSS data, 
while the Boötes and MWA analysis do include such corrections. Remaining discrepancies might 
be due to the $20$ per cent uncertainty of the LoTSS-DR1 flux density scale calibration \citep{LoTSS2019A}. 

Finally, we compare the LoTSS-DR1 data to two simulations of the radio sky, the 
SKA Design Study simulations (SKADS, \citealt{SKADS2008}) and the Tiered Radio
Extragalactic Continuum Simulations (T-RECS, \citealt{TRecs2018}), see Fig.\ \ref{fig:diffnumbercounts2}. 
We find that both simulations are in good agreement with LoTSS-DR1.
We also indicate the systematic uncertainty of the LoTSS-DR1 
flux density scale, discussed in detail in \citet{LoTSS2019A}, on the mean values of the differential 
source counts and show it as a grey band in the figure. Note that the flux density scale uncertainty 
is larger than the uncertainty from Poisson noise at most flux densities, except for a few bins at 
the highest flux densities.

The sample we choose for the SKADS simulations covers $100$~square degrees of the sky, with 
a minimum flux density of $1~\mu$Jy at $1.4$~GHz. It contains $6.1\times 10^6$ sources in total, 
which we consider at a frequency of $151$~MHz.
There is a small discrepancy in the flux density range from $3$ to 
$12$~mJy (see top panel of Fig.\ \ref{fig:diffnumbercounts2}), 
otherwise the agreement is excellent down to $0.7$~mJy. In the light of the already mentioned 
20 per cent error on the flux density calibration, the discrepancy does not seem to be significant. 

Three different settings are available from T-RECS for the two main radio source populations 
(active galactic nuclei and star-forming galaxies). For our source count comparisons we use the `wide' 
catalogue, which simulates a sky coverage of $400$~square degress with a lower flux density limit of 
$100$~nJy at $1.4$~GHz. The T-RECS `wide' catalogue does not include effects of 
clustering \citep{TRecs2018}, while the `medium' T-RECS catalogue does. We checked that this does 
not result in any significant differences for the differential source counts for the range of flux 
densities considered in this work. For all T-RECS catalogues frequency bands between 
$150$~MHz and $20$~GHz are provided. Here we use the flux densities at $150$~MHz. 
In Fig.~\ref{fig:diffnumbercounts2} the differential source counts of AGNs and SFGs are shown, 
as well as the sum of both populations. We find that T-RECS is in good agreement with the data of the masked LoTSS-DR1, except for a small discrepancy in the flux density range from $3$ to 
$12$~mJy.

\subsection{Consistency based on photometric redshift information}\label{sec:redshift}

\begin{table}
	\centering
	\caption{Number of sources of the masked (mask z) LoTSS-DR1 value-added 
		source catalogue for various flux density thresholds and for different values of minimum redshift $z$. 
		$N_z$ denotes the number of radio sources with redshift information `z\_best' and 
		$N$ is the total number of sources for the given cuts. Objects without redshift information 
		are included in $N$. There are 145\,839 radio sources without redshift estimate at any $S$ and 
		50\,358 radio sources with $S > 1$~mJy. Also shown is the fraction of sources with 
		redshift information $f_z = N_z/N$.}
	\begin{tabular}{lcccc}
		\hline\hline
		$z$ & $S_{min}$ & $N$ & $N_z$ 
		& $f_z$ \\
		& [mJy] & & & \\\hline
		\multirow{5}{*}{all} & 0 & 298\,950 &153\,111
		& 0.512 \\
		&1& 102\,370 & 52\,012 
		& 0.508 \\
		& 2& 50\,977&24\,420& 0.479 \\
		& 4 & 30\,372& 14\,506& 0.478\\
		& 8& 19\,499& 9591 & 0.492\\ \hline
		\multirow{2}{*}{> 0.2} & 0 &  & 130\,571 &  \\
		&1&  &  40\,295&  \\\hline 
		\multirow{2}{*}{> 0.5} & 0 &  & 81\,940 &  \\ 
		& 1&  & 26\,014 &  \\\hline 
		\multirow{2}{*}{> 1.0} & 0 & &18\,854 &  \\ 
		& 1 &  & 6651 & \\\hline 
	\end{tabular}
	\label{tab:redshiftcuts}
\end{table}

As already mentioned in the introduction, a large fraction of LoTSS-DR1 radio sources have identified 
infrared (72.7\%) and optical (51.5\%) counterparts, which allow for an estimate of a 
photometric redshift for around half of LoTSS sources \citep{LoTSS2019C}. Some of the 
identified objects also have spectroscopic redshift information available.  
Below we use the `z\_best' redshift information, which is the spectroscopic redshift when it is available 
and a photometric estimate in all other cases, from the LoTSS-DR1 value-added source 
catalogue to learn more about the contribution of local structure to the one- and two-point statistics.

The photometric redshifts in the catalogue are extracted from a combination of infrared/optical data from 
WISE/Pan-STARRS.  Due to missing Pan-STARRS information in the strip 
$55.0000~\text{deg} <$ Dec $< 55.2245$~deg and  RA $< 184.4450$~deg, we 
lack photometric redshifts from that strip. The only available data would be redshifts inferred 
from spectroscopic information of sources that match to a WISE catalogue source.
To account for that effect, we additionally mask that strip (see Fig. \ref{fig:pointings}), 
whenever we use redshift information and will denote this as `mask z'. 

Applying cuts in redshift rejects radio sources and the source density per cell 
decreases significantly.  In Table \ref{tab:redshiftcuts} we show how the total number of 
LoTSS-DR1 value-added sources changes after applying `mask z' for different 
minimal values of redshift, without and with a flux density threshold of $1$~mJy. 
For about $51\%$ of all radio sources redshift information is available and this number does not 
change significantly when we restrict the analysis to radio sources with flux densities above 1 mJy. 

\begin{figure}
\centering
\includegraphics[width=\linewidth]{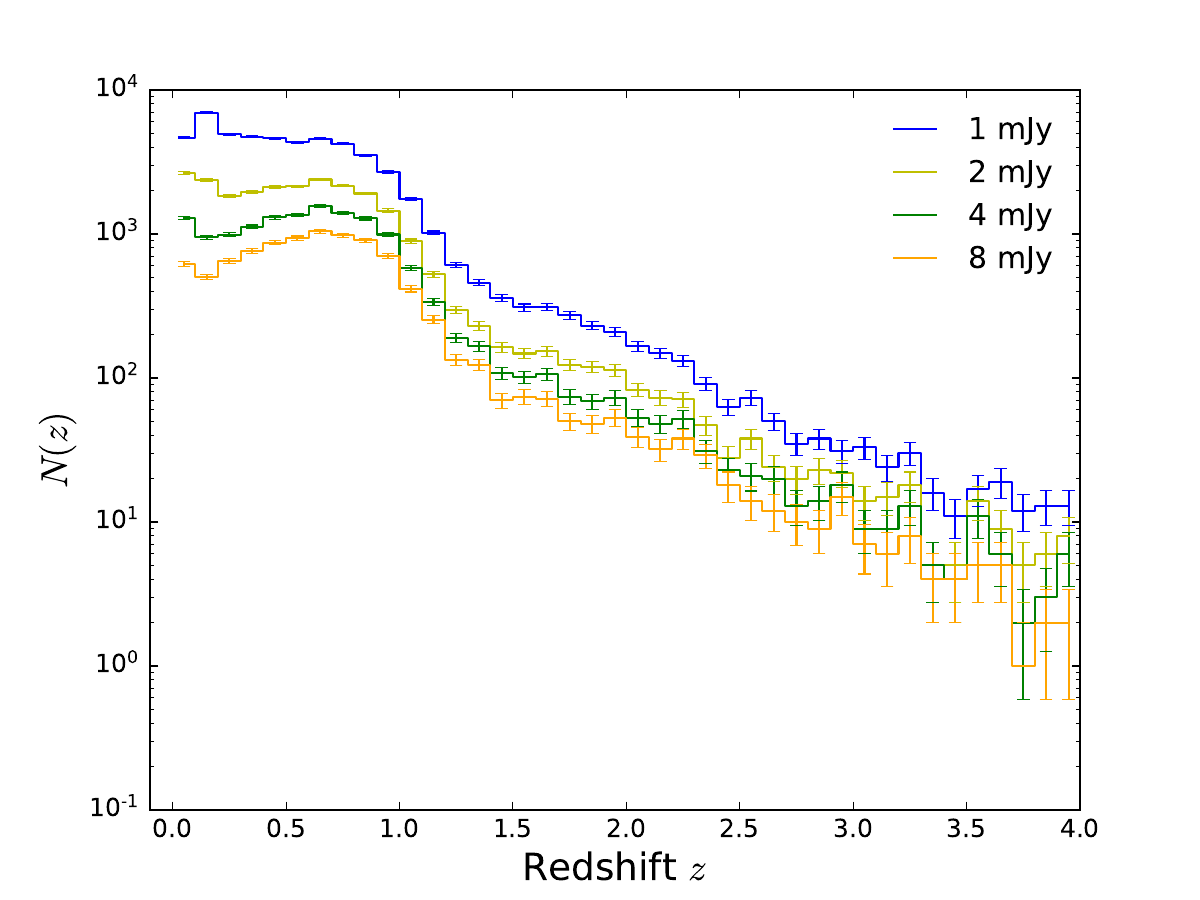}
\caption{Number of radio sources as a function of available $z$ for four different 
flux density thresholds, with error bars due to Poisson noise. Only sources with available redshift (`z\_best') of the LoTSS-DR1 value 
added source catalogue after applying `mask z' are considered here.}
\label{fig:zhistfluxthresholddndodz}
\end{figure}

\begin{figure}
\centering
\includegraphics[width=\linewidth]{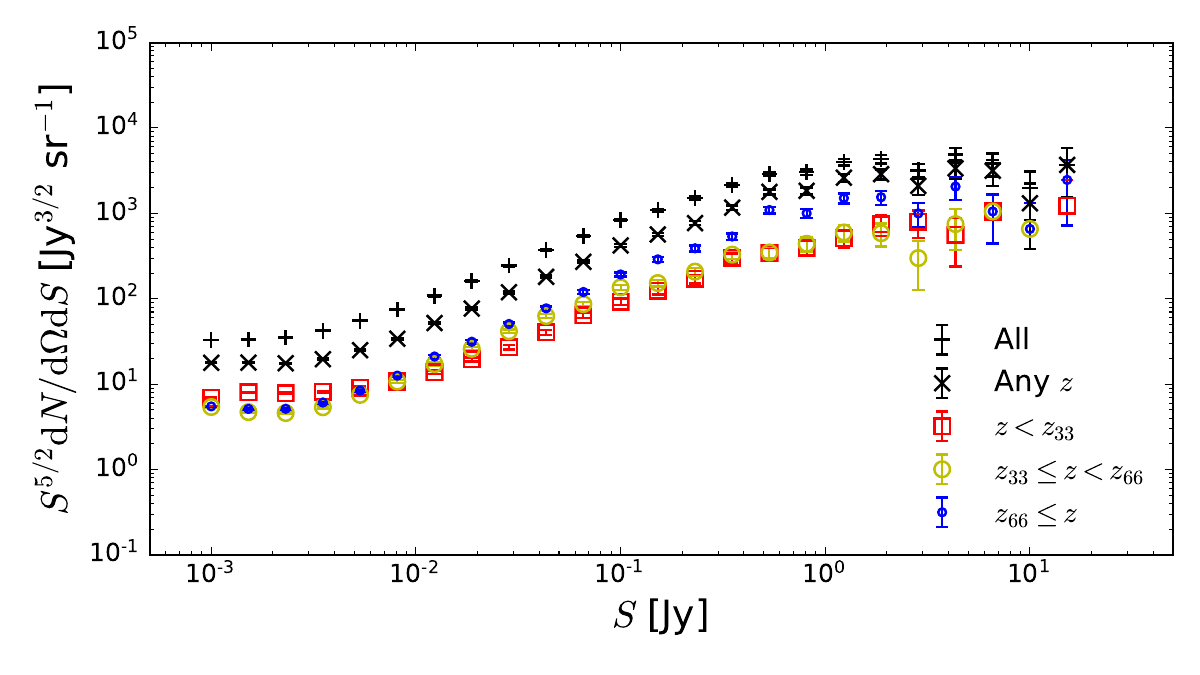}
\caption{Differential source counts of the LoTSS-DR1 value-added sources masked with `mask z' separated by 
redshift percentiles, $z_{33}=0.376$ and $z_{66}=0.705$. Additionally the differential 
source counts of all sources (`All') and of all sources with redshift information (`Any $z$') are shown. }
\label{fig:lotssdiffcountsredshift}
\end{figure}

The distribution of radio sources with available redshift estimate is shown in 
Fig.~\ref{fig:zhistfluxthresholddndodz} for the four samples with flux density thresholds of 
$1, 2, 4$ and $8$~mJy, respectively.  The brighter samples show the mode of the distribution at 
$z \approx 0.7$, while the $2$~mJy sample is bimodal and the $1$~mJy sample has its mode at 
$z \approx 0.1$. The median redshift increases continuously from $0.50$ for the $1$~mJy to $0.64$ for the $8$~mJy sample. This is in good qualitative agreement with the expectation (supported also by 
the simulations discussed above), that the brighter samples are 
dominated by AGNs at relatively high redshift while in the faintest sample SFGs at lower redshift start to dominate the statistics. First classifications of AGNs and SFGs in the LoTSS-DR1 catalogue have been done by \cite{Hardcastle2019} and \cite{Sabater2019}.
We additionally separated all sources with available redshift information after masking with `mask z'  by the $33$ and $66$ percentiles, which are:
\begin{equation}\label{eq:percentiles}
z_{33} = 0.376 \mbox{ and }z_{66}=0.705,
\end{equation}
respectively. From these three samples we inferred the differential source counts, which are presented in Fig. \ref{fig:lotssdiffcountsredshift}. These differential source counts support the above expectation, that the source distribution at fainter flux densities is dominated by objects at lower redshift and vice versa at brighter flux densities by objects at higher redshift.

Radio sources with redshift information are very likely (non-zero probability of misidentification) to be real sources and so we can consider that sample of 
radio sources as an independently confirmed sample. It is then interesting to compare its 
statistical properties with those of the sample without redshift information.  

\begin{figure}
\centering
\includegraphics[width=\linewidth]{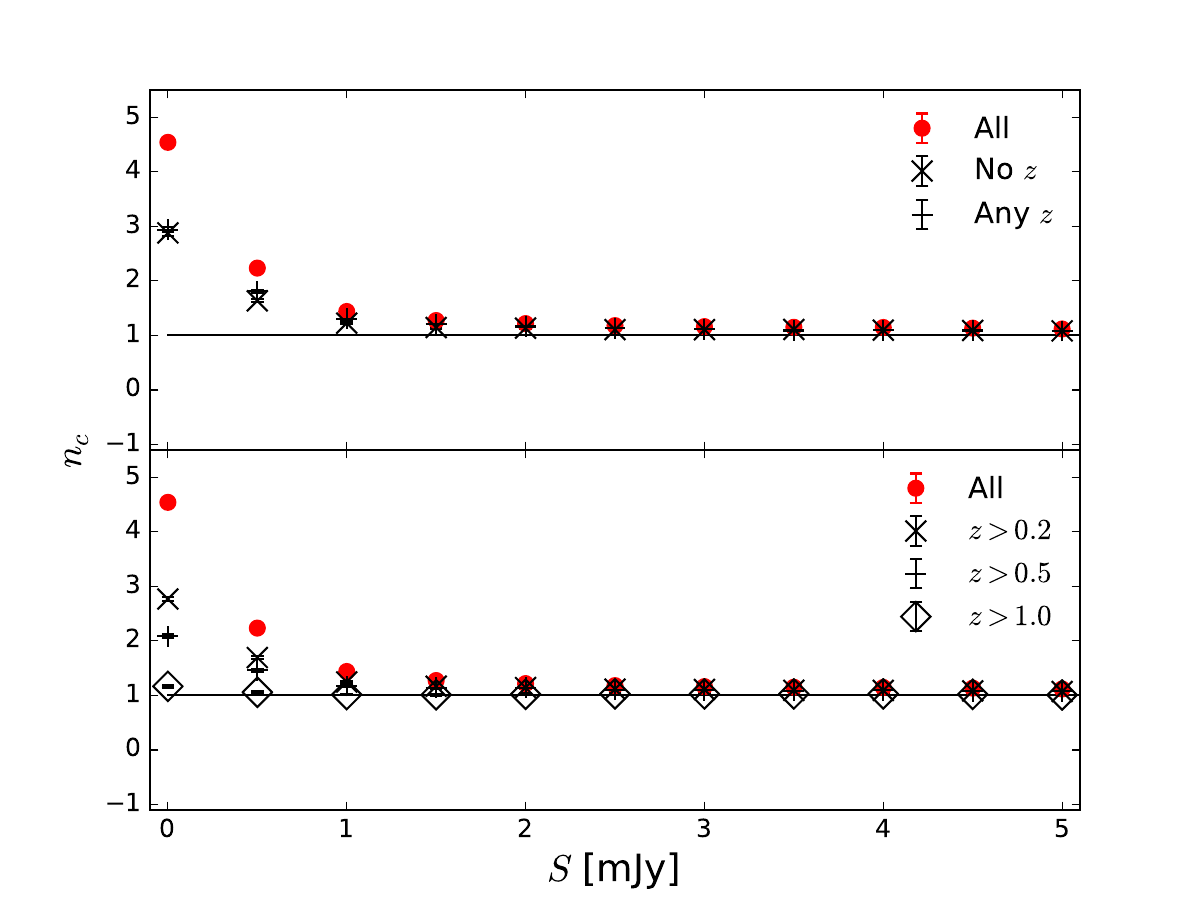}
\caption{Clustering parameter $n_c$ as function of flux density threshold and available redshift 
information based on  the value  `z\_best' from the LoTSS-DR1 value-added source catalogue after 
application of `mask z'. 
Top: We compare radio sources with and without redshift information and contrast them with the full sample. Bottom: Only objects with redshifts above the quoted value
are included in the respective data points. Error bars are computed from bootstrap sampling.}
\label{fig:photoz_variance}
\end{figure}

In Fig.\ \ref{fig:photoz_variance} we show the clustering parameter $n_c$ as a function of 
flux density threshold after applying `mask z'. In the top panel we compare the radio sources with redshift 
information to those without redshift information. We see that the values for $n_c$ agree very well 
with each other for all considered flux density thresholds. At flux densities below $1$~mJy, both sets of 
sources seem to cluster less than the sum of both sets.  

We also show in the bottom panel of Fig.\ \ref{fig:photoz_variance} how $n_c$ changes when 
we exclude all sources estimated to be below a certain 
redshift. 
Interestingly, we find that excluding radio sources from the local neighbourhood 
($z < 0.2$) decreases the clustering parameter $n_c$. The effect increases if we exclude 
radio sources from a larger volume and is strongest if we exclude all objects in the local Hubble 
volume ($z < 1$).
This effect is seen for all flux density thresholds, but is most prominent for thresholds below 
$1$~mJy. This is consistent with the expectation that there is more clustering in the 
late Universe, but a much more detailed study will be necessary to make quantitative statements, 
which we leave for a future work. We dismiss radio sources below $1$~mJy in the following 
section when we study the two-point correlation function. 

\begin{figure*}
\centering
\includegraphics[width=0.5\linewidth]{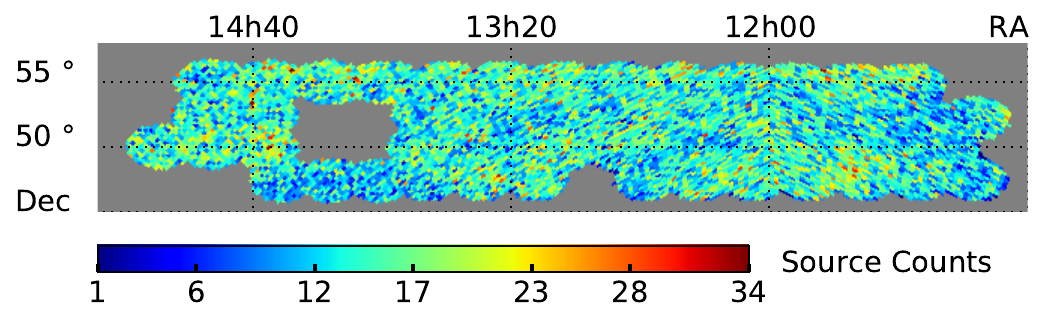}\includegraphics[width=0.5\linewidth]{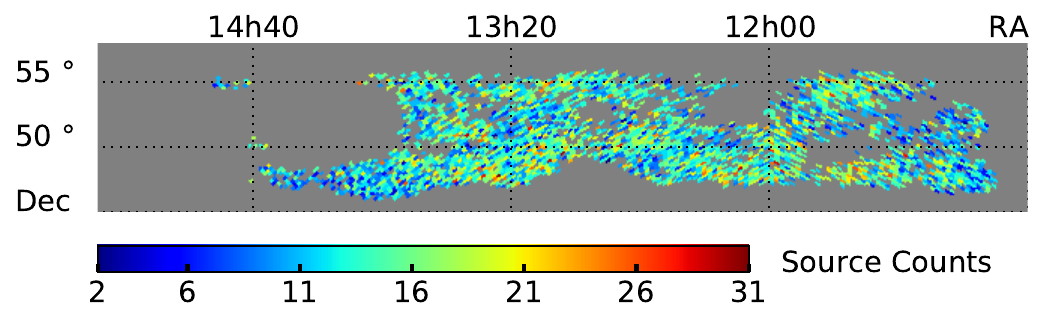}
\caption{Counts-in-cell map of the LoTSS-DR1 value-added source catalogue for  $S > 1.0$~mJy 
and after applying `mask d'  (left) and `mask 1' (right).} 
\label{fig:valueaddedcatalogue}
\end{figure*}

We conclude our study of the one-point statistics by pointing out that LoTSS-DR1 produces 
reliable radio source counts and 
shows statistical properties that are self-consistent and consistent with previous 
observations and simulations above integrated flux densities of $1$~mJy. The corresponding 
counts-in-cell map for `mask d' and `mask 1' with $S>1$~mJy is shown in Fig.~\ref{fig:valueaddedcatalogue}.

\section{Two-point statistics \label{sec:twopoint}}

\subsection{The angular two-point correlation function\label{sec:subtwopoint}}
In order to estimate the angular two-point correlation of radio sources we make use of the estimator proposed by \citet{LandySzalay1993}, 
\begin{equation}
\hat{w}(\theta) = \frac{DD - 2 DR + RR}{RR}, 
\label{eq:LSest}
\end{equation}
where $DD, DR$ and $RR$ denote the normalised pair counts at separation angle $\theta$ for data-data,
data-random and random-random source pairs (see App.~\ref{app:B} for details). 
The Landy-Szalay (LS) estimator has minimal bias and 
minimal variance and is claimed to be more robust than other estimators (see~\citealt{Kerscher2000} and App.~\ref{app:B}). 
Data points are taken from the LoTSS-DR1 value-added source catalogue and random points 
either from the mock catalogue (default) described in Sect.~\ref{sec:mocks}, or from a purely random sample. Data and random catalogues are 
masked alike. 

For a large enough random source catalogue, the expectation value of the LS
estimator is \citep{LandySzalay1993}:
\begin{equation}
\langle \hat{w}(\theta) \rangle = \frac{1+w(\theta)}{1 + w_\Omega} -1 \approx 
w(\theta) - w_\Omega, 
\end{equation}
where $w_\Omega = \int G_p(\theta) w(\theta) \mathrm{d} \theta$, with $G_p(\theta)$ being 
the normalized count of pairs of `atomic' cells (cells that are small enough to contain at most 
one point source) 
at separation $\theta$ in the analysed survey area. Thus the LS estimator (as well as all 
other estimators that have been proposed in the literature) is biased. The function 
$G_p(\theta)$ depends on the binning. 

The bias of the estimator is due to the so-called integral constraint, which is an effect of the finite survey area 
and reflects the fact that 
we cannot measure an unbiased estimate of the two-point correlation based on a single 
estimate of the total number of sources in the survey region. Given a model for
$w(\theta)$, we can estimate this bias from the random source catalogue via:
\begin{equation}\label{eq:bias}
w_\Omega = \frac{\sum_\mathrm{bins} RR(\theta) w(\theta)}{\sum_\mathrm{bins} RR(\theta)}.
\end{equation}

The variance of the estimator is \citep{LandySzalay1993}
\begin{align}
\mathrm{Var}[\hat{w}(\theta)] &= \left(\frac{1+w(\theta)}{1+w_\Omega}\right)^2
\frac{2}{N_d(N_d-1) G_p(\theta)} \\
& \approx \frac{2}{N_d(N_d-1) G_p(\theta)},  
\end{align}
where $N_d$ denotes the number of data points in the survey. The second expression 
holds for the assumption that the two-point correlation is small compared to unity. 
The factor $N_d(N_d - 1)/2$ scales the Poisson noise with the overall number of 
pairs and the factor $G_p(\theta)$ accounts for how many independent pairs can be 
probed at angular separation $\theta$.

For calculating the correlations we make use of the publicly available code {\sc TreeCorr}\footnote{\url{http://github.com/rmjarvis/TreeCorr}} in version $3.3$ \citep{TreeCorr2004}. 
{\sc TreeCorr} uses an algorithm that structures the sources in cells according to a 
logarithmic binning of cell separation. In that way the numerical problem 
of calculating the two-point correlations for 
objects in cells with $N_1$ and $N_2$ members
is reduced from scaling with $\mathcal{O}(N_1 N_2)$ to $\mathcal{O}(N_1 + N_2)$, which 
leads to a huge speed-up compared to a naive algorithm.
As it is advised to use mock catalogues that are much larger than the data catalogues, 
the computational time scales linearly with the number of mock sources considered. 
Using {\sc TreeCorr}, we fix the range to $0.1$~deg $\leq\theta\leq32$~deg with equal bin 
width of $\Delta\ln(\theta/1\, \mathrm{deg})=0.1$.
In order to account for the shot noise in samples with smaller numbers of sources, we increase the bin width by factors of two.
The bin centers are estimated by using the mean value of $\ln(\theta/1\, \mathrm{deg})$ for all 
pairs in the bin. The {\sc TreeCorr} parameter \texttt{bin\_slop} controls the accuracy of the
computation. It turns out that one must take care to change its default setting to obtain the 
required accuracy once the two-point correlations are at or below ${\cal O}(10^{-2})$, 
as discussed and demonstrated in some detail in App.~\ref{app:C}. 
\texttt{bin\_slop=0} gives the best possible result. It should also be stressed that for 
angles exceeding a few degrees it is important to compute angular distances on great
circles, which is achieved by setting the  {\sc TreeCorr} parameter \texttt{metric=`Arc'}. We have 
verified that using the Euclidean metric instead makes a noticeable difference at the largest angular scales accessible in LoTSS-DR1.

\begin{figure}
\centering
\includegraphics[width=\linewidth]{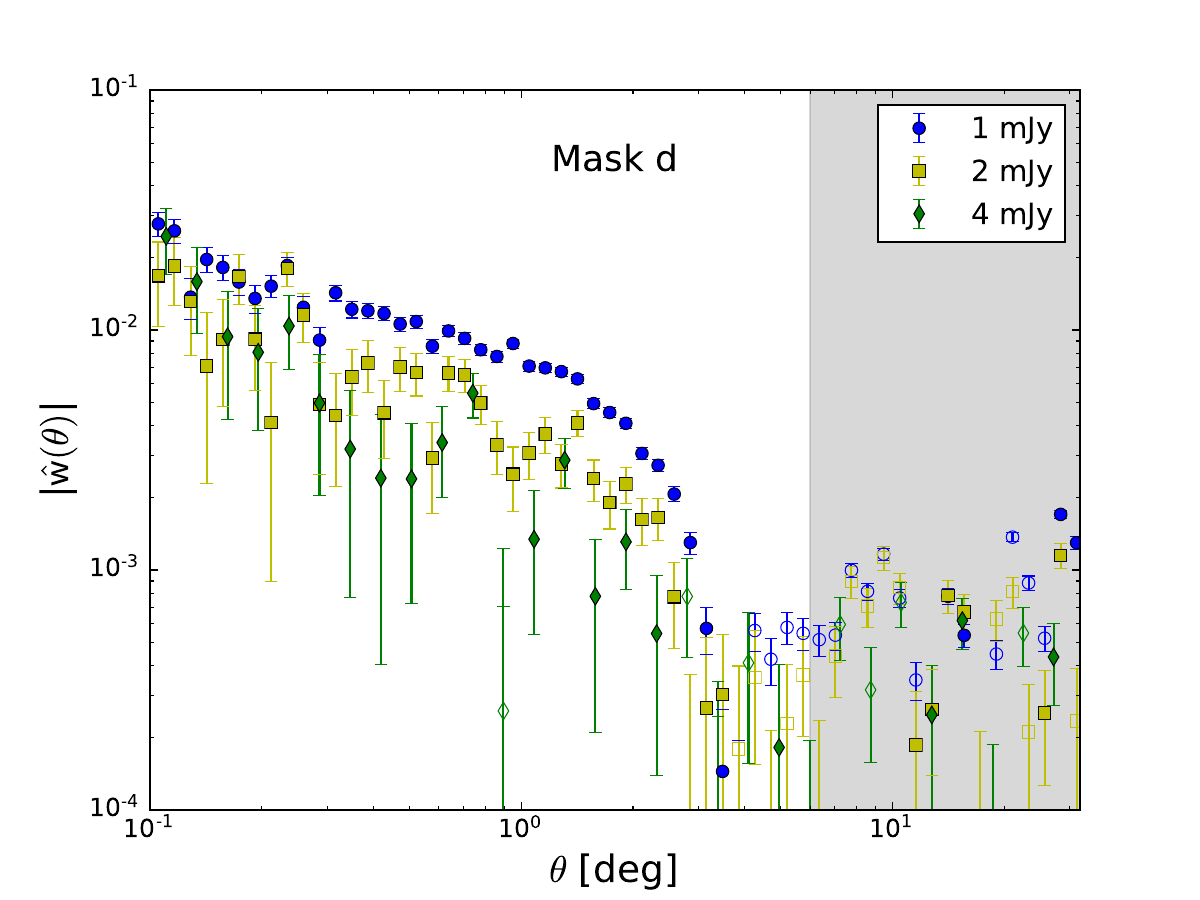} \\
\includegraphics[width=\linewidth]{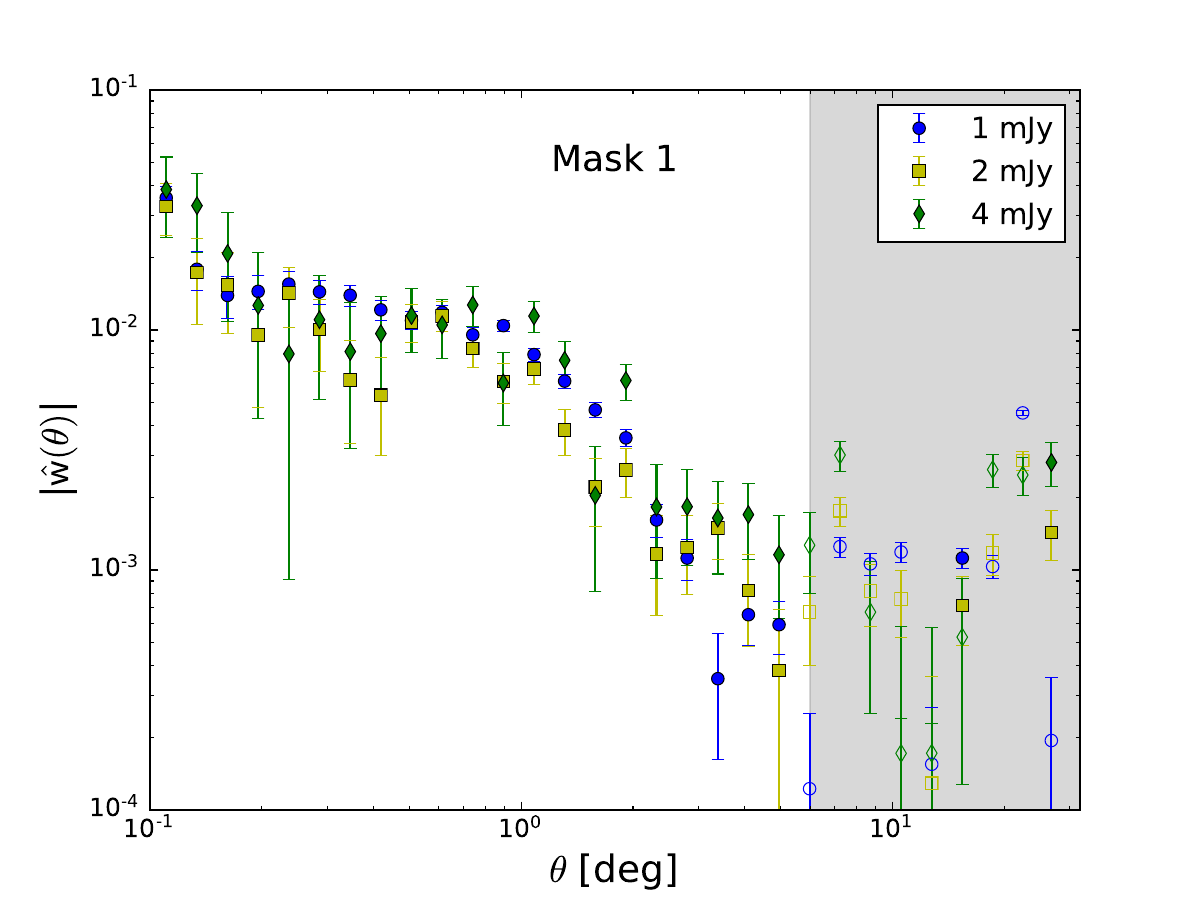}
\caption{Angular two-point correlation of sources from the LoTSS-DR1 value-added source catalogue
after masking with `mask d' (top) and `mask 1' (bottom) and at flux densities above $1, 2$ and $4$~mJy.
Positive and negative values are shown with full and open symbols, respectively. The grey shaded region indicates angular separations with decreasing number of weighted pair counts.}
\label{fig:2ptcorrelation_maskd}
\end{figure}

We base our analysis on the  LoTSS-DR1 value-added source catalogue. We start our analysis with `mask d' and flux density thresholds of $1$, $2$ and $4$~mJy.
At flux density thresholds larger than $1$~mJy we expect the point source completeness to be 
well above $99$ per cent. We also apply corresponding flux density thresholds on the mock 
catalogue (Sect.~\ref{sec:mocks}), which then contains $1\,923\,339$, $995\,218$, and $545\,520$
mock sources for `mask d' and $798\,490$, $412\,922$ and $226\,385$ mock sources for `mask 1', respectively.

The angular two-point correlation function $w(\theta)$ with statistical errors calculated by {\sc TreeCorr} is shown in Fig.~\ref{fig:2ptcorrelation_maskd}
for different flux density thresholds.
The error estimation of {\sc TreeCorr} is based on the Poisson noise in each separation bin.
We additionally tested error estimations in terms of bootstrapping and found no large difference in both estimations, see App. \ref{app:C} for details.
Note that previous radio continuum surveys showed larger bootstrap errors than Poisson errors, see \cite{Cress1996} for the FIRST survey. 
They found the Poisson error to be less than the bootstrap estimate by a factor of two for small scales around $\theta\sim0.05$~deg and even larger for increasing separations.
The geometry of the survey provides an increasing number of correlation weighted pair counts up to angular separations of $\theta < 6$~deg, at larger 
angular separations the weighted pair counts decrease and finally drop steeply at 30 deg. In the figures we shade angular scales $\theta>6$~deg in grey.

In the top panel of Fig.~\ref{fig:2ptcorrelation_maskd} we observe consistent behaviour for all three flux density samples above three degrees. Below three degrees the $1$~mJy sample is more correlated than the $2$ and $4$ mJy samples, which are more consistent. However, it can be seen that there are many angular bins in which the $4$ mJy sample shows a low value of the two-point correlation function. We believe this is likely as a result of having fewer sources in that sample. The bottom panel of that figure explores what happens if we restrict 
our analysis to the low-noise region of the survey after applying `mask 1' (see Sec.~\ref{sec:noise}). 
Now all three samples are more consistent with each other. However, the number of sources has been 
reduced by about a factor of two in each sample.

\begin{figure}
\centering
\includegraphics[width=\linewidth]{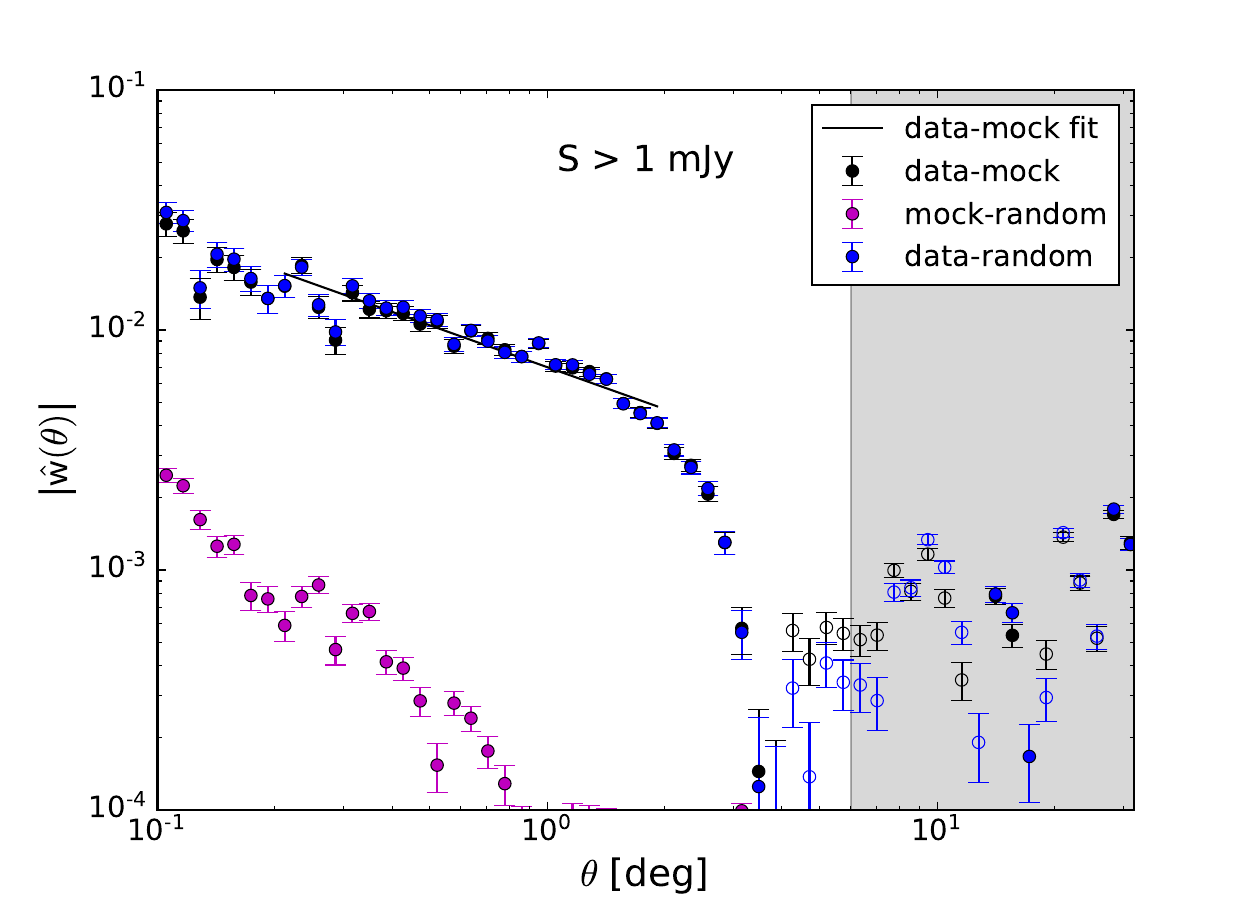} \\
\includegraphics[width=\linewidth]{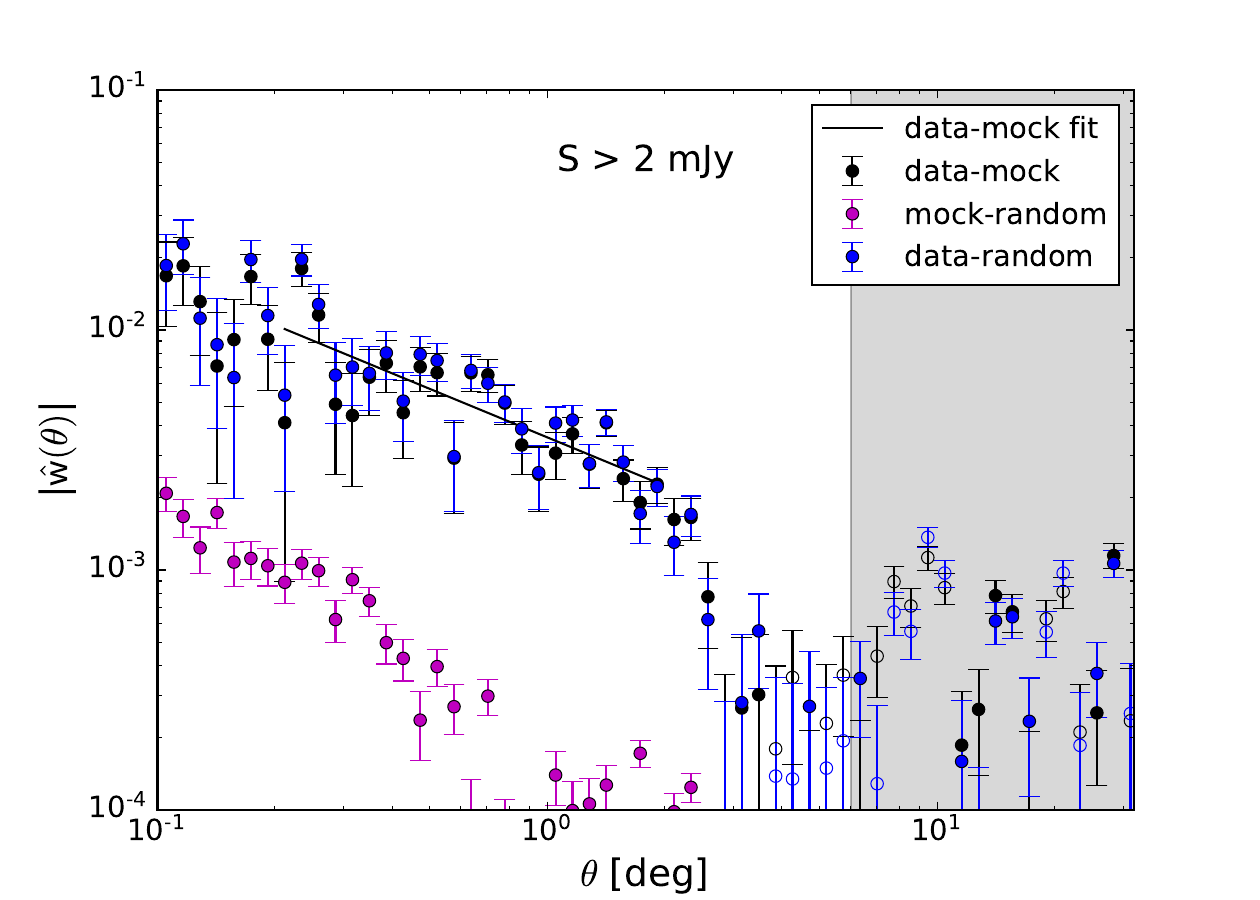} 
\caption{Comparison of two-point angular (auto-)correlation functions for `mask d' for different random catalogues: mock catalogue based on LoTSS local rms noise (data-mock), 
homogeneous random catalogue accounting for survey geometry only (data-random), and the correlation of the mock catalogue (mock-random) for flux densities above 1~mJy (top) 
and 2~mJy (bottom). We fit the data to the power-law model described in the text. Positive and negative values are shown with full and open symbols, respectively. The grey shaded region indicates angular separations with decreasing number of weighted pair counts.
}
\label{fig:datamockrandom}
\end{figure}

\begin{figure}
	\centering
	\includegraphics[width=\linewidth]{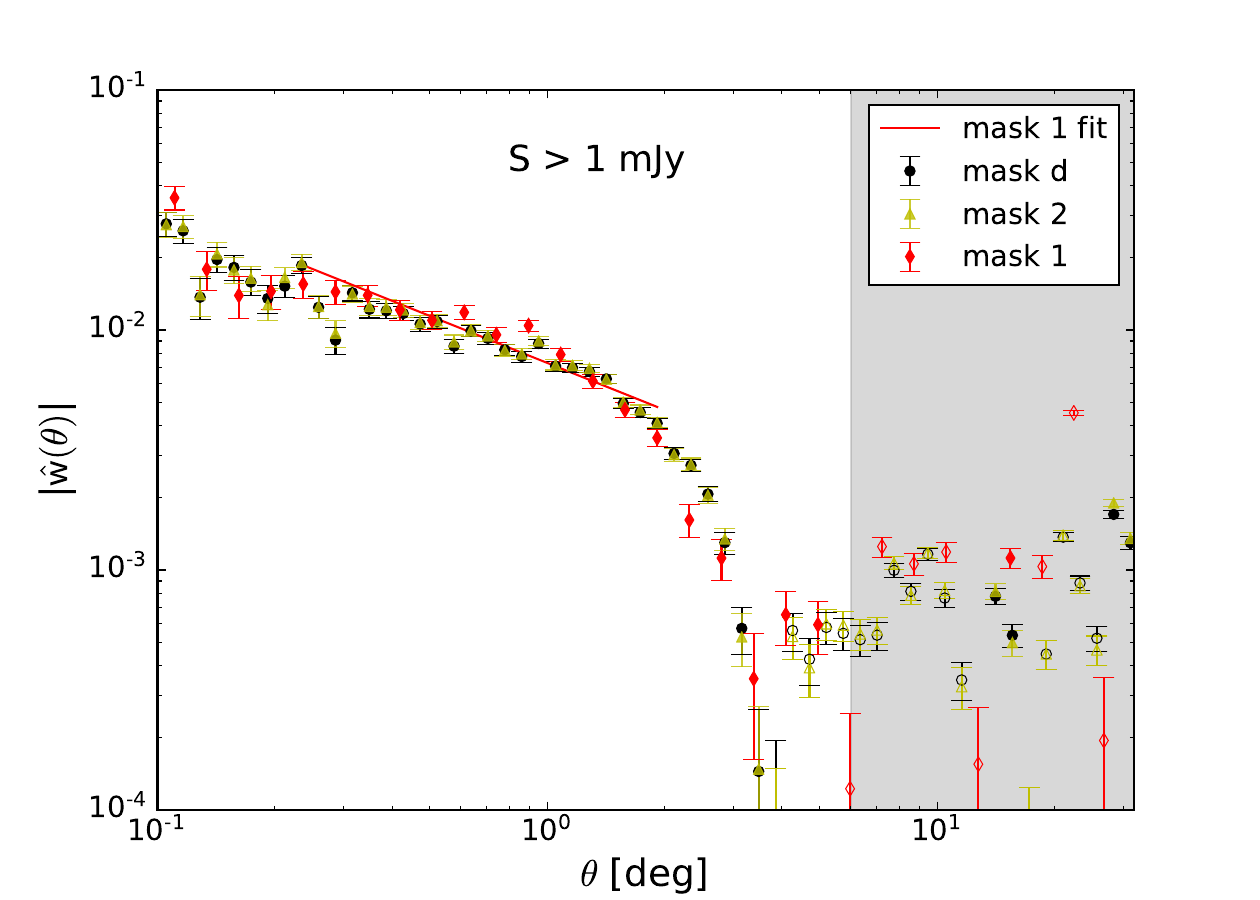}\\
	\includegraphics[width=\linewidth]{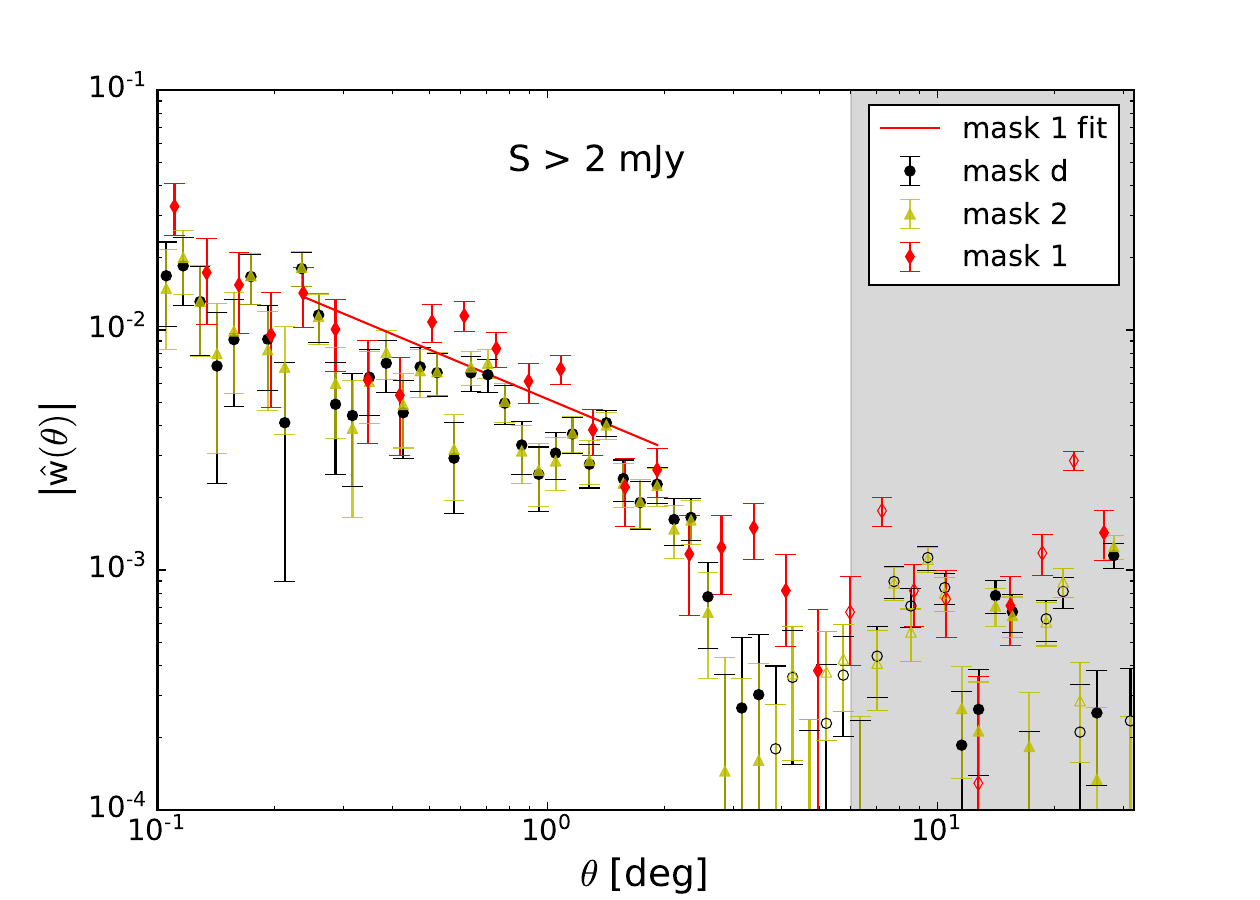}
	\caption{Angular two-point correlation from the LoTSS-DR1 value-added
	 source catalogue after masking with `mask d', `mask 1' and `mask 2' at flux densities above 
	 $1$~mJy (top) and $2$~mJy (bottom) for data-mock pairs; see caption of Fig.~\ref{fig:datamockrandom} for further details.
	 We fit the data of `mask 1' to the power-law model described in the text.
	Positive and negative values are shown with full and open symbols, respectively. The grey shaded region indicates angular separations with decreasing number of weighted pair counts.}
	\label{fig:2ptcorrelation_maskd12}
\end{figure}

The observed increase of correlation for decreasing flux density thresholds in `mask d', which is not observed in the low-noise region of `mask 1', is 
investigated further. Particularly, we ask if flux dependent correlation is related to the method of generating the mock catalogue, as it relies on the 
local noise patterns. To do so, we measure the correlation function of the mock catalogue itself, by comparing to a pure random sample (spatial 
Poisson process). In Fig.~\ref{fig:datamockrandom}, we see that there is almost a vanishing mock auto-correlation (denoted mock-random in the legend) 
above the typical size of an individual pointing (1.7~deg in radius), whereas for smaller angular separations the correlation is an order of magnitude smaller 
than that of the data sample.
We also show in Fig.~\ref{fig:datamockrandom} the data-mock and data-random 
(spatial Poisson process) auto-correlations. The data-mock and 
data-random results agree at all scales with small differences.
This also holds true for the three different masks `1', `2', and `d', which we have tested separately. 
The close similarity of results based on pure random samples and the mock catalogue shows that the flux density dependence
 of the observed correlations is not a result of how we generate the mock catalogue. 
 We also see from Figs.~\ref{fig:2ptcorrelation_maskd} and \ref{fig:2ptcorrelation_maskd12} that the 
reduced noise level of `mask 1' increases the correlation for the $2$ and $4$ mJy samples, but does not change the $1$ mJy sample significantly.

Whilst the procedure of generating mocks (Sect. \ref{sec:mocks}) does account in the large sense for the inhomogeneity of completeness (see e.g. Figures \ref{fig:completenesscell} and \ref{fig:mockcatalogues}), it may have completeness issues close to the 5$\sigma$ detection threshold. This could be due to a variety of reasons such as completeness when using \textsc{PyBDSF} to detect sources (which is not used for the randoms); the assumption of point sources when generating randoms and finally flux scale issues within the data. However, when applying flux thresholds that are significantly above the averaged $95\%$ completeness flux density of $0.39$~mJy, variations in completeness should not affect our results at a significant level.

\begin{figure}
\centering
\includegraphics[width= \linewidth]{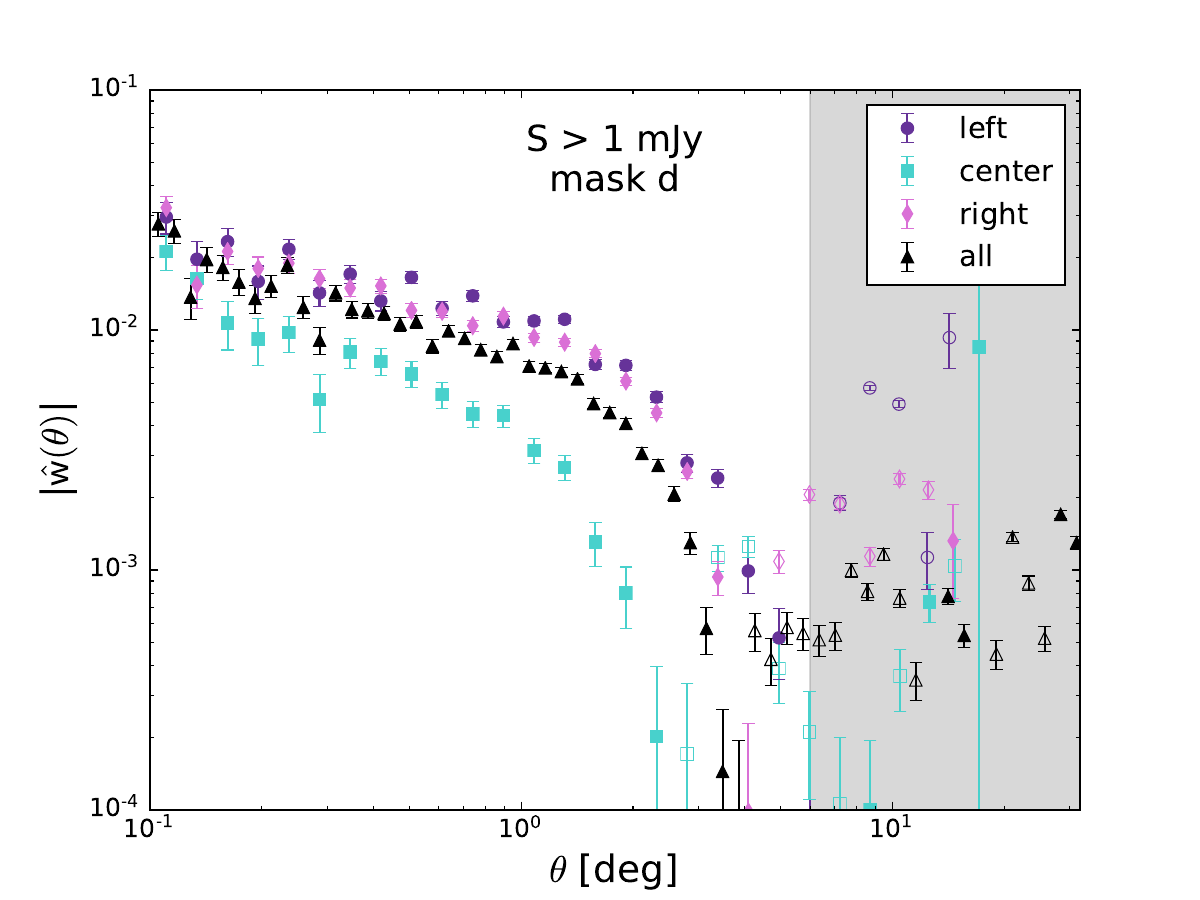}
\includegraphics[width= \linewidth]{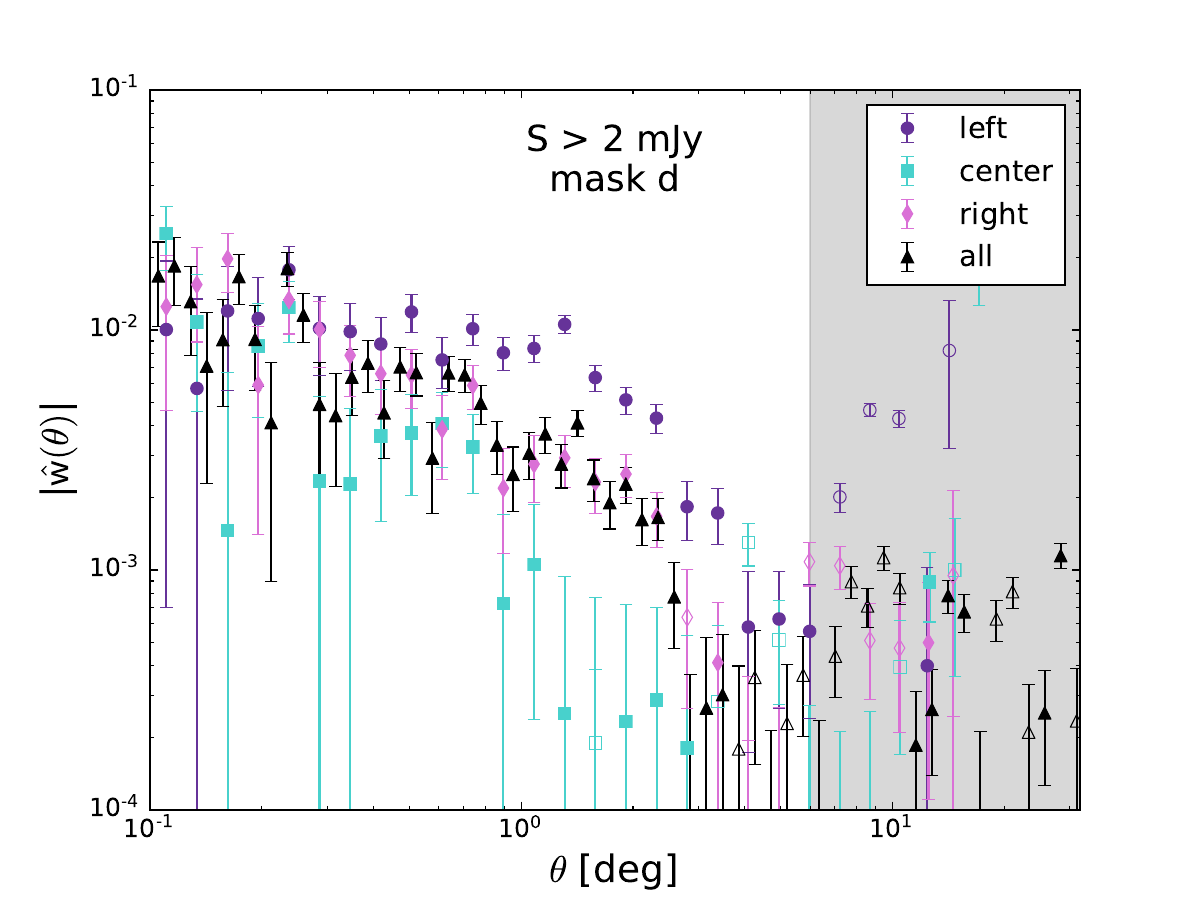}
\caption{Angular two-point correlation function of sources from the LoTSS-DR1 value-added source catalogue with `mask d' and flux density threshold of 1~mJy and 2~mJy, for 
three regions namely `Left', `Center', and `Right'. $\hat{w} (\theta)$  for the non-partitioned region with 1~mJy and 2~mJy threshold and mask d is also plotted. 
Positive and negative values are shown with full and open symbols, respectively. The grey shaded region indicates angular separations with decreasing number of weighted pair counts.}
\label{fig:jack-knife}
\end{figure}

\begin{figure}
\centering
\includegraphics[width= \linewidth]{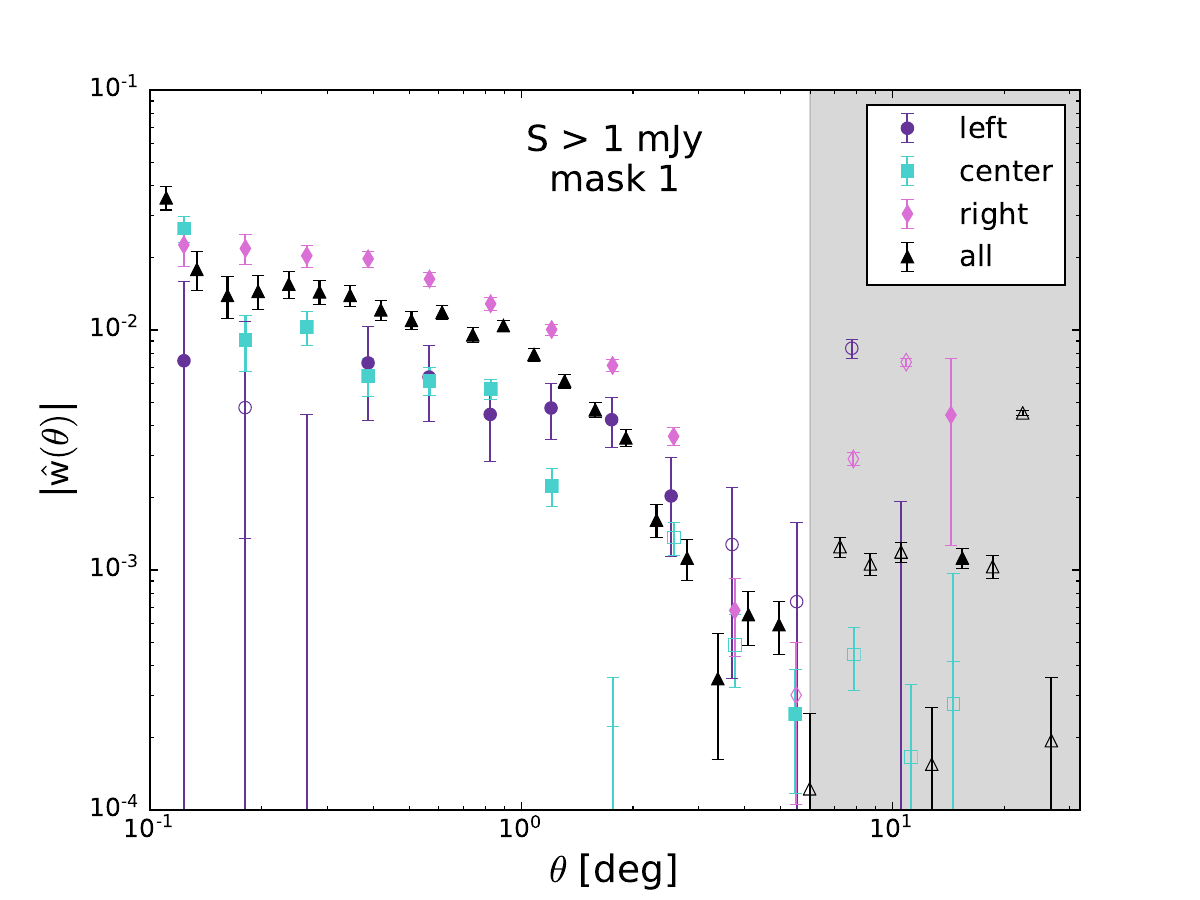}
\includegraphics[width= \linewidth]{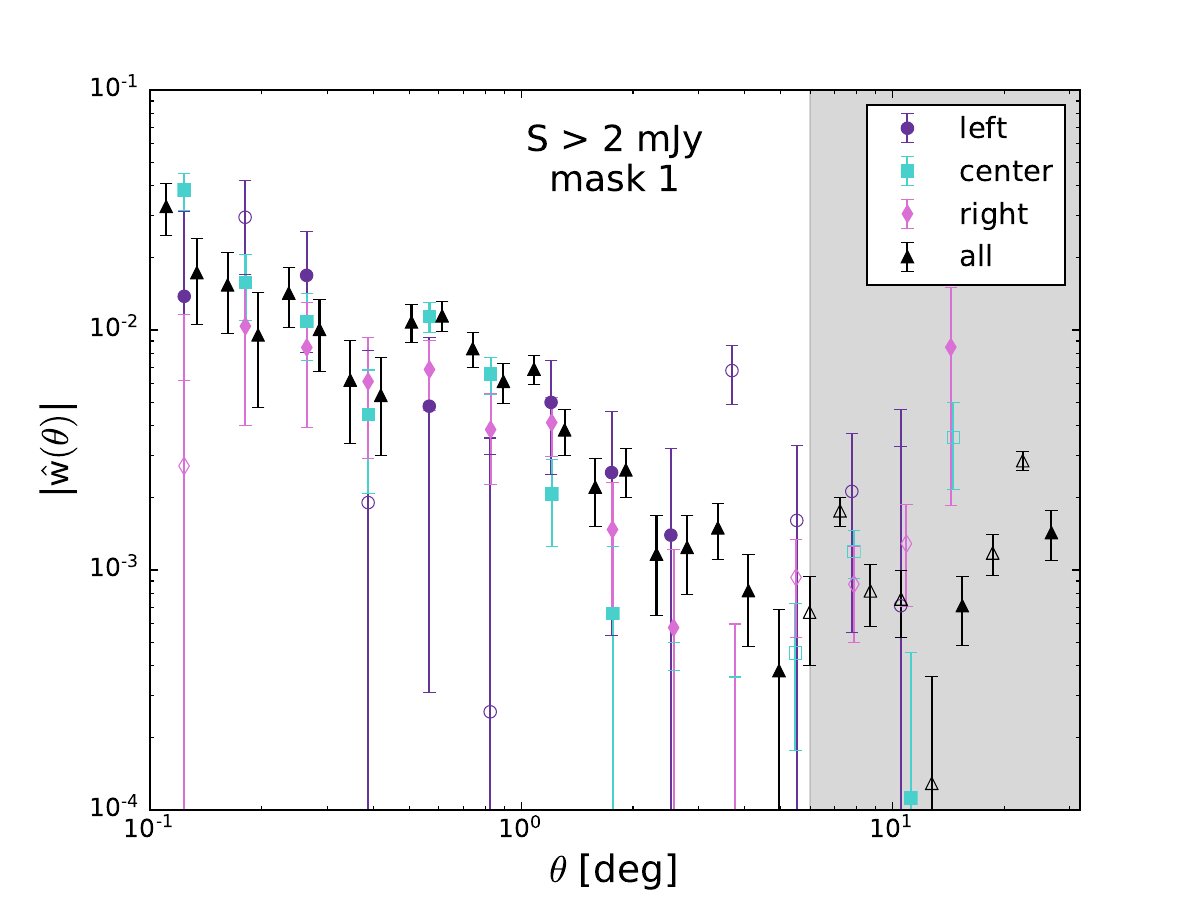}
\caption{Angular two-point correlation function of sources from the LoTSS-DR1 value-added source catalogue with `mask 1' and flux density threshold of 1~mJy and 2~mJy, for 
three regions namely `Left', `Center', and `Right'. $\hat{w} (\theta)$  for the non-partitioned region with 1~mJy and 2~mJy threshold and mask 1 is also plotted. 
Positive and negative values are shown with full and open symbols, respectively. The grey shaded region indicates angular separations with decreasing number of weighted pair counts.}
\label{fig:jack-knife_mask1}
\end{figure}

To further investigate the origin of the flux density dependence of the angular two-point correlation, we perform a jack-knife test and split up 
the survey into three regions on the sky, namely `Right', `Center', and `Left'. These lie within the following right ascension intervals: $[161,184]$, 
$(184,208]$, $(208,230]$~deg, respectively. We then compute $w (\theta)$ and errors as 
mentioned above and compare the results, shown in Fig.~\ref{fig:jack-knife} for the $1$ and $2$ mJy samples of `mask d' and in Fig.~\ref{fig:jack-knife_mask1} for `mask 1'. 
We observe for the $1$ mJy sample of `mask d' that the angular two-point correlation functions of the three regions agree at the smallest angular separations, but show significantly less correlation for the central region as compared to the left and the right region at scales around $1$ deg.
The reason for this discrepancy is not fully understood, we think it may be due to issues in the flux density calibration of individual pointings.

The hypothesis of a fluctuation in the flux density calibration is supported by the observed lack of source counts south of the unobserved hole in the 
HETDEX field. To see that we compare the LoTSS-DR1 radio source catalogue (Fig. \ref{fig:overview}) and the mock catalogue (Fig. \ref{fig:mockcatalogues}).
According to the mock catalogue, which is based on the local rms noise (Fig. \ref{fig:localrms}), we should see an overdense region, whereas the actual 
source counts reveal an underdensity. Also the completeness map (Fig. \ref{fig:completeness}) supports the findings from the mock catalogue. 
An underestimation of the flux scales in the corresponding pointings would give rise to exactly that effect. It would lead to smaller observed flux densities, 
which would lead to less observed sources close to the detection limit, but in terms of noise to cleaner and more complete regions. A simple model for 
the flux calibration assumes a linear relation between the true flux scale and the actual flux scale used in each pointing, $S_p = c_p S^\mathrm{true} + o_p$, where $c_p$ is fixed observing one or several calibrator sources for each particular pointing. 
For large enough flux densities, the offset $o_p$, which is expected to be at least of the order 
of the rms noise, is irrelevant, but becomes relevant close to the detection threshold. Consequently, samples with increased flux density threshold are less affected by flux density calibration offsets.

For the $2$~mJy sample of `mask d', the right region is consistent with the full 
sample, whereas the left region shows an increased correlation and the central region a decreased correlation. We note that the left region has the most complicated geometry. The interpretation of 
Fig.~\ref{fig:jack-knife} is complicated by different values of $w_\Omega$ for the three different regions, due to their different survey geometry and sky 
coverage. We conclude that the $2$ mJy sample shows a more self-consistent behaviour as compared to the $1$ mJy sample.  However, the differences in the angular two-point correlation function that occur in different regions within the field at $2$ mJy are not well understood. It is hoped that this will be reduced with the next data release of the LoTSS survey, where there will be a larger sky coverage and, if there are flux scaling issues, these flux scaling issues may be reduced. The results of the jack-knife test within different spatial regions are consistent with the idea that differences relate to the flux-density calibration of individual pointings, as the $2$ mJy sample is affected to a lesser extent than the $1$ mJy sample.

In the jack-knife of the 1~mJy `mask 1' sample we observe consistently less correlation in the left part of the survey than in the right part. 
The central region shows consistent behaviour at small scales with the left part and starts to deviate from it at scales $\sim1$~deg.
As seen previously for `mask d', the full sample shows correlation in between the three parts, as it is a combination of the three parts.  
Comparing the sky coverage of `mask d' and `mask 1' in Fig.~\ref{fig:mask123} the left part is affected most by cutting in the local noise per cell. 

For the $2$~mJy sample of `mask 1' we find consistent behaviour for all three parts and also consistent behaviour with the full sample.  Outliers and even negative correlations mostly seen in the left part can be explained by the highly decreased number of sources in this sample, which leads to higher contributions of shot noise.

Comparing the jack-knife test for `mask d' and `mask 1' we consider the results of `mask 1' to be more consistent and not as affected by flux calibration variations as in the case of `mask d', especially in the case of the 2~mJy sample. 
Therefore we will use the `mask 1' 2~mJy sample as the default sample for our future analysis.

\begin{table*}
	\centering
    \caption{Best-fit values of $w(\theta) = A (\theta/1\ \mathrm{deg})^{-\gamma}$ , fitted in the range $0.2\leq\theta\leq2.0$~deg and corresponding integral constraint $w_\Omega$ for the 
    LoTSS-DR1 value-added source catalogue after appropriate masking and for the 
    TGSS-ADR1 catalogue, with $68\%$ confidence intervals. For both catalogues various flux density thresholds are shown.}
    \begin{tabular}{cccccccc}
    	\hline\hline
    			\addlinespace[0.5ex]Survey & $S_{min}$ & $z$ & $A (\times 10^{-3})$ & $\gamma$& $w_\Omega(\times 10^{-3})$ &$\chi^2/$dof & $N$\\ 
        
         & [mJy]  & & &&& &\\\hline
		\addlinespace[1ex] LoTSS-DR1  & 1  & n.a. & $7.00^{+0.18}_{-0.18}$ & $0.58^{+0.04}_{-0.04}$& $1.9$ &4.47& 102\,940 \\
		\addlinespace[0.5ex] mask d& 2 & n.a. & $3.51^{+0.24}_{-0.25}$ & $0.74^{+0.10}_{-0.10}$&$ 0.7$ &1.88& 51\,288 \\
		\addlinespace[0.5ex]& 4 & n.a. & $1.97^{+0.27}_{-0.27}$& $0.78^{+0.21}_{-0.20}$&$0.4$ &0.68& 30\,556 \\
		\addlinespace[1ex]\hline 
		\addlinespace[1ex]  LoTSS-DR1 & 1  & n.a. & $7.20^{+0.42}_{-0.42}$ & $0.68^{+0.08}_{-0.08}$&$1.9$  &5.78& 40\,599 \\
		\addlinespace[0.5ex]   mask 1&2 & n.a. & $5.11^{+0.59}_{-0.60}$ & $0.74^{+0.16}_{-0.16}$&$1.2$ &2.70& 19\,719 \\
		\addlinespace[0.5ex] & 4 & n.a. & $7.45^{+0.95}_{-0.95}$ & $0.46^{+0.21}_{-0.20}$&$3.0$ &2.34& 11\,269 \\
         \addlinespace[1ex]\hline 
         		         \addlinespace[1ex] LoTSS-DR1 & 2 & Any $z$ & $6.58^{+0.42}_{-0.43}$ & $0.84^{+0.08}_{-0.08}$&$1.1$& 1.25& 24\,420 \\
         		\addlinespace[0.5ex]mask z& 4 & Any $z$ & $4.65^{+0.78}_{-0.80}$ & $0.84^{+0.25}_{-0.24}$& $1.2$&1.45& 14\,506 \\
         		\addlinespace[1ex]\hline 
         		\addlinespace[1ex] LoTSS-DR1 & 2 & Any $z$ & $6.68^{+0.93}_{-0.94}$  &$0.92^{+0.18}_{-0.18}$ & $1.2$&1.27 &9505  \\
         		\addlinespace[0.5ex]mask z1& 4 & Any $z$ & $6.48^{+1.18}_{-1.19}$&$0.64^{+0.28}_{-0.26}$ & $1.8$ &0.70&5432  \\
         		\addlinespace[1ex]\hline 
         		\addlinespace[1ex]  LoTSS-DR1 & 2 &$z<0.376$& $23.02^{+1.59}_{-1.59}$ & $0.82^{+0.09}_{-0.09}$&$4.1$  &2.13& 8\,430 \\
         		\addlinespace[0.5ex]  mask z   & 2 &$0.376\leq z<0.705$ & $6.50^{+0.10}_{-0.10}$ & $0.99^{+0.18}_{-0.19}$&$0.9$ &0.60& 7\,189 \\
         		\addlinespace[0.5ex]  & 2 &$0.705\leq z$& $7.30^{+0.90}_{-0.90}$ & $0.74^{+0.18}_{-0.17}$&$1.5$ &0.77& 8\,801\\\addlinespace[1ex]\hline
         		 \addlinespace[1ex]  LoTSS-DR1 & 2 &$z<0.376$& $21.50^{+4.65}_{-4.72}$  & $0.84^{+0.29}_{-0.29}$ & $4.4$  & 4.18&3420  \\
         		 \addlinespace[0.5ex]  mask z1   & 2 &$0.376\leq z<0.705$ & $4.77^{+2.09}_{-2.09}$ & $1.03^{+0.45}_{-0.52}$ & $0.5$ &0.35&2693  \\
         		 \addlinespace[0.5ex]  & 2 &$0.705\leq z$& $9.82^{+1.43}_{-1.43}$  & $0.38^{+0.26}_{-0.23}$ & $0.4$ &0.41&3399 \\\addlinespace[1ex]\hline
      \addlinespace[1ex]
        TGSS-ADR1  & 100 & n.a. & 10.16$^{+0.44}_{-0.44}$ & 0.59$^{+0.07}_{-0.07}$& $1.9$ &1.83& 219\,303\\
      \addlinespace[0.5ex]This work App.~\ref{app:A} & 200& n.a. & 11.65$^{+0.70}_{-0.70}$& 0.51$^{+0.11}_{-0.11}$& $2.7$&1.35& 119\,021\\\addlinespace[1ex]\hline
     \addlinespace[1ex]TGSS RB19 &  100& n.a. & 8.4 $\pm$ 0.1 & 0.72 $\pm$ 0.11& &-& 163\,654\\\addlinespace[1ex]\hline
    \end{tabular}
	\tablebib{       RB19:~\citet{RanaBagla2019}}
    \label{tab:BestFit}
\end{table*}

In order to ease the comparison between different samples and with angular two-point correlation functions published 
elsewhere \citep{Kooiman1995,Rengelink1999, BlakeWall2002, Overzier2003, Blake2004, RanaBagla2019, Dolfi2019}, 
we fit the data points in Fig.~\ref{fig:2ptcorrelation_maskd} to a power-law model of the 
form:
\begin{equation}\label{eq:powerlaw}
w(\theta) = A \left(\theta/1~\mathrm{deg}\right)^{-\gamma}. 
\end{equation}
The value of such a power-law fit for a cosmological analysis is limited, as it holds at best for a small range of angular scales. 
For fitting we make use of the publicly available Python {\sc LMFIT}\footnote{\url{http://lmfit.github.io/lmfit-py/index}} 
package \citep{lmfit2016}, where we used the default  Levenberg-Marquardt method. We fit the data points in the range of $0.2\leq\theta \leq 2.0$ deg. We then explicitly compute $w_{\Omega}$ using Eq. (\ref{eq:bias}) from the initial fit parameters and re-do the fitting but this time by selecting only those data points which are greater than $w_{\Omega}$. The entire process is re-iterated until stable values for $A$, $\gamma$ and $w_\Omega$ are obtained. Best-fit results obtained in such a manner are summarized in Table \ref{tab:BestFit} and are shown in Fig. \ref{fig:datamockrandom} and Fig. \ref{fig:2ptcorrelation_maskd12}.
This procedure is done for the 1, 2 and 4~mJy flux density thresholds with `mask d', as well as for  `mask 1'. 

We find that the values of $A$ and $\gamma$ depend on the detailed cuts and flux 
density thresholds applied. Higher flux density threshold means smaller correlation amplitude in case of `mask d', which is contrary to the findings of \cite{Wilman2003} in terms of the Bo\"otes Deep Field and \cite{RanaBagla2019} in terms of the TGSS-ADR1. This difference may reflect issues arising from flux scale issues which, as mentioned previously, may be affecting the results presented in this work and should be improved using the next data release of the LoTSS survey. Alternatively, this may reflect the changing distribution of populations at different flux density limits within our samples as we will have a larger fraction of SFGs at lower flux density limits. This could make direct interpretation of the clustering amplitude difficult as the SFG and AGN are thought to have different bias measurements \citep{Magliocchetti2017, Hale2018} and will have different redshift distributions.
In Table  \ref{tab:BestFit}, we also provide the goodness-of-fit in 
terms of $\chi^2$ over number of degrees of freedom. `mask 1' and `mask d' agree on goodness-of-fit and the best-fit values of 
$A$ and $\gamma$ for 1 and 2 mJy within three sigma. However, the two values for the $4$ mJy sample are inconsistent with each other, but 
the `mask 1' measurement suffers from a rather small number of sources. Based on the various tests reported above, 
we conclude that the most reliable measurement is obtained by `mask 1' for the 2 mJy sample: 
$A = (5.1 \pm 0.6)\times10^{-3}$ and $\gamma = 0.74 \pm 0.16$.

This result can be compared to angular two-point correlation functions reported in the literature. For the NVSS
$A =  (1.49 \pm 0.15)\times10^{-3},\gamma = 1.05 \pm 0.10$ from \citet{Blake2004} at $S_{\rm NVSS} > 10$~mJy and 
$A=(1.0\pm 0.2)\times10^{-3}$ for $\gamma=0.8$ from \citet{Overzier2003} at same flux density threshold.  
Results found for lower frequencies in the WENSS survey, e.g. $A=(2.0\pm0.5)\times10^{-3}$ for $\gamma=0.8$ from \citet{Rengelink1999} and $A=(1.01\pm0.35)\times10^{-3}, \gamma=1.22\pm0.33$ from \citet{Blake2004} at $S_{\rm WENSS} > 35$~mJy.  
Results in the same frequency range as LoTSS have been obtained from the TGSS-ADR1 catalogue and are shown in Table \ref{tab:BestFit}.
Thus the slope found in LoTSS-DR1 is consistent with the findings for TGSS-ADR1 at much higher flux densities, but differs from the slope found at 
higher frequencies. The amplitude found in LoTSS-DR1 is smaller than the one found for TGSS-ADR, but larger than the one from NVSS and WENSS.

\subsection{The angular two-point correlation function for redshift sub-samples}
\label{sec:tpcfredshift}

We further make use of the available redshift information in the LoTSS-DR1 catalogue, 
namely the `z\_best' values. We first divide the LoTSS-DR1 catalogue into two sub-samples 
based on the information whether a `z\_best' value for a given radio source is available or not. 
We then compute the angular two-point correlation function for the sub-sample with redshift information, called `Any $z$', which is shown 
in Fig.~\ref{fig:comparisonwthetanohalolensinglinear} for `mask z' (top). 
As the results of Sect. \ref{sec:subtwopoint} show more consistent results for `mask 1' than for `mask d', 
we additionally generate a redshift mask for the region of `mask 1', which is denoted as `mask z1'. 
The results for the angular two-point correlation with `mask z1' are shown in Fig.~\ref{fig:comparisonwthetanohalolensinglinear} (bottom).
Based on the strong difference between the angular two-point correlation of the 1 and 2~mJy samples of `mask d' we neglect the 1~mJy in the further analysis. 

Additionally we test different redshift subsamples defined in Eq. (\ref{eq:percentiles}) in 
Sect. \ref{sec:redshift},  where the survey is split up into three parts, namely 
$z_1$: $z < z_{33}$, $z_2$: $z_{33} \leq z < z_{66}$, and $z_3$: $z_{66} \leq z$. 
These parts are separated by the $33$ and $66$ percentiles, 
defined in terms of the survey without any flux density thresholds and are kept the same 
for higher flux density thresholds. The measured angular two-point correlations for a flux density threshold 
of $2$~mJy, masked with `mask z' for the three redshift bins are presented in 
Fig.~\ref{fig:comparisonwthetaredshiftbins}.
Due to the strongly decreased number of sources per bin in the $2$~mJy samples, we increased the bin width to $\Delta\ln(\theta/1\, \mathrm{deg})=0.4$.

Fitting a power law, as defined in Eq. (\ref{eq:powerlaw}), gives the results shown in 
Table \ref{tab:BestFit}. We can see that the goodness-of-fit is close to one, except for the first redshift bin. 
We see stronger correlation for most of the redshift bins, which is expected as there is less smearing. The exponent $\gamma$
and the amplitude $A$ are larger as compared to the best-fit LoTSS-DR1 2~mJy `mask 1' sample and to the NVSS values.
And the amplitudes increase further if we consider individual bins in redshift as compared to the study that 
includes any value of the redshifts. 

However, a disclaimer is in order: We did not estimate and propagate errors on the redshift estimation. 
Thus the error bars shown assume that the redshift estimates used here are exact. 
We expect that the errors for the `Any $z$' sample are 
nevertheless realistic, as only the fact is used that those sources have optical and infrared counterparts and the photometric 
redshift estimator found a solution. 
But when we split up the radio sources with photometric redshift into three bins, the reliability of the
redshift estimate becomes an issue. It is well known that there is a finite and non-negligible probability that AGNs from bin $z_3$ would be 
misestimated and end up as sources in bin $z_1$, see \citet{LoTSS2019C}. Propagating this effect through our analysis pipeline 
and correcting for it was beyond the scope of this work. 
Since only half of the LoTSS-DR1 radio sources have redshift information available, it is currently impossible to measure the bias evolution of the complete sample. We also note that due to only half of the sources having redshifts available, there will be underlying selection effects in these sub-samples that may not necessarily represent the full sample as a whole.

\subsection{Comparison of angular two-point correlations to expectation of cosmological standard model}

\begin{figure}
\centering
\includegraphics[width=\linewidth]{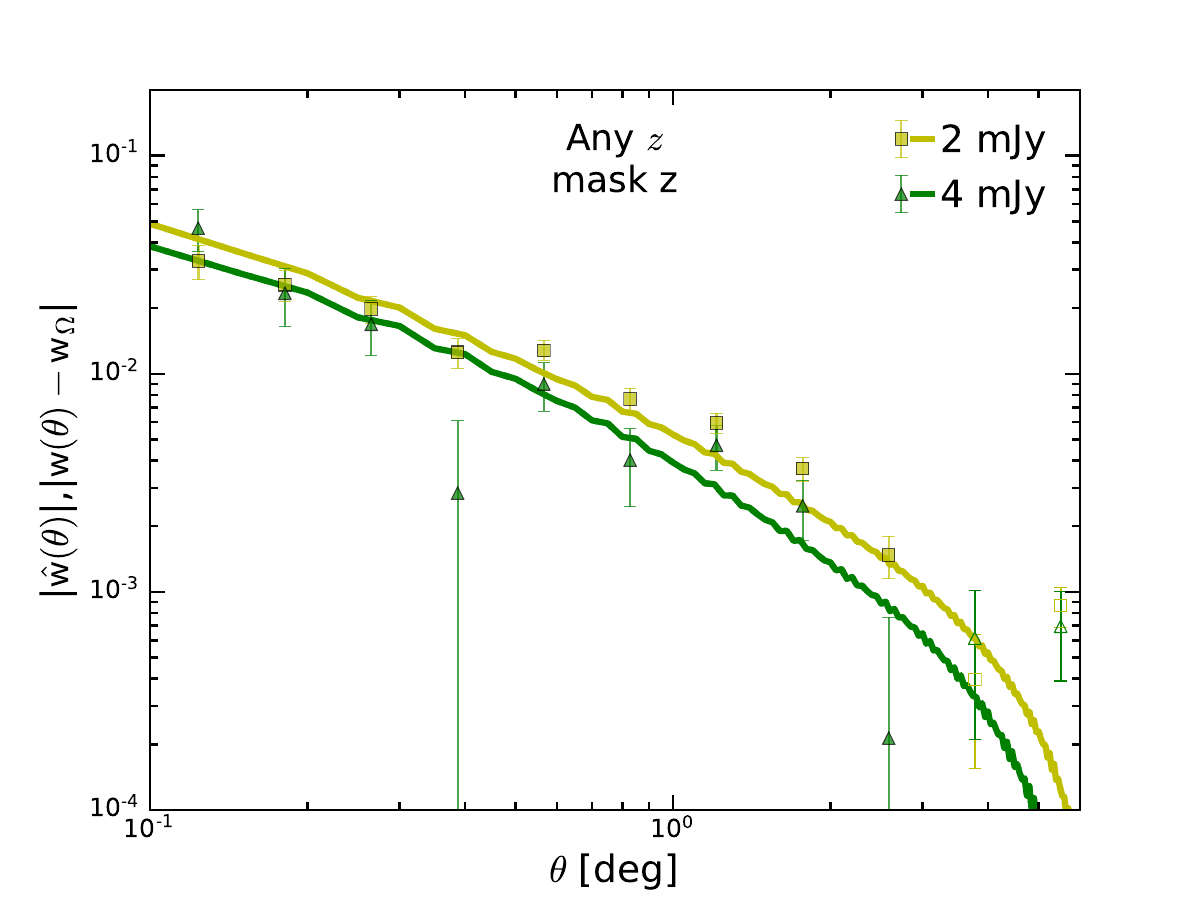}
\includegraphics[width=\linewidth]{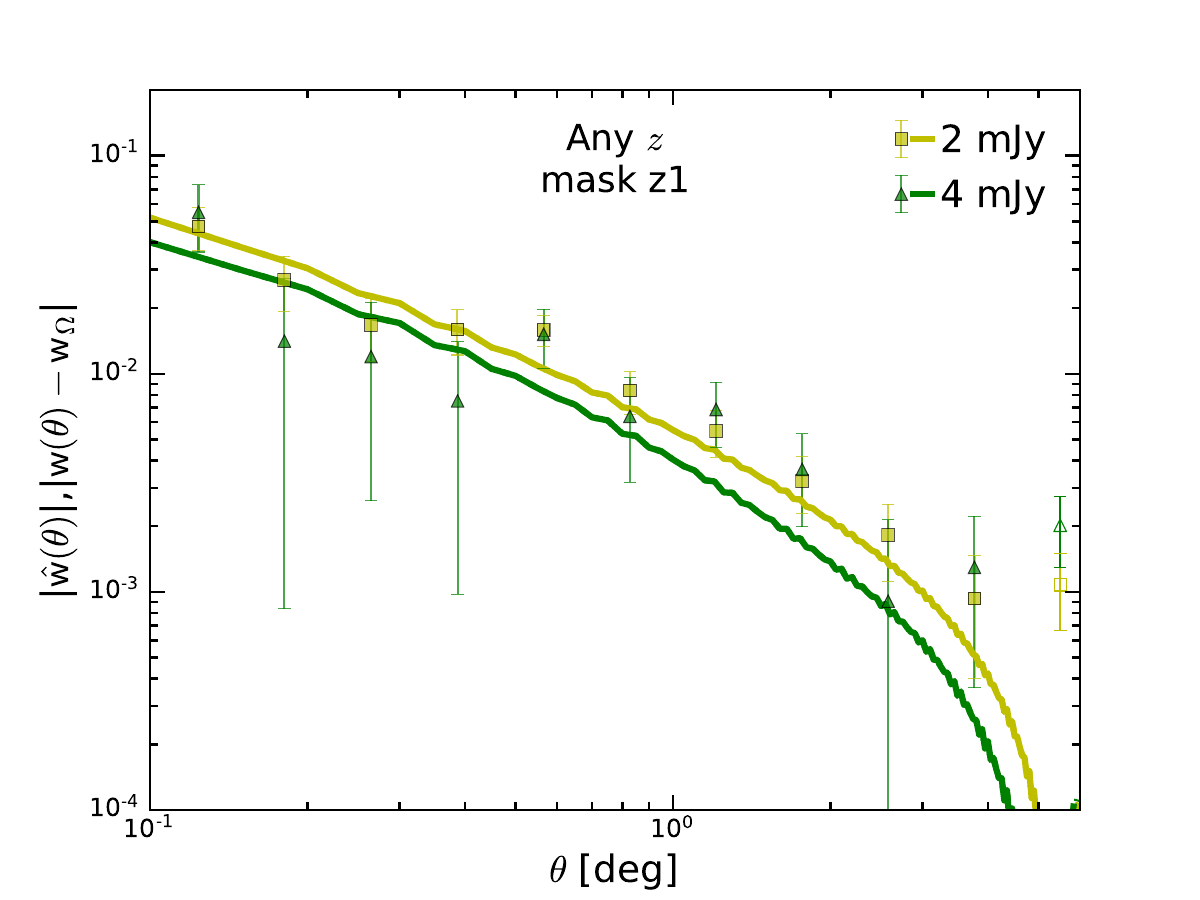}
\caption{Comparison of the angular two-point correlation function estimated from the LoTSS-DR1 value-added source catalogue for 
radio sources with redshift information and theoretical expectations (solid lines) for the best-fit $\Lambda$CDM cosmological 
parameters from Planck, generated using {\sc CAMB sources} with halofit and $b(z)$ from Eq. (\ref{eq:bias}). The integral constraint $w_\Omega$ is computed for the expectations and subtracted from them. Positive values are shown with full symbols and solid lines, whereas negative values are shown with open symbols and dashed lines.}
\label{fig:comparisonwthetanohalolensinglinear}
\end{figure}

\begin{figure}
\includegraphics[width=\linewidth]{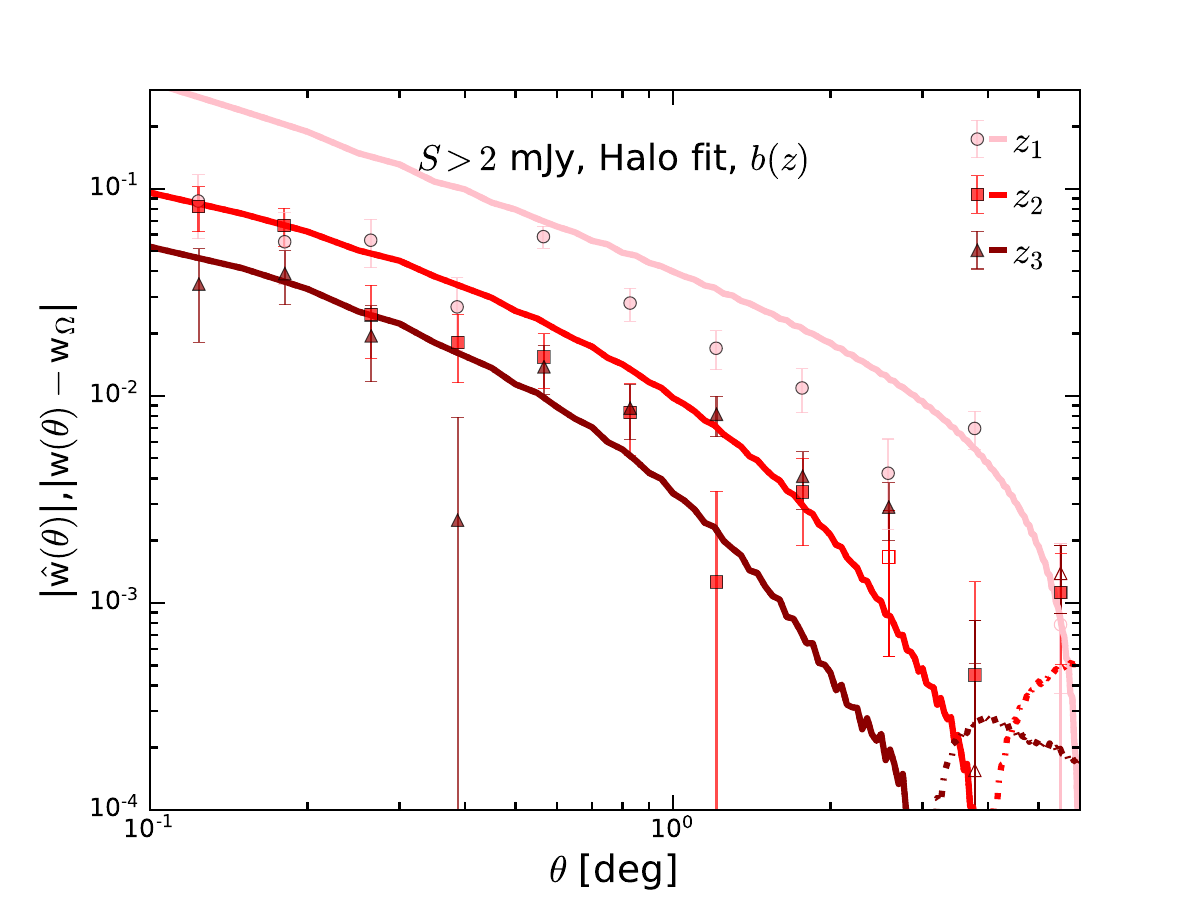}\\
\includegraphics[width=\linewidth]{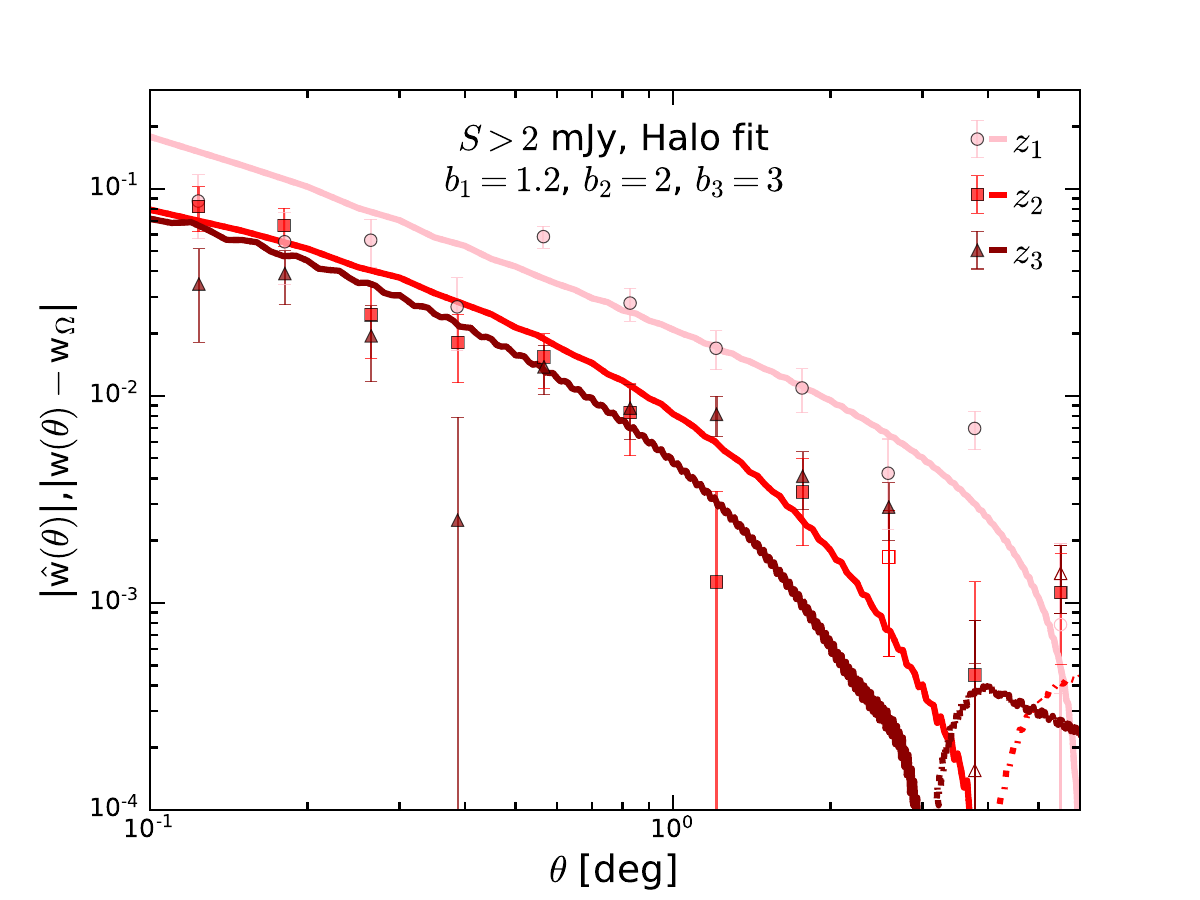}
\caption{Angular two-point correlation function for three redshift bins $z_1,z_2$ and $z_3$ for a flux density threshold of  $2$~mJy. The lines show the expectations for the cosmological standard model.  Both panels use the Halofit option of {\sc CAMB sources}, which accounts for the non-linear evolution of large scale structure. In the top panel we use the bias function of Eq. (\ref{biasTiwari}), whereas we use a piecewise constant bias in the bottom panel. The integral constraint $w_\Omega$ is computed for the expectations and subtracted from them. Positive values are shown with full symbols and solid lines, whereas negative 
values are shown with open symbols and dashed lines.}
\label{fig:comparisonwthetaredshiftbins}
\end{figure}

In order to compare our measured angular two-point correlation function to expectations, 
we rely on the publicly available {\sc CAMB sources}
code \citep{CAMBsources2011} to calculate the angular power spectrum $C_l$ for $2\leq l \leq 4000$. 
From this power spectrum we infer the two-point correlation function $w(\theta)$ by using Eq. (\ref{eq:twopointfunction}). 
In doing so we assume a vanishing monopole and dipole. The theoretical monopole vanishes by definition and the theoretically 
expected dipole is the sum of a structure dipole and the kinematic dipole \citep{EllisBaldwin1984} caused by the proper motion of the Solar system.
We have checked with a simulation that the survey area of LoTSS-DR1 would pick up that dipole at a level that is about an order of 
magnitude below the actually observed signal and we thus neglect the dipole contribution in this analysis (see App. \ref{app:dipole} for further details). The dipole contribution will become more 
important at larger angular separations for larger survey areas \citep{Bengaly2018}.

In order to predict the angular two-point correlation, we have to specify a cosmological model, the redshift distribution of the observed sources and how well they trace 
the underlying matter density distribution, which is expressed as a bias function. 
We fix the cosmological parameters to the best-fit $\Lambda$CDM cosmology of the Planck 2018 analysis
\citep{PlanckF2018, PlanckA2018}, which are the Hubble rate today ($H_0$), the dimensionless, Hubble independent baryon density ($ \Omega_b h^2$) and cold dark matter density ($\Omega_c h^2$) with $h=H_0/(100\text{ km s}^{-1}\text{ Mpc}^{-1})$, the primordial amplitude of curvature perturbation ($A_s$) and the spectral index of curvature perturbation ($n_s$) with their recent best-fit values:
\begin{align*}
& H_0 = 67.32 \text{ km s}^{-1}\text{ Mpc}^{-1},\\
& \Omega_b h^2 = 0.022383,\; \Omega_c h^2 = 0.12011, \\
& \ln(10^{10}A_s) =3.0448,\; n_s =0.96605.
\end{align*} 
The optical depth, which is usually also reported, is of no concern for the prediction of the angular power spectrum of matter.
The redshift distribution of radio sources is estimated from the histogram of the measured photo-$z$ from the LoTSS-DR1 value-added source catalogue, shown 
in Fig.~\ref{fig:zhistfluxthresholddndodz}, which is used as source window function for the three different flux density threshold samples. For the galaxy bias, $b(z)$,
we use a parametrisation introduced by \cite{NusserTiwari2015,TiwariNusser2016} as a fit to NVSS data:
\begin{equation}
b(z) = 1.6 + 0.85z+0.33z^2 ,
\label{biasTiwari}
\end{equation}
which was adapted by \cite{Bengaly2018} and \cite{Dolfi2019} in the context of TGSS data.
The {\sc CAMB} package allows to include the effects of gravitational lensing and it allows users to include effects of non-linear structure formation via its halo-fit option \citep{Takahashi2012,Mead2015}. 
Additionally we do not use the Limber approximation, which is per default used for $l> 100$. 
An inappropriate application of the Limber approximation gives rise to ringing phenomena in $w(\theta)$ that depend on the details of the binning of the redshift distribution function.
We make use of the cubic-spline interpolation of {\sc CAMB} to generate a smooth window function from the observed redshift distribution for sources with $z\leq2$.

In Fig. \ref{fig:comparisonwthetanohalolensinglinear} we show the two-point correlations from radio sources with available redshift information for the 2 and 4~mJy flux density thresholds and compare them to the predictions of non-linear theory, including the Halofit and count lensing options of the {\sc CAMB} 
package and the bias from Eq. (\ref{biasTiwari}). In order to account for the integral constraint, we calculate it using Eq. (\ref{eq:bias}) 
with the random-random pairs obtained by {\sc TreeCorr} and the expectation from {\sc CAMB} in order to subtract it from the expectation.
We find reasonable agreement for angular separations below a few degrees, for the $2$ and $4$~mJy samples of `mask z', as well as of `mask z1'.
 \emph{The agreement between the theoretical expectations and the results for the $2$ and $4$ mJy samples is remarkable as we did not adjust any model parameter.}

Besides varying the flux density threshold, we can also put the data into several redshift bins, as done previously in Sect. \ref{sec:redshift} and \ref{sec:tpcfredshift}, 
which allows us to test the bias model in more detail.
We compare two scenarios for the 2~mJy `mask z' sample only, as the angular two-point correlation behaves similarly for `mask z' and `mask z1'.
For the first scenario we use of the Halofit option of {\sc CAMB}, together with the bias function $b(z)$ and include the effect of lensing.
We see in the top-left panel of Fig. \ref{fig:comparisonwthetaredshiftbins} that the {\sc CAMB} predictions for redshift bin $z_1$ overestimate 
the amount of correlation 
while we obtain a reasonable agreement for the $z_2$ bin at smallest angular scales and for $z_3$ below $\sim$ 0.8 degrees. A possible explanation is that the bias function (\ref{biasTiwari}), which is based on 
NVSS data, overestimates the amount of bias at lower redshifts for a population mix that includes many more SFGs compared to NVSS. 

In order to test this hypothesis, we compute a second scenario, where we use a constant bias $b(z_1) = b_1 = 1.2$ for the $z_1$ bin, make use of the 
Halofit option of {\sc CAMB} and include lensing. Doing so, the expectation of the first redshift bin is in better agreement below 2~deg with the 
estimated two-point 
correlation function. This indicates that LoTSS-DR1 radio sources at small photometric redshift are almost unbiased tracers of the large scale structure, 
which is to be expected if the sample is dominated by SFGs (which are thought to or assumed to have smaller bias, see e.g. the models used in \citet{SKADS2008} and results from \citet{Hale2018}). This also may relate to selection effects of which sources have associated redshifts, which may preferentially select low redshift SFGs over higher redshift AGN.
We also use a piecewise constant bias of $b_2=2$ and $b_3=3$ for the redshift bins $z_2$ and $z_3$ respectively, which also improve the match of the expectations to the estimated two-point correlation function.

A more detailed study including the precise measurement of the bias functions and cosmological parameters like e.g. $\sigma_8$ are beyond the scope 
of this work, as a good understanding of the uncertainties of the photometric redshift distribution is needed to do so.

\section{Conclusions \label{sec:conclusions}}

We have presented the first statistical analysis of the spatial distribution of radio sources from 
LoTSS, based on the observation of 424 square degrees of the sky. We did so in order to 
characterise the global properties of the survey, check the quality of the LoTSS-DR1 catalogues 
and test whether upcoming data releases will provide promising opportunities to probe cosmology. 
We achieved all three of those goals. 

The data quality was assessed by a suite of tests on top 
of those already presented in \citet{LoTSS2019A} and \citet{LoTSS2019B}. We measured the 
point-source completeness of the survey and found it to be complete to better than 99 per cent 
above a flux density of $1$ mJy. 
We demonstrated that in the mean, source counts are independent from the 
distance from the pointing centre out to an angular separation of approximately $1.6$~deg, which corresponds 
almost to the average effective pointing radius of 1.7 deg, although showing sizeable variation in the counts 
between pointings, see Fig.~\ref{fig:countspointing}. We also showed that source counts 
around the five brightest objects (i.e. $> 10$ Jy) in the LoTSS-DR1 value-added source catalogue
do not show a statistically significant deficit of sources, though they are at the lower end of the 
spread of source counts. Combined with our results for point-source completeness we conclude that LoTSS-DR1 
allows us to probe the radio sky over more than four orders of magnitude in flux density. We 
also demonstrated that the statistical moments of the counts-in-cells distribution of the 
LoTSS-DR1 value-added source catalogue with only the five most 
incomplete pointings and a handful of pixels with less than 5 sources removed, are in 
excellent agreement  with those from the LoTSS-DR1 radio source catalogue masked with our most 
aggressive noise mask that restricts the analysis to low-noise cells (below the median cell-averaged 
rms noise). This assures us of the excellent quality of the pipeline described in detail in 
\citet{LoTSS2019B} and \citet{LoTSS2019C} to construct the value-added source catalogue.

The next step was to measure the statistical moments of the distribution of the radio sources 
in various aspects. We tested if the counts-in-cell tests show any indication of clustering or if 
they agree with a Poisson distribution, the most naive expectation for any sky survey. 
We can exclude with very high confidence that the counts-in-cell are consistent with a 
Poission distribution. The counts-in-cell statistics show a clear signature of clustering, quantified 
by the clustering parameter $n_c$, which is a proxy for the number of cluster members. 
Comparing the radio source catalogue and the value-added source catalogue, in which many 
multi-component sources have been identified and assigned to a single radio source and many 
artefacts have been removed, shows a significant difference in $n_c$ and also in higher statistical 
moments. 
Note that as one increases the flux density threshold, the deviation from a Poisson distribution becomes smaller. 
However, for $S > 10$~mJy and the available 424 square degrees of survey 
area the counts per cell become so small that measuring a deviation from a Poisson distribution is difficult.

For the value-added source catalogue we showed that the simplest compound Poisson distribution in
which each cluster contains a random number of objects that are again Poisson distributed fits the 
data very well. A possible explanation for that finding is that this is due to multiple component 
radio sources, but here the reader should be aware that the statistical test is not able to 
distinguish real multi-component sources, e.g. lobes of radio galaxies, from a SFG with a 
radio artefact in its vicinity or a group of SFGs. As the deviation 
from the Poisson distribution is strongest below flux densities of 1 mJy, it is unlikely that all of the 
clustering is due to real radio sources, at least some of that clustering might still be due to artefacts, 
as presumably the reliability of the LoTSS-DR1 value-added source catalogue reduces when 
the noise level is approached. This hypothesis is also supported by the reduction of $n_c$ when 
only sources with photometric redshift information are used in the data analysis. 
Presumably, fluctuations of the flux density scale between individual pointings also give rise to deviations from a Poissonian source 
distribution. A significant increase of the number of pointings in DR2 will allow us to study this issue in detail.
The clearly detected deviation from the Poisson distribution suggests an additional contribution to the observed large variance of counts of sub-mJy radio sources at higher frequencies, additional to cosmic variance and sample variance \citep{Heywood2013}. 

We further studied the differential source counts of LoTSS-DR1 and compared them to 
other data at low radio frequencies. They are in good agreement above 1 mJy and follow 
the expectations from the SKADS and T-RECS simulations. The photometric redshift estimates for 
about half of all radio sources obtained from 
crossmatching with optical and infra-red observations allow us to get a first impression of the 
redshift distribution of the LoTSS-DR1 sample. It will be important to figure out how representative 
they are also for the other half of the sample. An important step forward in that respect will be the WEAVE-LOFAR survey, which will obtain about a million spectroscopic redshifts.

We also estimated the angular two-point correlation of the LoTSS-DR1 value-added sources in Sect.~\ref{sec:twopoint}. 
Different flux density thresholds and masking strategies lead to slightly varying results.
We conclude that the 2~mJy sample from low-noise regions (mask 1) is the most reliable sample. 
We find less correlation than in the analysis of TGSS-ADR data \citep{RanaBagla2019, Dolfi2019} on all 
scales accessible by LoTSS-DR1, see also Fig. \ref{fig:tgssfitloglog}.

We finally used the distribution of photometric redshifts for about half of all LoTSS-DR1 value-added sources to also compare 
to the Planck 2018 best-fit cosmology, using an off-the shelf bias model and a piecewise constant bias model.
For the angular scales below 6~degrees we find relatively good agreement between our measurements and the 
expectation (no fitted parameters) for the 2 and 4~mJy samples.
A more detailed comparison that also makes use of binned redshift information reveals problems with the bias model of \cite{TiwariNusser2016}
especially at low redshifts, 
which is likely due to the fact that we do not account for the difference between AGNs and SFGs in that analysis.

To conclude, we recover that the radio sky is statistically isotropic at better than one per cent at angular scales above 1 deg and we 
find that large-scale structures as probed with a subset of LoTSS-DR1 sources that have photometric redshifts and a flux density limit of $2$~mJy,
are relatively consistent with the Planck 2018 best-fit cosmology at angular scales below 6~degrees (see 
Fig.~\ref{fig:comparisonwthetanohalolensinglinear}). 
A measurement of cosmological parameters was beyond the scope of this 
work. A next step will be to improve the consistency of the flux density calibration and to quantify and estimate the errors of the measured 
distribution of photometric redshifts. With those two elements improved, in combination with a vastly improved imaging pipeline for DR2 and 
a much larger sky coverage of $5700$~square degrees, we expect that we will start to be able to make interesting cosmological 
tests and measure cosmological parameters based on LoTSS radio sources.

\begin{acknowledgements}
	
This paper is based on data obtained with the International LOFAR Telescope (ILT) under project codes LC2\_038 and LC3\_008.

LOFAR \citep{LOFAR2013} is the Low Frequency Array designed and constructed by ASTRON. It has observing, data processing, and data storage facilities in several countries, which are owned by various parties (each with their own funding sources), and which are collectively operated by the ILT foundation under a joint scientific policy. The ILT resources have benefited from the following recent major funding sources: CNRS-INSU, Observatoire de Paris and Université d'Orléans, France; BMBF, MIWF-NRW, MPG, Germany; Science Foundation Ireland (SFI), Department of Business, Enterprise and Innovation (DBEI), Ireland; NWO, The Netherlands; The Science and Technology Facilities Council, UK; Ministry of Science and Higher Education, Poland; The Istituto Nazionale di Astrofisica (INAF), Italy.

This research made use of the Dutch national e-infrastructure with support of the SURF Cooperative (e-infra 180169) and the LOFAR e-infra group. The Jülich LOFAR Long Term Archive and the German LOFAR network are both coordinated and operated by the Jülich Supercomputing Centre (JSC), and computing resources on the supercomputer JUWELS at JSC were provided by the Gauss Centre for Supercomputing e.V. (grant CHTB00) through the John von Neumann Institute for Computing (NIC).

This research made use of the University of Hertfordshire high-performance computing facility and the LOFAR-UK computing facility located at the University of Hertfordshire and supported by STFC [ST/P000096/1], and of the Italian LOFAR IT computing infrastructure supported and operated by INAF, and by the Physics Department of Turin university (under an agreement with Consorzio Interuniversitario per la Fisica Spaziale) at the C3S Supercomputing Centre, Italy.

This research made use of Astropy, a community-developed core Python package for astronomy \citep{Astropy2013} hosted at http://www.astropy.org/, and of the astropy-based reproject package (\url{http://reproject.readthedocs.io/en/stable/}).
Some of the results in this paper have been derived using the healpy \citep{healpy2019} and HEALPix \citep{Healpix2005} package.
This research made use of DS9 \citep{Ds9}, TOPCAT \citep{TOPCAT}, matplotlib \citep{matplotlib}, NumPy \citep{NumPy}, SciPy \citep{SciPy}, TreeCorr \citep{TreeCorr2004} and tqdm \citep{tqdm}. 

We thank Mike Jarvis for providing details on {\sc TreeCorr} and Anna Bonaldi for discussions 
on T-RECS results. We thank Aritra Basu for useful comments and discussions.

MB, NB, TMS and DJS acknowledge the Research Training Group 1620 `Models of Gravity', 
supported by Deutsche Forschungsgemeinschaft (DFG) and by the German Federal Ministry 
for Science and Research BMBF-Verbundforschungsprojekt D-LOFAR IV (grant number 05A17PBA).
CLH acknowledges funding support from the Science and Technology Facilities Council (STFC) for a PhD studentship [ST/N504233/1].
DB acknowledges funding from STFC grant [ST/N000668/1].
The Leiden team acknowledges support from the ERC Advanced Investigator programme NewClusters 321271.
MJH acknowledges support from STFC [ST/R000905/1].
PNB and JS are grateful for support from the UK STFC via grant ST/R000972/1.
GJW gratefully acknowledges support of an Emeritus Fellowship from The Leverhulme Trust.
WLW also acknowledges support from the CAS-NWO programme for radio astronomy with project number 629.001.024, which is financed by the Netherlands Organisation for Scientific Research (NWO).

\end{acknowledgements}

\bibliographystyle{aa} 
\bibliography{aa}


\begin{appendix} 

\section{Masking of the TGSS-ADR1 radio source catalogue and comparison of angular two-point correlation function \label{app:A}}

\begin{figure}
	\centering
	\includegraphics[width=0.9\linewidth]{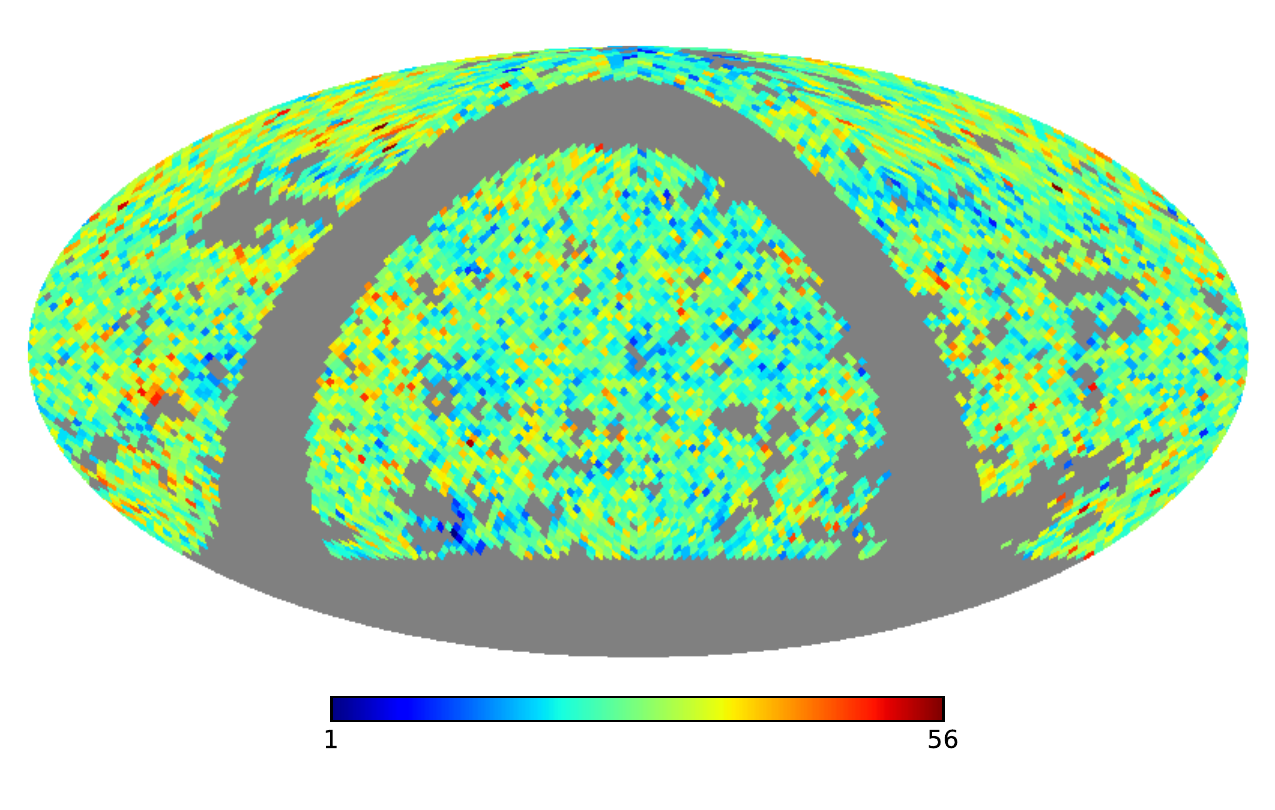}
	\caption{Source count map of the TGSS-ADR1 radio source catalogue with a flux density threshold of $100$~mJy shown in equatorial coordinates and Mollweide projection, the cell size is given by $N_\mathrm{side}=32$.}
	\label{fig:TGSS}
\end{figure}

In order to compare the LoTSS-DR1 value-added catalogue with the TGSS-ADR1 source catalogue
\citep{TGSS2017}, 
it is necessary to also define a mask for TGSS-ADR1. This mask and the source counts per cell at 
$S > 100$~mJy of the TGSS-ADR1 source catalogue are shown in Fig.~\ref{fig:TGSS}, with a sky coverage fraction after masking of  $f_{sky}\simeq 0.64$. 
Since the surface density of sources from TGSS-ADR1 is significantly smaller than from 
LoTSS-DR1 catalogues, we decided to use $N_\mathrm{side} = 32$. Grey regions in Fig.~\ref{fig:TGSS} 
are masked due to a galaxy cut ($|b|\leq 10$~deg), unobserved regions, incompletely 
observed {\sc HEALPix} cells at the boundaries of the survey (Dec $<-53$~deg), 
missing pointings, and cell averaged local noise above $5$~mJy/beam. 
The rms noise is stated to typically deviate between $2.5$ and $5$~mJy/beam, with a median of $3.5$~mJy/beam \citep{TGSS2017}.

The corresponding differential source count is shown in Fig.~\ref{fig:diffnumbercounts} 
and compared to our results from LoTSS-DR1. Above $100$~mJy both source counts agree 
very well. This also confirms the estimates of completeness in \citet{TGSS2017}. Figure 10 of that 
work shows a plot of the completeness of the TGSS-ADR1 source catalogue which we read off to 
be 95 \% at a flux density threshold of $100$~mJy. This completeness estimate is based on 
the detection fraction, which is the fraction of TGSS source counts and SKADS source counts and the completeness is stated to be 50 \% at 25~mJy.

\begin{figure}
\centering
\includegraphics[width=\linewidth]{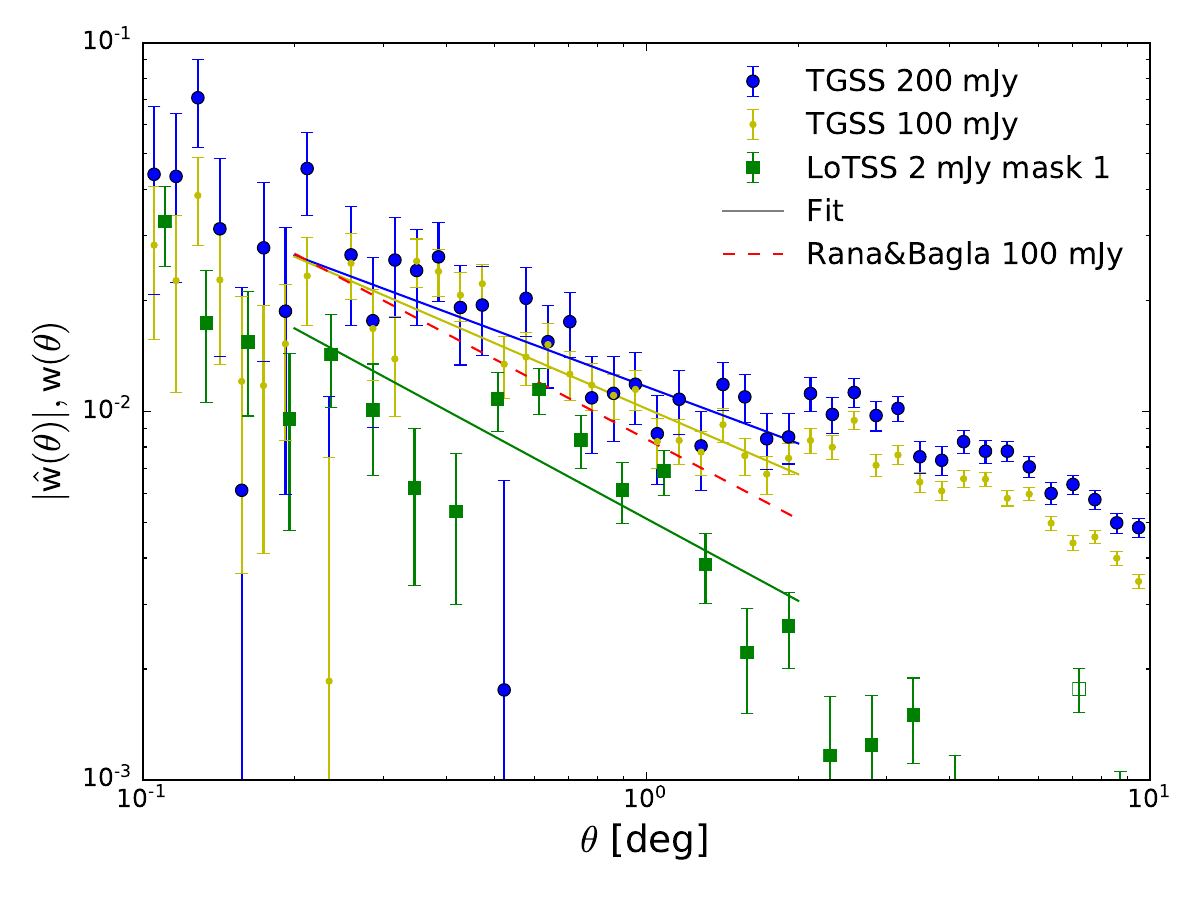}
\caption{Comparison of the two-point correlation function $w(\theta)$ for the TGSS-ADR1 
source catalogue for different flux density thresholds and for the LoTSS-DR1 value-added source 
catalogue. The errors shown are estimates by means 
of {\sc TreeCorr} and represent just statistical errors. 
We fit $w(\theta)$ by a power-law in the range $0.2$~deg $ \leq \theta 
\leq 2$~deg.}
\label{fig:tgssfitloglog}
\end{figure}

We also compare our results for the angular two-point
correlation function for the LoTSS-DR1 $2$~mJy `mask 1' sample (see Sect.~\ref{sec:twopoint}) to 
the masked TGSS-ADR1 at flux density thresholds of $S_{min} = 100$ and $200$~mJy 
(see Fig.~\ref{fig:tgssfitloglog}).
We made use of the error estimations computed by {\sc TreeCorr} using the same settings as in 
Sect.~\ref{sec:twopoint}. The different flux density thresholds give self-consistent results and show 
stronger angular correlations than found from LoTSS-DR1.
For separations between one and 10~degrees, the results of LoTSS and TGSS differ significantly.
Additionally, we fit with a power-law model $w(\theta)=A(\theta/1~\mathrm{deg})^{-\gamma}$ in linear space in the range $0.2$~deg $ \leq \theta 
\leq 2$~deg using {\sc LMFIT}. The results of the fit are included in Table \ref{tab:BestFit}. 
We find somewhat larger angular correlations compared to the results in 
\citet{RanaBagla2019}, which is probably due to the fact that we include cells with averaged 
rms noise of up to 5~mJy/beam, whereas \citet{RanaBagla2019} exclude all cells in 
$N_\mathrm{side}=1024$ resolution that exceed averaged rms noise of 4~mJy/beam. Thus we 
keep more radio sources for the analysis, as can be seen in Tab.~\ref{tab:BestFit}.
Another approach to estimate the angular two point correlation function of the TGSS-ADR1 catalogue was presented by \citet{Dolfi2019}. 
They fitted a power law to small angular seperations 
$\theta \leq 0.1$ deg only and thus no quantitative comparison is shown here. 
To produce a reference catalogue of the TGSS, they masked regions and sources with greater 
rms noise than 5~mJy/beam, declination $< -45$ deg and Galactic latitude $|b|<10$ deg in 
a resolution of $N_\mathrm{side}=512$ and included sources with flux density 
$S \in [200,1000]$~mJy. They find a smaller amplitude $A=(6.5\pm 0.6)\times10^{-5}$,  but much 
steeper slope $\gamma=2.87\pm 0.02$ at $\theta < 0.1$ deg. At $\theta > 0.1$ deg, they also find a 
much flatter slope and see an excess of correlation with respect to the NVSS catalogue, 
but this result is just shown in a figure without quantifying the excess by a number.

\section{Comparison of different estimators for $w(\theta)$ \label{app:B}}

Several estimators have been suggested in the literature for the determination of the 
two-point correlation function. All those estimators are based on counting pairs per bin in 
angular separation $\theta$ and bin width $\Delta \theta$. These pairs are denoted by 
\begin{align}
DD(\theta) &= \frac{\textrm{number of data-data pairs at } \theta \pm \Delta \theta/2}{N_d (N_d - 1)/2},  \\
DR(\theta) &= \frac{\textrm{number of data-random pairs at } \theta \pm \Delta \theta/2}{N_d N_r},  \\
RR(\theta) &= \frac{\textrm{number of random-random pairs at } \theta \pm \Delta \theta/2}{N_r (N_r - 1)/2} ,
\end{align}
where $N_d$ and $N_r$ are the numbers of radio sources (data) and random (or mock) sources 
respectively.

\begin{figure}
\includegraphics[width=\linewidth]{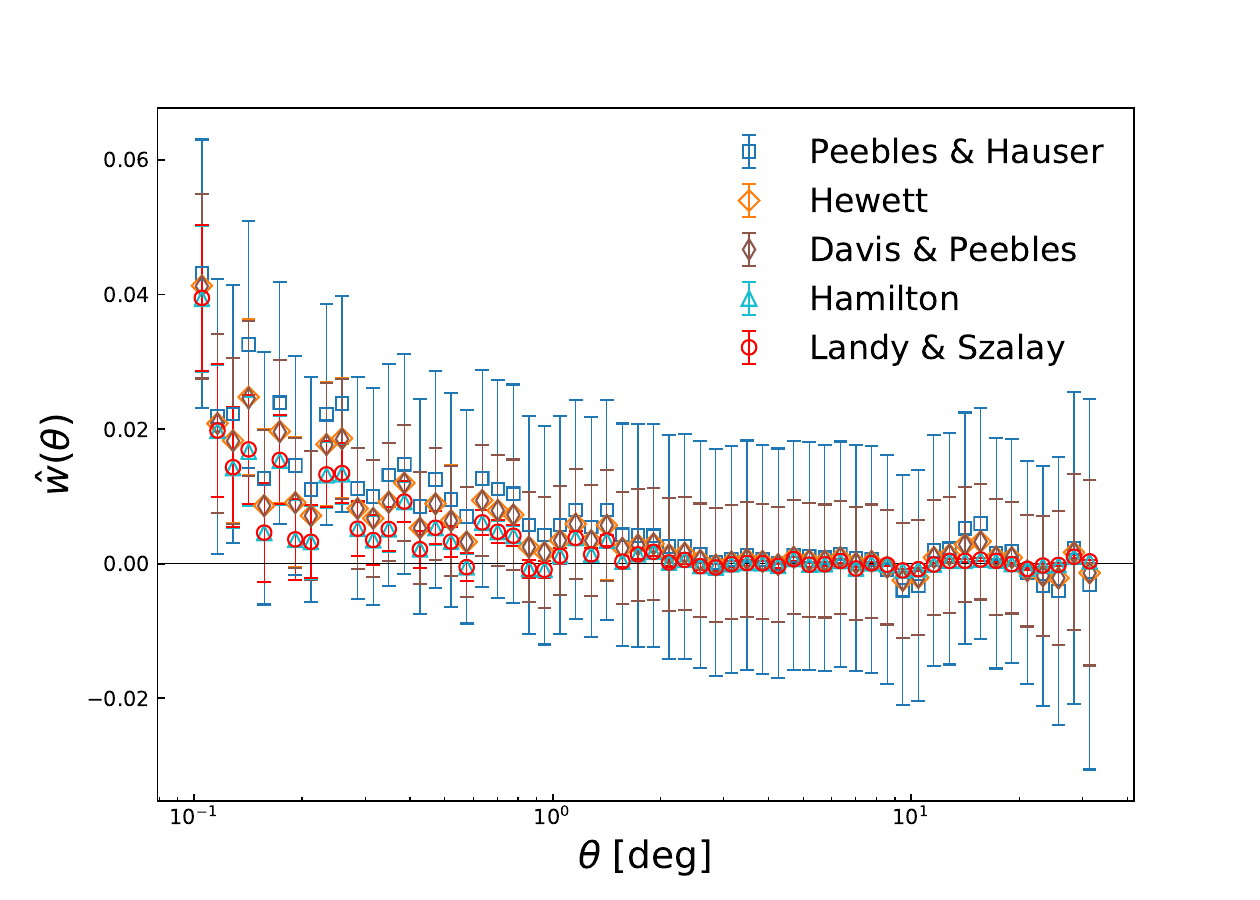}
\caption{Comparison of five different estimators of the angular two-point correlation function 
$w(\theta)$. We evaluate it for the LoTSS-DR1 value-added source catalogue with $S > 4$~mJy 
after applying `mask d'. Here we compare to a truly random catalogue with $N_r = 20 N_d$, rather
than to the mock catalogue of Sect.~\ref{sec:mocks}. 
The errors are obtained via the variances from Table \ref{tab:estimators}.
 \label{fig:estimators}}
\end{figure}

We have written a brute force code to determine $DD, DR$ and $RR$ exactly and to allow us 
to compare the performance of those estimators for the particular LoTSS-DR1 survey 
geometry and to test the accuracy of the software package {\sc TreeCorr} (see App. \ref{app:C}). As a brute force computation of the two-point correlation function is numerically expensive 
(the estimation of the two-point correlation scales with $N_r^2$ and the estimation of its 
variance scales with $N_r^3$), we restricted this tests to the $S> 4$~mJy sample of the 
LoTSS-DR1 value-added source catalogue with `mask d', which had $N_d = 30\,556$ and we used 
$N_r =  20 N_d$ sources from a purely random sample. We also investigated how the performance of different estimators scales for smaller random samples.  

Fig.~\ref{fig:estimators} shows the results for the following estimators:
\begin{align}
\hat w_\mathrm{PH} &\equiv \frac{DD - RR }{RR} & \text{\cite{PeeblesHauser1974}}, \\
\hat w_\mathrm{Hew} &\equiv \frac{DD - DR }{RR} & \text{\cite{Hewett1982}},\\
\hat w_\mathrm{DP} &\equiv \frac{DD - DR }{DR} & \text{\cite{DavisPeebles1983}},\\
\hat w_\mathrm{Ham} &\equiv \frac{DD \times RR - DR^2}{DR^2} & \text{\cite{Hamilton1993}},\\
\hat w_\mathrm{LS} &\equiv \frac{DD - 2DR + RR }{RR} & \text{\cite{LandySzalay1993}},
\end{align} 
For most data bins we find that $|\hat w_\mathrm{PH}| > |\hat w_\mathrm{Hew}| \approx   
|\hat w_\mathrm{DP}| >  |\hat w_\mathrm{Ham}| \approx |\hat w_\mathrm{LS}|$. 

The expected biases and variances of the five estimators are tabulated in Table~\ref{tab:estimators}. The results are expressed in terms of the quantities 
\begin{align}
p &= \frac{2}{N_d (N_d - 1)}\left(\frac{1}{G_p} - 2 \frac{G_t}{G_p^2} + 1\right) \approx \frac{2}{N_d (N_d - 1)}\frac{1}{G_p}, \\
t &=  \frac{1}{N_d}\left(\frac{G_t}{G_p^2} - 1\right),
 \end{align}
where $G_p$ is the fraction of pixel pairs separated by a given angular separation for pixels small enough such that they contain at most a single source. 
$G_t$ is the fraction of triplets given one source at the center and two other at a given angular separation, respectively.

\begin{table}
\centering
\caption{Bias and variance of the five considered estimators of the angular two-point correlation function for the case $N_r \gg N_d$ and assuming $|w(\theta)|$ and $|w_{\Omega}|$ (see Eq. (\ref{eq:bias})) are both small compared to unity.}
    \begin{tabular}{ccc}
    	\hline
	\hline
    	 estimator & bias & variance \\ \hline
    	 
         \addlinespace[1ex] PH & $- w_\Omega$ & $p + 4 t$ \\
          Hew &  $- w_\Omega$ & $p + t$ \\
          DP & $- w_\Omega - t $ & $p + t$ \\
          Ham & $- w_\Omega - t $ & $p$ \\
          LS & $- w_\Omega$ & $p$\\ 
          \hline
      \end{tabular}
    \label{tab:estimators}
\end{table}

Between $2$ and $8$~degrees, all estimators give very similar results, c.f. Fig.~\ref{fig:estimators}.
However, for separations larger than $10$ degrees, the PH estimator shows a significant deviation compared to the results obtained with the Landy \& Szalay estimator and with the Hamilton Estimator. 
The shown errors are underestimates since they are obtained via the variances from Table~\ref{tab:estimators}.
Hence, we assume that $N_r\gg N_d$ and $|w(\theta)|$ and $|w_\Omega |$ (see Eq. (\ref{eq:bias})) are both small compared to unity.

\begin{figure}
\includegraphics[width=\linewidth]{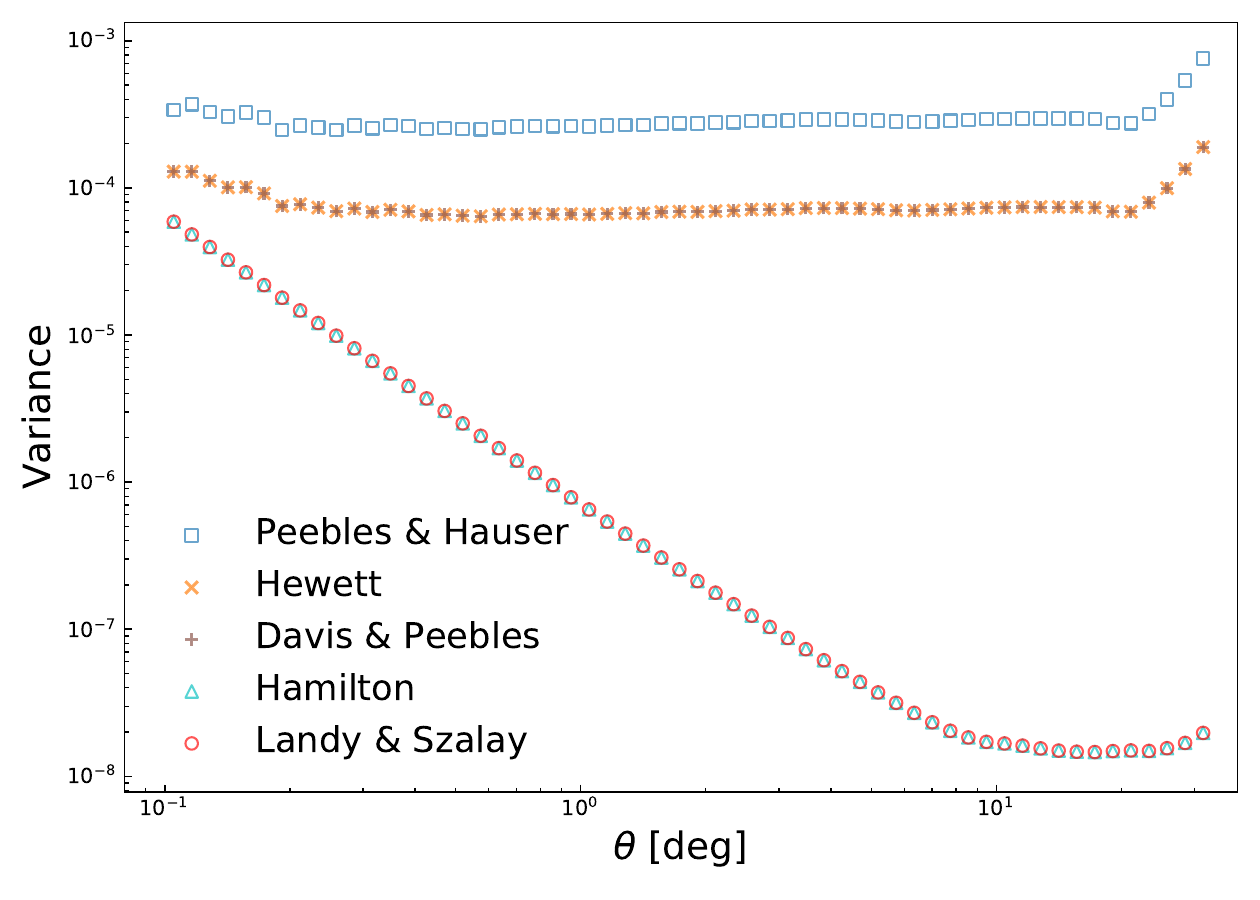} 
\caption{Estimated variance of different estimators for the LoTSS-DR1 
value-added source catalogue at $S > 4$~mJy.
The Landy \& Szalay and Hamilton estimators have identical variance as well as the estimators by Hewett and Davis \& Peebles.}
\label{fig:estimator_variance}
\end{figure}
Fig.~\ref{fig:estimator_variance} shows this expected variance for the different estimators.
Since $G_p$ is equivalent to $RR$, the random pair counts resulting from the use of the random catalogue with $620\, 440$~points can be used to estimate $G_p$.
However, the estimation of $G_t$ scales with $N_r^3$.
Due to the necessary computing time, $G_t$ is estimated via $12$ runs with $3\, 000$~points each.
It can be seen that the contribution of $t$ to the variance is significant at all angular scales and dominates over the contribution of $p$.

Our findings confirm previous studies of the performance of different estimators
\citep{Pons-Borderia1999, Kerscher2000, Vargas-Magana2013}, including the estimators studied 
in this work. These previous studies showed that the LS estimator operates best in almost 
every application, especially for wide separation ranges extending to the large scales, a 
typical feature of current surveys. However, the previous studies had their focus on the study of the 
three dimensional two-point correlation and investigated them in the context of galaxy redshift surveys rather than in the context of radio continuum surveys.

\begin{figure}
\includegraphics[width=\linewidth]{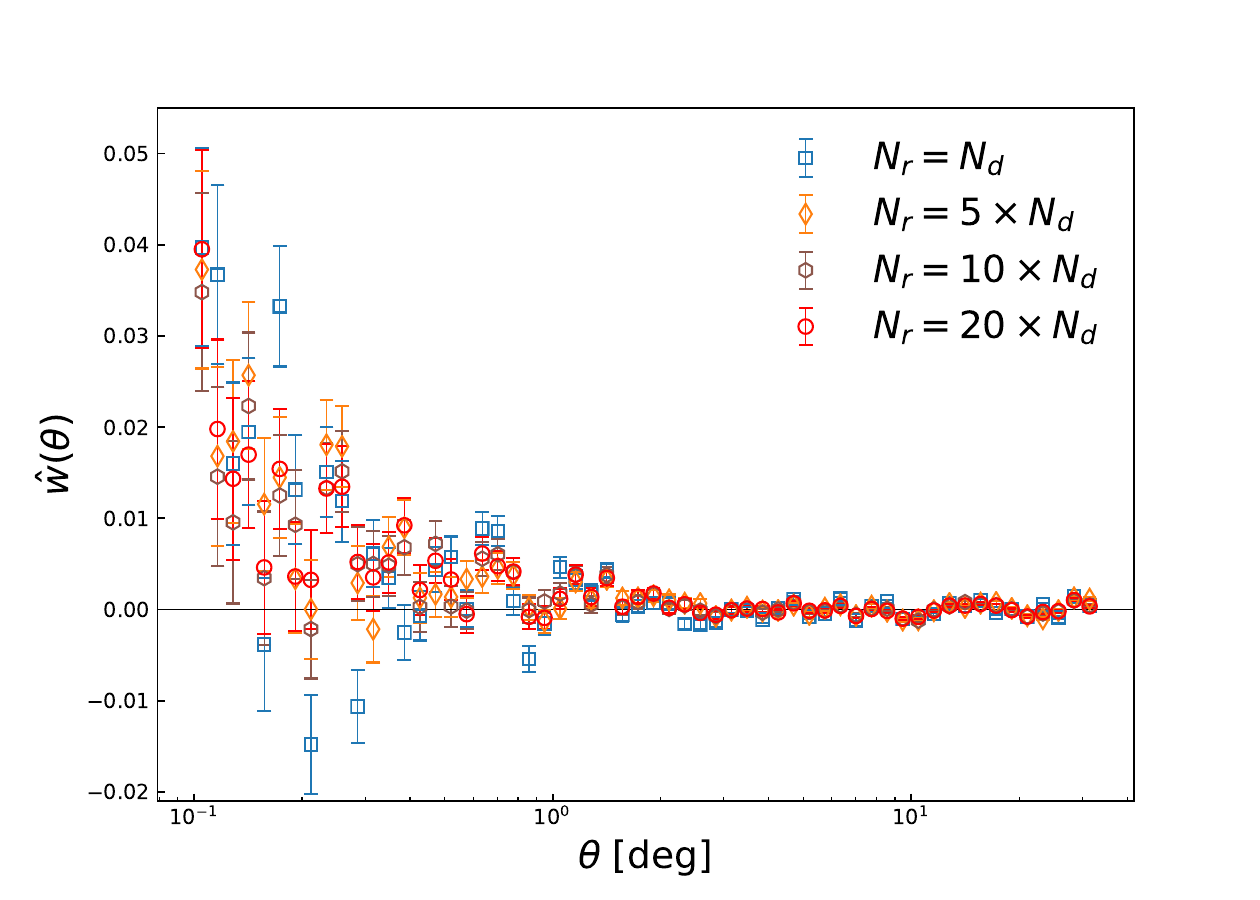} 
\caption{Comparison of the results for the two-point correlation function using the LS estimator 
and various sizes of random catalogues. For separations larger than $1$~degree smaller random 
catalogues give results very similar to those from large random catalogues with $N_r=20N_d$. The errors are estimated from Table \ref{tab:estimators}.}
\label{fig:dependence_nr}
\end{figure}

We also investigated how the results for the Landy and Szalay estimator depend on the sample 
size of the random catalogues. As can be seen in Fig.~\ref{fig:dependence_nr}, there is 
more fluctuation if small random samples are used. Especially at angular separations above 1 deg, 
$N_r = 5 N_d$ does already give rise to reliable estimates.
Therefore, it would be a computational advantage to calculate the two-point correlation function with small random catalogues (but $N_r > N_d$) if only large separations are of interest.

\section{Testing of the TreeCorr Software Package \label{app:C}}

The {\sc TreeCorr} software package (version 3.3) provides various parameters for setting options that 
enhance the accuracy of its computations. By default, {\sc TreeCorr} uses metric distances which 
are only accurate for small separations but are fast to calculate. In this work we also examine 
the two-point correlation function for larger separations up to $32$~degrees. Therefore, 
great-circle distances are used to obtain accurate distance measurements on larger scales.
{\sc TreeCorr} takes this option via \texttt{metric=`Arc'}, which is used throughout the following analysis.

\begin{figure}
\centering
\includegraphics[width=\linewidth]{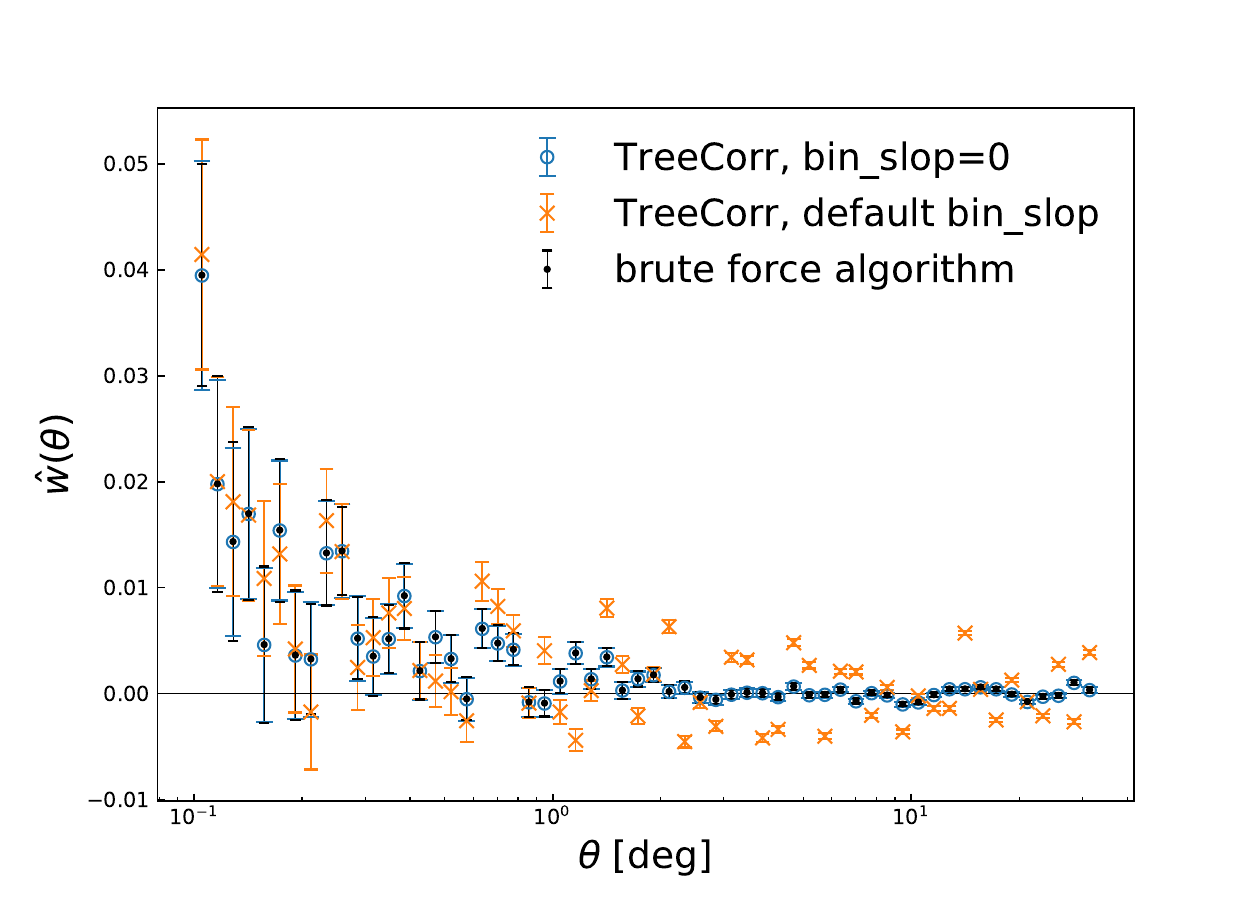}
\caption{Test of the accuracy of {\sc TreeCorr}. We compare the {\sc TreeCorr} default settings 
(orange crosses) and the best possible {\sc TreeCorr} precision (\texttt{bin\_slop=0}) to the results from 
an exact brute force code (black dots). \label{fig:TreeCorr_test1}}
\end{figure}

Furthermore, the accuracy of the algorithm depends on the configuration parameter 
\texttt{bin\_slop}. This parameter controls the accuracy of {\sc TreeCorr} to put pairs in the 
correct angular bin when identifying the many `trees'. 
For the chosen bin width of $\Delta\ln(\theta/1\, \mathrm{deg}) =0.1$ the default value is $1$.
If this parameter is set to zero, as we do for the analysis presented in Sect. \ref{sec:twopoint}, 
{\sc TreeCorr} should give the most accurate result (more information can be found in the 
{\sc TreeCorr}-documentation\footnote{\url{http://rmjarvis.github.io/TreeCorr/html/index.html}}).
Fig. \ref{fig:TreeCorr_test1} shows the values for the two-point correlation function when calculated 
from an exact brute force code (documented in \citet{BiermannThesis}), 
calculated with {\sc TreeCorr}'s default value for \texttt{bin\_slop} and using the best possible 
{\sc TreeCorr} precision, i.e. \texttt{bin\_slop=0}.
By eye, the most precise {\sc TreeCorr} results are indistinguishable from the exact values.
In contrast, the default settings give results that lead to misestimates that are of the order 
of the expected signal at angular scales of $1$ deg. It is clear that the accuracy of the estimates 
should be at least an order of magnitude better than the expected signal. The analysis of NVSS \citep{NVSS1998}
data suggests, that it should be at the level of $10^{-3}$ to $10^{-4}$ at angular scales above 1 deg
\citep{BlakeWall2002, Overzier2003}. Hence, it is essential to modify the default settings of {\sc TreeCorr} to 
calculate accurate results.

Using a smaller value for \texttt{bin\_slop} extends the computing time, obviously.
Using the default accuracy yields a computing time of a few seconds\footnote{16GB RAM, 2.4 GHz Quad-Core Intel Xeon CPUs, using 7 cores},
when using the $4$~mJy flux density threshold and a random catalogue containing 
$N_r=20 N_d$ points, whereas using the brute force setting (\texttt{bin\_slop=0})  takes about 
$70$~minutes. However, obtaining the most accurate results possible with 
{\sc TreeCorr} is still roughly $12$ times faster than using our own brute force algorithm that we used
for the purpose to check the performance of {\sc TreeCorr}.

\begin{figure}
	\centering
	\includegraphics[width=\linewidth]{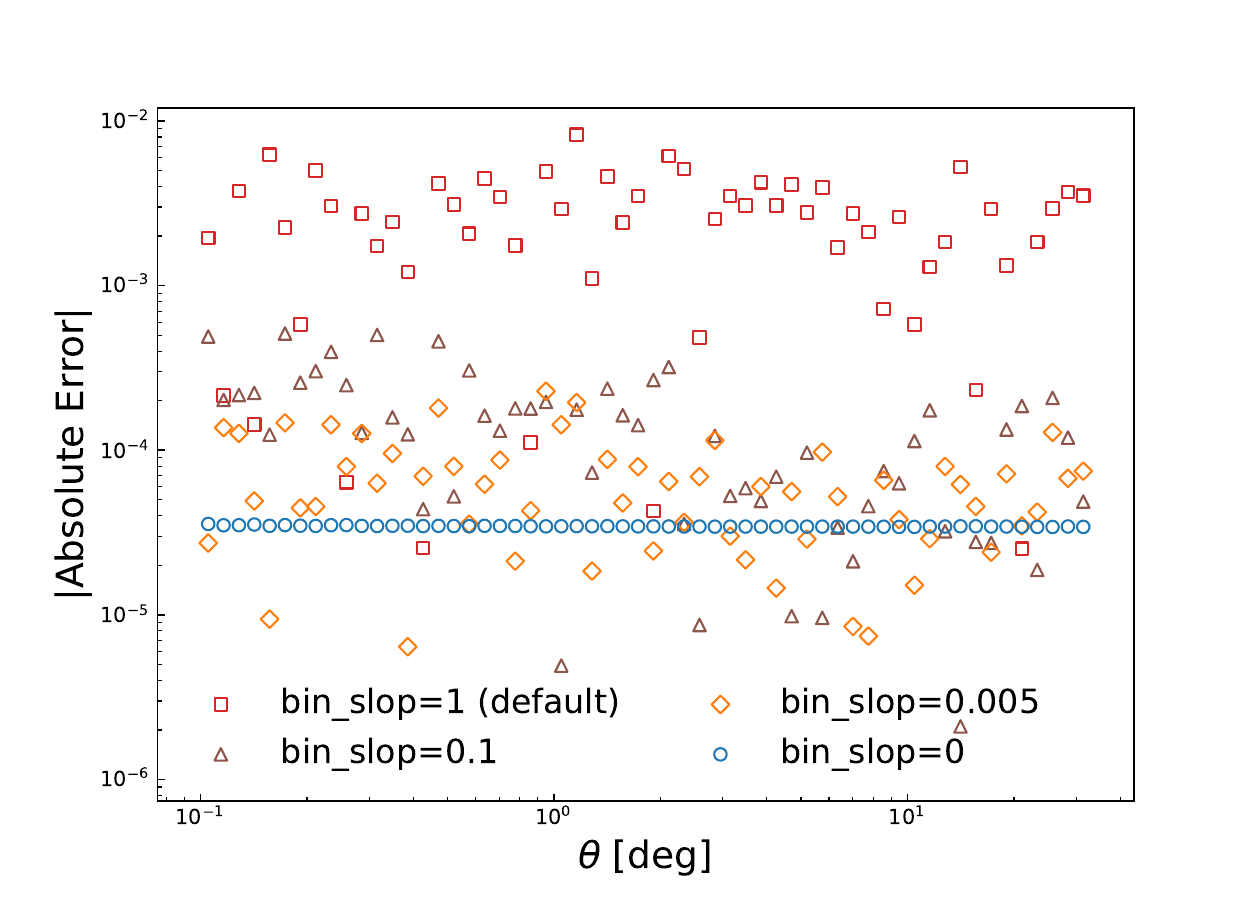}\\
	\includegraphics[width=\linewidth]{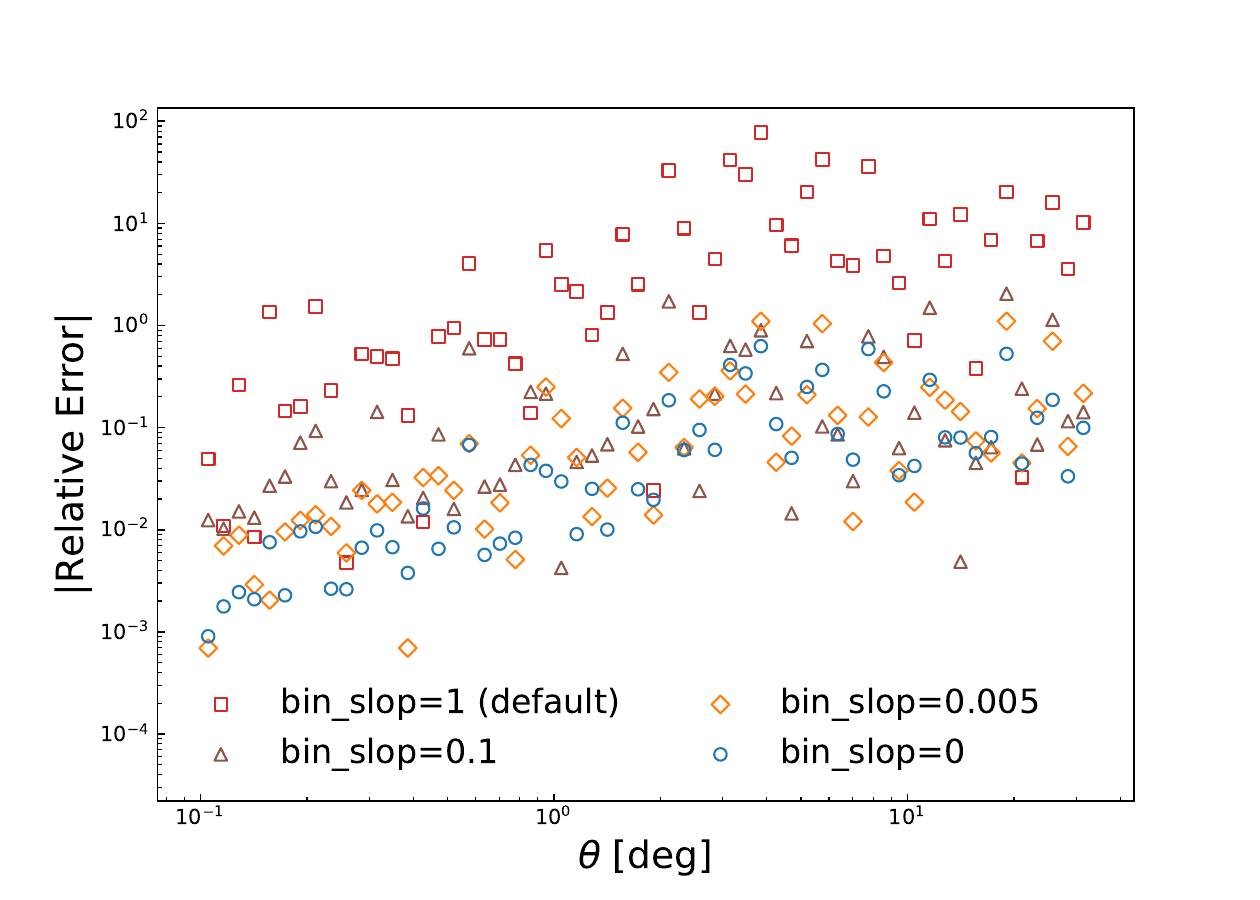}
	\caption{Top: Mean absolute error of {\sc TreeCorr} with respect to the brute force algorithm. We compare the {\sc TreeCorr} default settings 
		(red boxes), two small values for \texttt{bin\_slop} (brown triangles: $0.1$, orange diamonds: $0.05$) and the best possible {\sc TreeCorr} precision (\texttt{bin\_slop=0}, blue circles). Bottom: Relative error of {\sc TreeCorr} with respect to the brute force algorithm, calculated via Eq. (\ref{eq:relativeerror}). \label{fig:TreeCorr_test2}}
\end{figure}

Nonetheless, it is relevant to test other settings for {\sc TreeCorr}'s accuracy since, 
using lower flux density thresholds results in a higher number of sources and larger mock 
catalogues as mentioned in Sect. \ref{sec:onepoint} and \ref{sec:twopoint}. Hence, the 
computational time increases significantly. Note also that a brute 
force estimate of the variance of $\hat w$ scales as $N_r^3$, which poses substantial 
computational challenges for small flux density thresholds in upcoming data releases (we expect 
to lower our flux density threshold below $1$ mJy  for a cosmology analyis of LoTSS-DR2). 
Fig. \ref{fig:TreeCorr_test2} shows the absolute error of TreeCorr results, with respect to the brute force algorithm and using values for the \texttt{bin\_slop} of $1$, $0.1$, $0.05$ and $0$. 
Setting the value of  \texttt{bin\_slop} to $0.05$ is $\sim 9$ times faster and using a value of 
$0.1$ is $\sim 24$ faster than using the most exact settings.
Brute force settings for {\sc TreeCorr} yield an absolute error of about $3.5\times 10^{-5}$ as it is almost constant over the considered separations.
The origin of this constant offset is not further examined.
It could be either caused by limitations of {\sc TreeCorr} or of the brute force algorithm or of both algorithms.
A small value for the \texttt{bin\_slop} can give more precise results in some cases, i.e. for some separations, however, the absolute error shows clear fluctuations.
Using \texttt{bin\_slop=0.1} could result in absolute errors as high as the correlation function at separations of $3$ deg and larger.
Additionally we show the relative error of {\sc TreeCorr} with respect to the brute force algorithm in Fig. \ref{fig:TreeCorr_test2}.
The relative error is calculated via:
\begin{equation}\label{eq:relativeerror}
\Big| \frac{\hat{w}_{\text{boot}} (\theta)-\hat{w}_{\text{\sc TreeCorr}} (\theta)}{\hat{w}_{\text{boot}} (\theta)}\Big|
\end{equation}

\begin{figure}
	\centering
	\includegraphics[width=\linewidth]{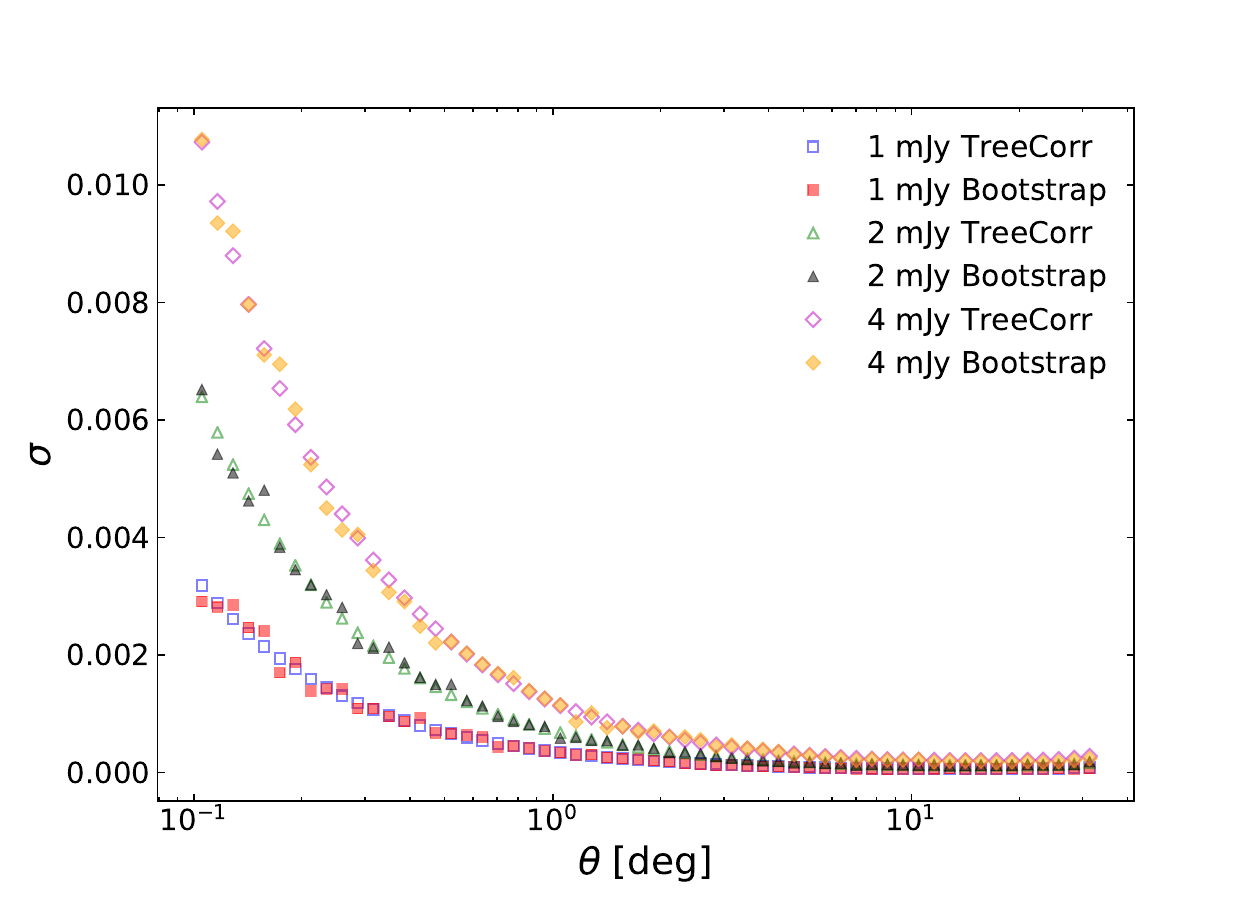}
	\caption{Comparison of errors calculated by {\sc TreeCorr} using \texttt{bin\_slop}$=0$ and by means of 100 bootstraps for the LoTSS-DR1 value-added catalogue after masking with `mask d'.}
	\label{fig:booterrors}
\end{figure}

In order to test the error computation by {\sc TreeCorr} we additionally estimate the error in our measurement of $ w (\theta)$ for each bin via bootstrap re-sampling method as described by \cite{Ling1986}. 
For this we use 100 pseudo-random samples, of the same size as the original catalogue, generated by randomly choosing sources with replacement from the original catalogue. 
We then compute the angular two-point correlation function ($\hat{w}^{i}_{\text{boot}} (\theta)$) for each sub-sample and the bootstrap errors as the standard deviation given by the following equation.,
\begin{equation}
\sigma_{w}(\theta) = \sqrt{ \frac{1}{N-1} \sum_{i=1}^{N}\left(\hat{w}^{i}_{\text{boot}} (\theta) - w_{0} (\theta)\right)^{2}}.
\end{equation}
where $ w_{0} (\theta)$ is the mean value for the sub-samples and $N$ the total count.
Both error estimations for the LoTSS-DR1 value-added catalogue, masked with `mask d' and flux density thresholds of 1, 2 and 4~mJy are shown in Fig. \ref{fig:booterrors}.
The error estimate by {\sc TreeCorr} using \texttt{bin\_slop}$=0$ and bootstrapping agree within all three flux density thresholds and the difference between both is in maximum of order $4\times 10^{-4}$ in the range $0.1\leq\theta\leq 2$ deg and of order $10^{-5}$ for larger separations.
Therefore we decide to use error estimates done by {\sc TreeCorr} with \texttt{bin\_slop}$=0$ in our analysis.

The above findings are valid when using LoTSS-DR1 data and may vary for different surveys.

\section{Kinematic radio dipole}\label{app:dipole}

Following \cite{EllisBaldwin1984}, the kinematic radio dipole, which is due to the proper motion of the Solar system, 
contributes to the source counts per solid angle with a Doppler shift of the emitted radiation from a source and the aberration of the observed 
source positions, i.e.
\begin{equation}
\label{eq:boostedcounts}
\left(\frac{\text{d}N}{\text{d}\Omega}\right)_{\mathrm{obs}} = 
\left(\frac{\text{d}N}{\text{d}\Omega}\right)_{\mathrm{rest}} \left[1+[2+x(1+\alpha)]\beta\cos\theta\right],
\end{equation}
with $\beta =v_\odot/c$ and $x$ defined as:
\begin{equation}
x\equiv - \frac{\text{d}\ln N}{\text{d}\ln S}.
\end{equation}
The amplitude of the kinematic dipole is given by
\begin{equation}
d=[2+x(1+\alpha)]\beta,  
\end{equation}
and $\theta$ measures the angle between the position of a source and the direction of Sun's proper motion. 

\begin{figure}
\centering
\includegraphics[width=\linewidth]{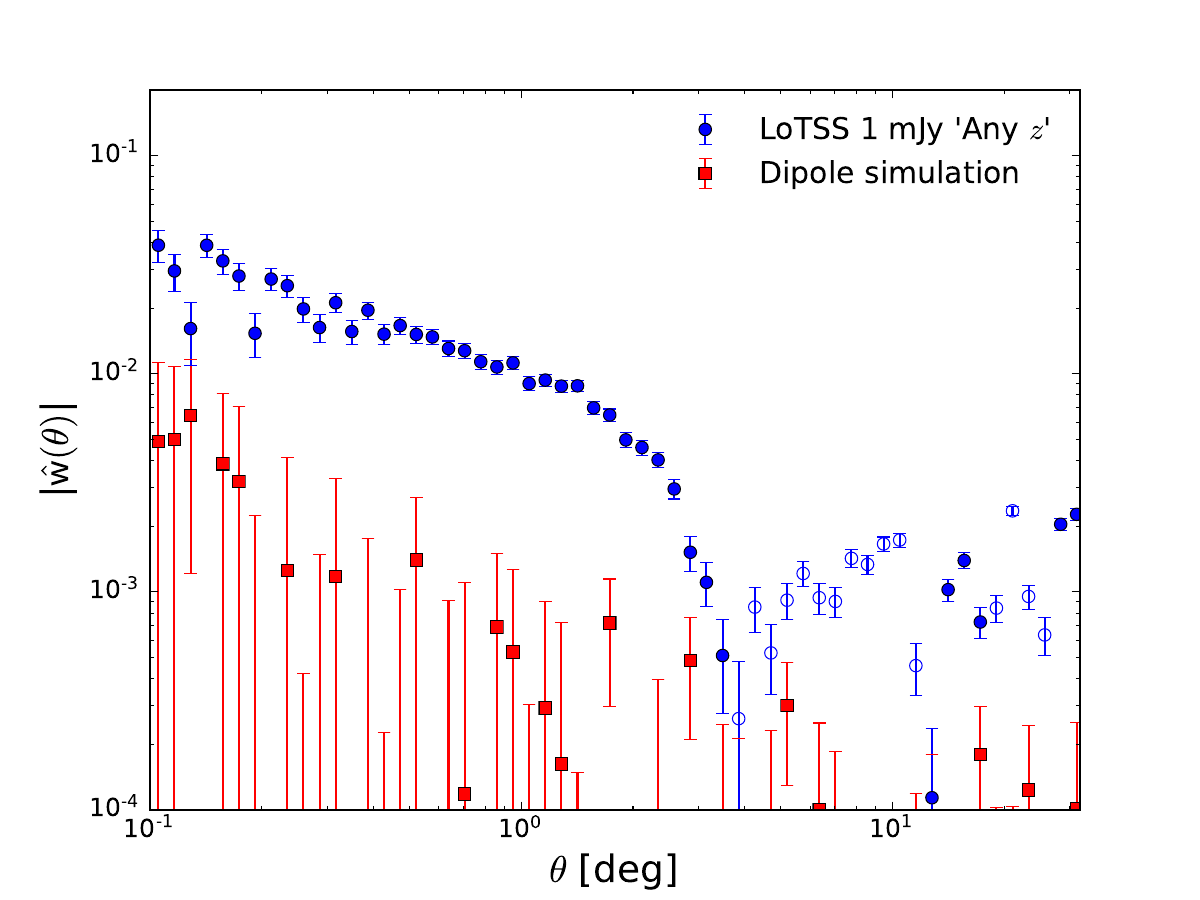}
\caption{Comparison of the two-point correlation function $w(\theta)$ for the 1~mJy `Any $z$' sample of the LoTSS-DR1 value added source catalogue and 
a simulated sky with contribution from a kinematic dipole. Negative values are shown with open symbols.}
\label{fig:dipolecomparisonlog}
\end{figure} 

To estimate the contribution of the kinematic radio dipole to the angular two-point correlation function we follow the procedure of 
\citet{Rubart} and first generate a sky of random sources with associated random flux densities. 
The spherical coordinate positions $(\Phi, \Theta)$ of simulated sources are drawn randomly by:
\begin{align}
\Phi&= 2\pi\cdot \texttt{random(0,1)}\\
\Theta &= \arccos(1-2\cdot\texttt{random(0,1)})
\end{align}
Using this definition, we already fulfil the convention of Co-Latitude necessary for {\sc HEALPix}.
Additionally we generate random flux densities:
\begin{equation}
S = S_0(1-\texttt{random(0,1)})^{-x}.
\end{equation}
We fix $S_0 = 0.9$ mJy, such that we can apply a flux density threshold after boosting of 1 mJy.  

We then calculate boost and aberration for each individual source, where we use the latest findings of Planck \citep{PlanckA2018}.
They infer the proper motion of the Sun to be $\mathrm{v_\odot} = 369.82 \pm 0.11$~km/s towards 
$(167.942\pm0.011,\ -6.944\pm0.005)$~deg in Equatorial coordinates (J2000), which results in a kinematic radio dipole amplitude of:
\begin{equation}
d = 4.63 \times10^{-3},
\end{equation}
where we assumed typical values of $x=1$ and $\alpha = 0.75$ for the boosting.
 
From this boosted simulation we estimate the angular two-point correlation using the same settings as described in Sect. \ref{sec:tpcf_intro}, where the total number of simulated sources is fixed to the amount of sources in the LoTSS-DR1 `Any $z$' sample, together with a pure random sample.
The results of this estimation is compared in Fig. \ref{fig:dipolecomparisonlog} to the LoTSS-DR1 `Any $z$' sample with a $1$~mJy flux density threshold. 
We see an order of magnitude lower correlation than observed in the actual data sample and therefore neglect the dipole term in our theoretical expectation.

\end{appendix}

\end{document}